\documentclass[showpacs,amsmath,amssymb,aps,twocolumn,superscriptaddress,prx]{revtex4-2} 
\usepackage{mathtools}
\usepackage{algorithm}
\usepackage{algpseudocode}
\usepackage[pdftex]{graphicx} \graphicspath{{}}
\usepackage{grffile} 

\usepackage{comment}

\usepackage{xspace}

\usepackage{float}
\usepackage{dcolumn}
\usepackage{bm}
\usepackage[pdftex,colorlinks=true]{hyperref}
\hypersetup{
	colorlinks=true,
	linkcolor=blue,
	urlcolor=cyan,
}
\usepackage{ulem}

\usepackage{color}
\definecolor{ForestGreen}{RGB}{34, 139, 34}

\newcommand{\HS}{HS\xspace}

\begin{document}

\title{Engineering Hierarchical Symmetries}

\author{Zhanpeng Fu}
\affiliation{Max Planck Institute for the Physics of Complex Systems, N\"{o}thnitzer Str.~38, 01187 Dresden, Germany}
\affiliation{Zhiyuan College, Shanghai Jiao Tong University, 200240, Shanghai China}
\affiliation{Institute for Advanced Study, Tsinghua University, Beijing 100084, People’s Republic of China}
	
\author{Roderich Moessner}
\affiliation{Max Planck Institute for the Physics of Complex Systems, N\"{o}thnitzer Str.~38, 01187 Dresden, Germany}

\author{Hongzheng Zhao}
\email{hzhao@pku.edu.cn}
\affiliation{Max Planck Institute for the Physics of Complex Systems, N\"{o}thnitzer Str.~38, 01187 Dresden, Germany}
\affiliation{School of Physics, Peking University, 100871, Beijing  China}
 
\author{Marin Bukov}
\affiliation{Max Planck Institute for the Physics of Complex Systems, N\"{o}thnitzer Str.~38, 01187 Dresden, Germany}
	
\date{\today}

\begin{abstract}
The capacity to custom tailor the properties of quantum matter and materials is a central requirement for enlarging their range of possible functionalities. A particularly promising route is the use of driving protocols to engineer specific desired properties with a high degree of control and flexibility.  Here, we present such a program for the tunable generation of sequences of symmetries on controllable timescales. Concretely, our general driving protocol for many-body systems generates a sequence of prethermal regimes, each exhibiting a lower symmetry than the preceding one.  We provide an explicit construction of effective Hamiltonians exhibiting these symmetries, which imprints emergent quasiconservation laws hierarchically, enabling us to engineer the respective symmetries and concomitant orders in nonequilibrium matter. We provide explicit examples, including spatiotemporal and topological phenomena, as well as a spin chain realizing the symmetry ladder $\text{SU(2)}{\rightarrow}\text{U(1)} {\rightarrow} \mathbb{Z}_2{\rightarrow} E$.
Our results have direct applications in experiments with quantum simulators. 
\end{abstract}

\maketitle
\let\oldaddcontentsline\addcontentsline
\renewcommand{\addcontentsline}[3]{}


\section{Introduction}

Symmetry is ubiquitous in nature, and it underpins intriguing and fundamental phenomena including the existence of conservation laws, integrability, and the classification of phases of matter and  transitions between them~\cite{landau2013statistical,chaikin1995principles}; it is a crucial component of a plethora of topological phenomena~\cite{pollmann2012symmetry,chen2013symmetry}. 
Therefore, exploring protocols to engineer a desired symmetry and control its breaking, as well as investigating emergent phenomena associated with engineered symmetries, has attracted long-standing interest in both fundamental physics~\cite{anderson1952approximate,dine1981simple,brauner2010spontaneous,castelnovo2012spin,abanin2015exponentially,kuwahara2016floquet,bertini2021finite} and quantum engineering~\cite{martinez2016real,ji2020quantum,jepsen2021transverse}.

Recently, time-dependent protocols were proposed to Floquet-engineer symmetry as an emergent phenomenon~\cite{oka2019floquet,schweizer2019floquet,geier2021floquet,petiziol2022non,kalinowski2023non,jin2023fractionalized,sun2023engineering}, leading to the discovery of nonequilibrium phases of matter~\cite{kitagawa2010topological,potter2016classification,titum2016anomalous,khemani2016phase,else2016floquet,yao2017discrete,else2020long,dumitrescu2022dynamical,zhang2022digital,mi2022time}. However, little has been known about how to engineer sequences of different symmetries in a simple and controlled setting, which is a question of considerable importance for a variety of reasons. 
In statistical physics, symmetries can significantly impact how a system reaches thermal equilibrium~\cite{d2016quantum,vidmar2016generalized,hainaut2018controlling,haldar2018onset,agrawal2022entanglement,murthy2023non,kranzl2023experimental}. Moreover, temporal sequences with specific symmetry content can be used to stabilize order in Floquet-engineered matter~\cite{luitz2020prethermalization,beatrez2023critical}. 
They can also give rise to an interesting interplay of spontaneous with explicit symmetry breaking: From a practical perspective, engineered time-dependent symmetries can potentially enhance the control over wanted or unwanted spontaneous symmetry-breaking (SSB) processes on real quantum devices~\cite{brauner2010spontaneous,trenkwalder2016quantum,garcia2019spontaneous,kokail2019self,choi2020robust,tran2021faster,richter2021simulating,keenan2023evidence,zhao2023making}.

In this work we study the engineering of hierarchical symmetries (\HS) in a time-dependent setup: We investigate whether or not, and under which conditions, a sequence of emergent symmetries can be engineered to occur hierarchically in time, in a controllable way.
Realizing such HS in time-dependent systems is a demanding challenge for the following reasons: 
(1) \textit{A priori}, explicit symmetry-breaking processes do not, in general, preserve any subgroup structure, and they introduce transitions among all possible symmetry sectors; 
(2) because of the absence of energy conservation in time-dependent systems, heating can further speed up the destruction of manifestations of symmetries, in particular quickly degrading any features sensitive to symmetry, e.g., melting any spontaneous symmetry-breaking order.

Here, we propose a way to overcome these difficulties and construct a generic protocol to realize HS in driven many-body systems; it applies to any hierarchical symmetry group structure, irrespective of the specific microscopic details of the underlying model. It is explicit in that we provide a general scheme for realizing any sequence of \HS. In addition, this construction is not limited to Floquet systems; it also applies to more general time dependence, e.g., quasiperiodically~\cite{zhao2019floquet,else2020long,lapierre2020fine,wen2021periodically,mori2021rigorous,long2022many,nathan2022topological,he2023quasi}, and even some randomly~\cite{zhao2021random,guarnieri2022time}, driven systems.

\begin{figure*}[t!]
    \centering
    \includegraphics[width = \linewidth]{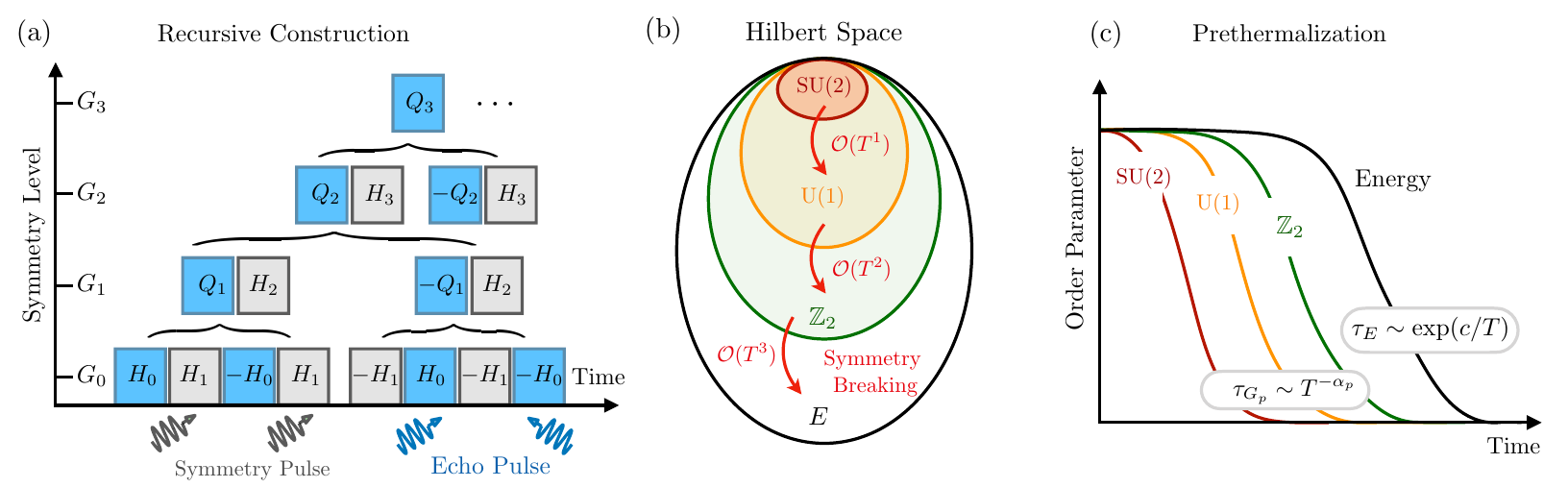}
    \caption{
    (a) Hierarchical symmetry breaking engineered via a recursive construction, where the ``symmetry pulse'' imposes a higher symmetry structure while the ``echo pulse'' cancels unwanted symmetry-breaking processes order by order.
    (b) Hilbert space showing a paradigmatic example of the symmetry ladder $\text{SU(2)}\rightarrow\text{U(1)} \rightarrow \mathbb{Z}_2\rightarrow E$, with $E$ denoting the trivial group. The symmetry-breaking process can be parametrically suppressed by using a smaller driving period $T$. (c) Schematic for the sequence of prethermal steady states, exhibiting a lower symmetry than the preceding one. Their lifetimes scale algebraically with $T$, while heating in energy can be exponentially suppressed.
    }
    \label{fig:scheme}
\end{figure*}

The key conceptual ingredient is a recursive time-dependent ansatz, in which unwanted processes, breaking a desired higher symmetry explicitly, cancel hierarchically in the high-frequency regime, cf.~Fig.~\ref{fig:scheme}(a). Therefore, different symmetry-breaking effects only become noticeable beyond a sequence of long timescales, which leads to a corresponding sequence of prethermal steady states with controllable lifetimes, each exhibiting a lower symmetry than the preceding one, cf.~Figs.~\ref{fig:scheme}(b) and 1(c) for an example. 
Our protocol thus also allows us to imprint emergent quasiconservation laws hierarchically. Further, in conjunction with the process of spontaneous symmetry breaking, our scheme enables the engineering of different types of prethermal nonequilibrium order within the same time evolution.

Our account is structured as follows. We first present a definition of hierarchical symmetry, and discuss its realization using nonequilibrium drives in Sec.~\ref{sec:2}. This approach establishes the conceptual framework and as a result identifies the central ingredients of a pair of protocols, one of which is of operational simplicity and efficiency, while the other  is of complete generality. 
In Sec.~\ref{sec:applns}, we lay out an intuitive picture of our central result, illustrated by three explicit applications:
(i)~a spin chain with three, engineered, distinct, prethermal steady states, characterized by continuous non-Abelian SU(2), Abelian U(1), and discrete $\mathbb{Z}_2$ symmetry, respectively;
(ii)~a quantum clock model featuring a dynamical crossover between prethermal steady states without an equilibrium counterpart that exhibit $\mathbb{Z}_4$ and $\mathbb{Z}_2$ time crystalline order; and
(iii)~a free fermion system supporting a change in topology between a topological insulator (TI) and a higher-order topological insulator (HOTI) upon breaking time-reversal symmetry, as exemplified by the dynamical reduction of edge modes to corner modes in successive prethermal steady-states. 
In Sec.~\ref{sec:Exp}, we discuss a potential experimental realization.
We close with a summary of our results and an outlook for future applications in Sec.~\ref{sec:outlook}. Copious technical details are covered in the appendixes.

\section{Implementing Hierarchical Symmetries}
\label{sec:2}

\subsection{General recursive driving protocol}

We first present our most fundamental result, namely, an explicit general scheme for obtaining the \HS for symmetry ladders as mentioned above.
We use the example of periodic drives, but the constructions proposed below are not limited to Floquet systems with step drives, they also apply to continuous drives
\footnote{The proof in \ref{sec:sm1} relies on the Baker-Campbell-Hausdorff (BCH) formula which is generalized to continuous drives via the Floquet-Magnus expansion. Since in the latter, the commutator structure decouples from the time-ordered integrals, HS can also be engineered for continuous drives.}
and even quasiperiodically or randomly driven systems, as long as the dynamics can be approximated in a perturbative expansion by an effective Hamiltonian.

Consider a family of periodically driven systems whose evolution operator over one period, $U_F$, is defined by concatenating $l$ (different) time-evolution operators:
\begin{equation}
\label{eq:U_F}
    U_F  = U_l U_{l-1} \cdots U_1 \equiv e^{-iTQ}.
\end{equation}
Whenever the drive frequency $\omega=2\pi/T$ is large compared to the typical local energy scales, periodically (and even randomly) driven systems exhibit a long-lived prethermal plateau~\cite{bukov2015universal,mori2016rigorous,abanin2017effective,else2020long,zhao2021random} as a result of energy quasi-conservation. The dynamics can be approximated by a static effective Hamiltonian $Q_{[M]}$, obtained by means of the inverse-frequency expansion (IFE)~\cite{bukov2015universal}:
\begin{equation}
\label{eq:Q}
    Q_{[M]} = \sum_{m=0}^{M}Q^{(m)}, \qquad Q^{(m)} \propto T^m,
\end{equation} 
with $M$ denoting the truncation order. Hence, generic ergodic systems evolve into a prethermal metastable state, described by the generalized canonical ensemble (GCE), 
$
    \hat{\rho}_{\mathrm{GCE}}{\sim}{\exp{(-\sum_\alpha \lambda_\alpha C_\alpha)}},
$
with conserved quantities $C_\alpha$ associated with $Q_{[M]}$, and the Lagrange multipliers $\{\lambda_{\alpha}\}$ fixed by the initial state~\cite{vidmar2016generalized,kuwahara2016floquet}. We aim to construct a generic protocol that inscribes a structure of hierarchical symmetries into $Q_{[M]}$, one corresponding to each order $Q^{(m)}$ of the IFE.

Concretely, consider a finite set of Hamiltonian generators $H_{n}, H_{n-1}, \dots, H_0$ and an associated ladder of symmetry groups $G_{n}{\supset} G_{n-1}{\supset}{\cdots} {\supset} G_0$. Each Hamiltonian $H_{p}$ preserves the corresponding symmetry group $G_p$ so that $[H_{p}, S_{q}] {=} 0$ for all generators $S_q$ of $G_q$ with $q{\le}p$; equivalently, each Hamiltonian $H_p$ breaks only one subsymmetry of the symmetry ladder, reducing the symmetry group $G_{p+1}$ to $G_{p}$.

We realize such a symmetry hierarchy dynamically by imprinting it iteratively in the structure of the effective Hamiltonian~\eqref{eq:Q}, order by order in the IFE. The Floquet unitary at (hierarchy) level $n$ is constructed recursively as
\begin{eqnarray}
\label{eq:general_seq}
     U_{F,n} &=& e^{-il_{n-1} T Q_{n-1} } e^{-i T H_n} e^{+i l_{n-1} T Q_{n-1} } e^{-iT H_n} 
\end{eqnarray}
where $l_{n} {=} 3{\times}2^{n}{-}2$ is the length of the drive sequence and we set $Q_0 {=} H_0$, cf.~Fig.~\ref{fig:scheme}(a). The effective stroboscopic generator at level $n$ is defined via the relation, $U_{F,n} {\equiv} e^{-i l_n T Q_n}$ \footnote{Note the difference in notation between $Q_{[M]}$ which denotes the effective IFE Hamiltonian truncated to order $M$, and $Q_n$ which conserves the symmetry group $G_n$.}. 
If $H_0$ preserves no symmetry, i.e., $G_0$ corresponds to the trivial group, $U_{F,n}$ also preserves no symmetry.
In Sec.~\ref{sec:sm1}, we prove, by induction, that the corresponding IFE approximation at order $m$, $Q_n^{(m)}{\propto}T^m$ [cf.~Eq.~\eqref{eq:Q}] preserves the symmetry group $G_{n-m}$, $(n{-}m{\geq}0)$
and explicitly breaks all higher symmetries up the ladder. 
The key ingredient of this construction is that, in Eq.~\eqref{eq:general_seq}, the prefactors in front of the two $Q_{n-1}$ operators differ by a sign, ensuring the exact cancellation of $G_n$ symmetry-breaking terms in the leading order (time average) $Q_n^{(0)}$.

To observe the effect in the dynamics, note that if we start from a symmetry-broken initial state, i.e., the order parameters, or if conserved quantities associated with each symmetry exhibit nonzero expectation values, these symmetries will be revealed in the dynamics of the system via the occurrence of a hierarchical series of prethermal plateaus, cf.~Fig.~\ref{fig:scheme}(c).  
Only when these conserved quantities have nonzero expectation values in the initial state can we track their long-time decay towards the trivial value at infinite temperature; hence, this defines a requirement for the initial states we will consider below.

Dynamically, we expect that different prethermal plateaus exhibit hierarchical lifetimes that can be systematically prolonged by simply increasing the driving frequency. For a small driving period, the corresponding timescales are well separated since symmetry breaking occurs hierarchically, i.e., order by order in the IFE (see Sec.~\ref{sec:applns} for a more detailed discussion).

\subsection{Illustration for small hierarchical symmetry groups}
Let us explicitly illustrate the mechanism behind \HS using a concrete example for small $n$. For $n{=}1$, we have 
\begin{equation}
\label{eq.Uf1}
    U_{F,1} {=} e^{-iH_0T}e^{-iH_1T}e^{iH_0T}e^{-iH_1T}.
\end{equation}
The effective Hamiltonian to leading order consists of $Q_1^{(0)} {\sim} H_1$, which preserves $G_1$, and $Q_1^{(1)} {\sim} T[H_1, H_0]$ which reduces $G_1$ to $G_0$. Notice that the opposite signs in the prefactors in front of the two $H_{0}$ generators in the drive sequence ensure the exact cancellation of the $G_1$-breaking terms $H_1$ in the leading order $Q_1^{(0)}$; they only become effective at $\mathcal{O}(T)$ in $Q_1^{(1)}$. 

At hierarchy level $n{=}2$, we introduce a new Hamiltonian $H_2$ that preserves $G_2$ (and hence also its subgroups $G_1$ and $G_0$). The new time-evolution operator is 
$
    U_{F,2} {=} e^{-iQ_1l_1T}e^{-iH_2T}e^{iQ_1l_1T}e^{-iH_2T}.
$
Now, $Q_2^{(0)} {\sim} H_2$ preserves $G_2$, while $Q_2^{(1)}{\sim} T[H_2, H_1]$ reduces $G_2$ to $G_1$; finally, $Q_2^{(2)}$ contains a term
$T^2[H_2,[H_1,H_0]]$, which explicitly breaks $G_1$ to $G_0$. 

Note that, in practice, $e^{iQ_{1}l_{1}T}$ can be implemented by reversing the order of the temporal sequence of Eq.~\eqref{eq.Uf1} and conjugating each individual driving element (i.e., going backward in time); this operation is generally accessible on current quantum computing platforms~\cite{mi2022time}. In Appendix~\ref{sec.manipulate}, we present a generalization of our drive protocol amenable to Trotterization--a widely used technique to implement Hamiltonian dynamics on quantum computing platforms.

This construction can be performed recursively for higher $n$; remarkably, for each successive order of the IFE of the \HS protocol in Eq.~\eqref{eq:general_seq}, $Q_n^{(m)}$ breaks only the corresponding successive subgroup, as desired. Note that we do not make any assumptions about the microscopic details of the generators $H_n$; hence, our construction is completely general and applies to any hierarchical symmetry group structure, making it widely applicable.

\subsection{Sequence length and shortening}
Because of its recursive character, the generic driving sequence~\eqref{eq:general_seq} is exponentially long, $l_n{\sim} 2^n$, in the number $n$ of elementary unitary operators of the form $\exp(-i c H_k)$. Its appeal lies in its complete generality. 
Naturally, when considering applications to real physical systems (Sec.~\ref{sec:applns}), it is worthwhile to consider ways to shorten this sequence. Indeed, one can anticipate that the algebraic structure of the drive Hamiltonians may allow further reduction of the protocol length. 

As a concrete illustration of this possibility, consider three Hamiltonians $H_{2,1,0}$ corresponding to $G_2{\supset}G_1{\supset}G_0$; if, in addition, they obey the relation $[H_0,H_1+H_2]{=}0$, then the shorter protocol 
\begin{equation}
\label{equ:evo}
    U_F {=} (e^{-i H_0T}e^{-i H_1T})e^{-i H_2T}(e^{iH_0T}e^{iH_1T})e^{-iH_2T} {=} e^{-i T Q}
\end{equation}
defines an effective Hamiltonian, such that $Q_2^{(0)}{=}H_2$ has the symmetry group $G_2$, 
$Q_2^{(1)}$ reduces $G_2$ to $G_1$, 
and $Q_2^{(2)}$ explicitly breaks $G_1$ to $G_0$.
An interesting open problem of little practical importance is whether generic \HS can be realized using a recursive sequence of subexponential length.

A simple realization in spin-$1/2$ chains, discussed in Appendix~\ref{sec:sm6}, corresponds to the symmetry ladder $\text{U(1)} {\rightarrow} \mathbb{Z}_2 {\rightarrow} E$, with $E{=}\{id\}$ the trivial group. In the next section, we use this idea to implement an even more exotic four-step \HS protocol featuring a non-Abelian symmetry.

\subsection{Absence of symmetry in the instantaneous Hamiltonians}

In all examples discussed above, although the exact Floquet operator does not preserve any symmetry, some of the driving Hamiltonians preserve a high-level symmetry, e.g., $H_n$ in Eq.~\eqref{eq:general_seq}. While this requirement is essential for the construction of the general recursive protocol, \HS can also occur even when the instantaneous Hamiltonians $H_n$ preserve no symmetry at {\it any} time. In Sec.~\ref{sec:Exp} and in Appendix~\ref{sec:Ap_F}, we provide explicit examples to demonstrate this phenomenon. This feature is important for practical implementations, especially on real quantum platforms that do not provide direct access to evolution generated by the desired symmetric Hamiltonians. For example, Rydberg atom or superconducting qubit platforms only realize SU(2) symmetric Hamiltonians indirectly, e.g., via Floquet engineering~\cite{xu2018emulating,geier2021floquet,scholl2022microwave}.

\section{\label{sec:applns}Applications}
In this section, we present applications of hierarchically engineered symmetries for three different concepts or phenomena--non-Abelian symmetry, spatiotemporal order, and topological properties.  

\subsection{Implementation of hierarchical Abelian and non-Abelian symmetries}
\label{sec:SU2HSB}

\begin{figure}[t!]
	\centering
	\includegraphics[width= 1.0\linewidth]{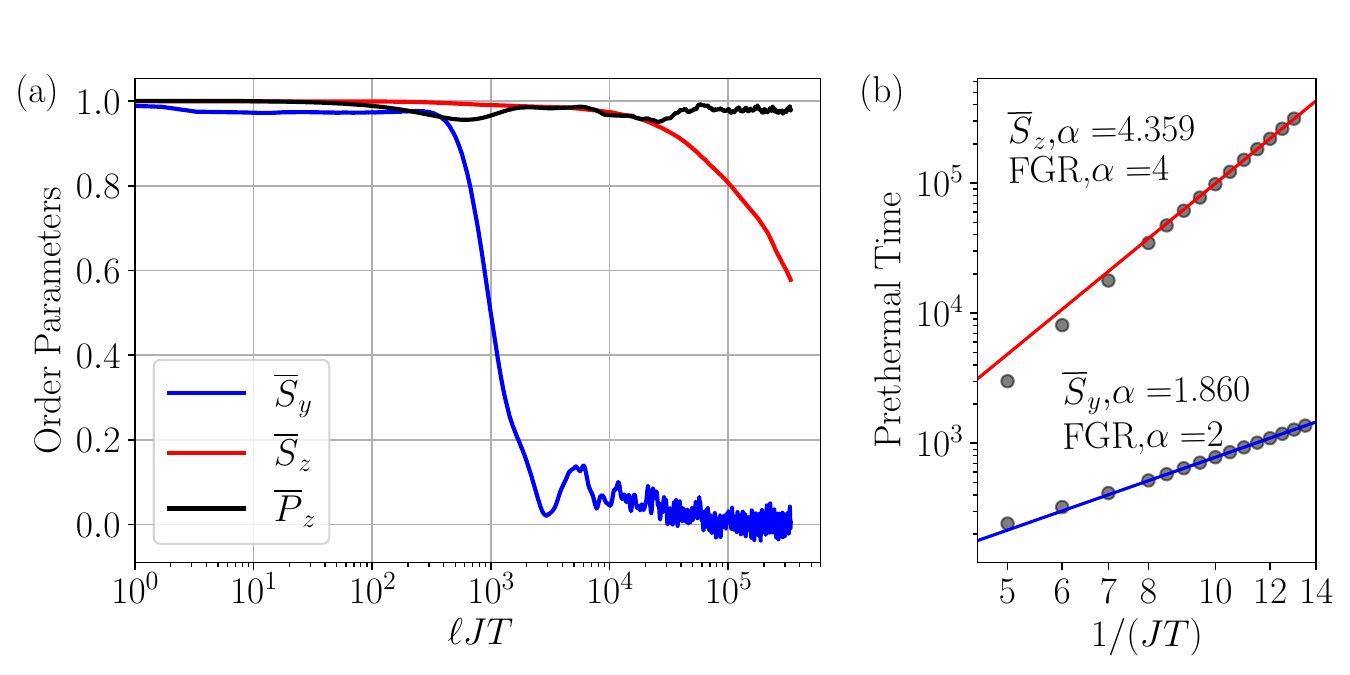}
	\caption{{Dynamical detection of $\text{SU(2)}\rightarrow\text{U(1)} \rightarrow \mathbb{Z}_2\rightarrow E$ \HS.}
 (a) Dynamics of order parameters for the hierarchical quasiconservation laws. Observables are normalized with respect to their initial values such that they all start from unity. Different lifetimes suggest that \HS emerge at different timescales. The driving period is $JT {=} 1/13.5$. (b) Lifetimes for each quasiconservation law prolonged parametrically as $T^{-\alpha}$ in the high-frequency limit. The scaling exponent $\alpha$ follows the prediction of Fermi's golden rule. For the $\mathrm{SU(2)}$ and $\mathrm{U(1)}$ plateaus, we take data points in the range $1/(JT) \in [10, 13.5]$ and $1/(JT) \in [9.5, 13.5]$, respectively, to perform the fitting.
 We use $\delta_x/J {=} 10, \epsilon/J {=} 6$ and the coupling strength $J'/J{=}5$. The system size is $L {=} 16$.
 The numerical simulations are performed using exact diagonalization; we use $30$ random realizations of the driving protocol and the initial state to compute the ensemble average.
 }
\label{fig:SU2}
\end{figure}

We uncover the effects of \HS via numerically simulating the dynamics of a paradigmatic many-body spin system with a rich emergent hierarchical symmetry structure $\text{SU(2)}{\rightarrow}\text{U(1)} {\rightarrow} \mathbb{Z}_2{\rightarrow} E$. 
To demonstrate that \HS can occur beyond time-periodic systems, we consider a system driven by a fully random sequence built out of two possible unitaries $U_{\pm}$:
\begin{eqnarray}
\label{eq.specialconstruction}
    U_{\pm} &\!\!\!=\!\!\!& U(-H_0^{\pm},H_1,H_2,H_0^{\pm},-H_1,H_2,H_3|T/14){\times}\\
    &&U(-H_2,H_1,-H_0^{\pm},-H_2,-H_1,H_0^{\pm},H_3|T/14), \nonumber
\end{eqnarray}
where we define
$
    U(D_{1},\dots, D_{l}| T) {=} 
    e^{-iD_{1}T}\cdots e^{-iD_{l}T}
$
for simplicity, 
with many-body spin-${1}/{2}$ generators
\begin{eqnarray}
\label{eq.spinmodel}
     &&H_3 {=} J\sum_{\langle i,j\rangle}\sigma_i^x\sigma_j^x+\sigma_i^y\sigma_j^y+\sigma_i^z\sigma_j^z,\nonumber\\
     &&H_2 {=} J'\sum_{\langle i,j\rangle}\sigma_i^x\sigma_j^x+\sigma_i^y\sigma_j^y-\sigma_i^z\sigma_j^z,\\
     &&H_1 {=} -J'\sum_{\langle i,j\rangle}\sigma_i^y\sigma_j^y-\sigma_i^z\sigma_j^z,\;
     \nonumber\\
     &&H_0^{\pm} {=} (\delta_x\pm \epsilon)\sum_{i}\sigma_i^x.\nonumber
     \label{eq.su2drive}
\end{eqnarray}
The spins interact with their nearest neighbors with strengths $J$ and $J'$, and a uniform $x$ field of amplitude $\delta_x\pm \epsilon$ is applied randomly in time as the protocol sequence grows.

One can derive two different effective Hamiltonians $Q_{\pm}$ for $U_{\pm}$, which coincide up to order $\mathcal{O}(T^2)$, $Q_{+}^{(m)}{=}Q_-^{(m)}$ for $m{=}0,1,2$; this happens since, in the driving protocol, the only difference occurs through $H_0^{\pm}$, whose effect is suppressed to order $m\leq2$ by the particular construction of Eq.~\eqref{eq.specialconstruction}. 
To leading order, summing up all generators, it is easy to see that $Q_{\pm}^{(0)}{\propto} H_3$ reproduces the Heisenberg model, preserving the highest symmetry SU(2); moreover, using IFE and the property $[H_0^{\pm},H_1{+}H_2] {=} 0$, one can show that $Q_{\pm}^{(1)}=\mathcal{O}(T)$ reduces SU(2) to U(1); in turn,
$Q_{\pm}^{(2)}=\mathcal{O}(T^2)$ further reduces U(1) to a $\mathbb{Z}_2$ symmetry generated by the parity operator $P_z {=} \prod_{i}\sigma^z_i$ \footnote{Note that there is another $\mathbb{Z}_2$ generated by $P_x=\prod_{i}\sigma^x_i$ which will not be broken by the drive. However, since this is not a subgroup of the $\mathbb{Z}_2$ generated by $P_z$, we do not consider it in the present \HS example.}, cf.~Fig~\ref{fig:scheme}(b) and Appendix~\ref{sec:sm6}.
This $\mathbb{Z}_2$ symmetry itself is weakly but explicitly broken by higher-order terms. Consequently, it is expected that, if we start from an SU(2)-broken initial state, the quasiconservation laws associated with the above symmetries will persist with different lifetimes in the high-frequency (or small-$T$) regime.

To verify this expectation, we first prepare the initial state $|\psi(0)\rangle$ as a Haar random state  in the $z$-magnetization sector containing $N_{\downarrow}{=}6$ down spins out of all $L{=}16$ sites; it is then rotated around the $x$ axis by
{$\prod_{i}e^{-i\pi/16\sigma_i^x}$}, resulting in an ordered state with nonzero initial magnetization along the $z$ and $y$ directions~\footnote{Such an initial state is a high-temperature state with respect to $Q_{\pm}^{(0)}$ as verified by the energy density which is close to zero.}. If the SU(2) symmetry is preserved, both $S_y {=} \sum_{i}\sigma_i^y$ and $S_z {=} \sum_{i}\sigma_i^z$ are quasiconserved quantities. As illustrated in Fig.~\ref{fig:SU2}(a), for a fixed period $JT{=}0.1$, their normalized expectation values, $\Bar{S}_{y/z}(t){=}\langle\psi(t)| S_{y/z} |\psi(t)\rangle/ \langle\psi(0)| S_{y/z}|\psi(0) \rangle$, indeed remain almost unchanged until a long timescale $J\ell T\approx 5{\times} 10^{2}$. Then, the system exhibits a noticeable decay in $\Bar{S}_{x}$, indicating the explicit breaking of SU(2) by the higher-order terms in the IFE. 
By contrast, the quasiconservation of $\Bar{S}_{z}$ corresponding to the U(1) symmetry is more robust and only exhibits roughly $20\%$ deviation from the initial unit value around $J\ell T{\approx} 5{\times} 10^{4}$, when the $y$ magnetization completely vanishes.

To detect the preservation of the $\mathbb{Z}_2$ symmetry, we measure the normalized expectation value of the parity operator 
$\Bar{P}_z(t)$, which remains close to its initial value throughout the entire time evolution that we can numerically simulate.

The dynamics, constrained by all of these emergent symmetries, can be stabilized by using a higher drive frequency, and consequently, the lifetime of each quasiconservation law can be parametrically prolonged. We define the lifetimes $\tau_y$ and $\tau_z$ as the time when the magnetizations $\langle S_y(t)\rangle $ and $\langle S_z(t)\rangle $ drop below the threshold values
$e^{-1}$ and $e^{-0.45}$, respectively~\footnote{These threshold values are chosen for numerical simplicity and the scaling exponent of the prethermal lifetime in the high-frequency regime does not depend on these specific choices.}. As shown in Fig.~\ref{fig:SU2}(b), 
for moderate driving frequencies, i.e., $1/T\approx 5J$, we already observe a noticeable separation in the prethermal lifetime in $\bar{S}_z$ and $\bar{S}_y$ by 1 order of magnitude in time. 
In fact, albeit away from the perturbative high-frequency regime, such a clear hierarchy is robust, and it exists even when the driving frequency is comparable to local energy scales, cf.~Sec.~\ref{sec:Exp}. 

In the high-frequency limit, both timescales follow an algebraic scaling in the form of $\tau\sim T^{-\alpha}$ with the scaling exponent $\alpha\approx 2$ for $\Bar{S}_y$ and $\alpha\approx 4$ for $\Bar{S}_z$, following the Fermi's golden rule (FGR) prediction, cf.~Appendix~\ref{sec:FGR}. 
This scaling prediction is not limited to this specific initial state, as long as the initial state has a finite energy density (or finite local temperature), such that it is sufficiently far from the ground state. In this case, the system can quickly reach a local thermal ensemble. One can use FGR to theoretically estimate the damping rate of the order parameter or its autocorrelation function.
We mention in passing that, for low-energy initial states, one can observe oscillatory behavior in local observables, with a damping rate that significantly deviates from the FGR prediction, originating from the emergence of massive Goldstone modes~\cite{hou2024floquet}.

Since the $\mathbb{Z}_2$-breaking perturbations of $\mathcal{O}(T^3)$ are extremely weak, it is challenging to determine the concrete scaling law for the lifetime of $\Bar{P}_z$. In Appendix~\ref{sec:sm6}, we consider another \HS example to illustrate a simpler hierarchy $\text{U(1)}{\to}\mathbb{Z}_2{\to}
E$, where we show that the decay of $\Bar{P}_z$ can also be suppressed by using a shorter driving period. However, since $\Bar{P}_z$ is a nonlocal operator, its decay may not be described by FGR, 
and determining the lifetime of its prethermal plateau is an open problem.

Although our numerical simulations are only able to reach the long times required to observe the plateau decay for small enough quantum spin systems, we show that the finite-size effects are negligible for quantum spin chains of size larger than $L\ge 12$ spins in Appendix~\ref{sec:fin_size}. We also verify that \HS exists in large classical systems of hundreds of spins; see Ref.~\cite{hou2024floquet}.

\HS stabilize quasiconservation laws for both Abelian and non-Abelian, continuous and discrete symmetries with the corresponding timescales parametrically under control. 
In the following, we will go beyond this and demonstrate how \HS can be harnessed to engineer different types of nonequilibrium order, connected by dynamical crossover regimes and corresponding to the hierarchical symmetry groups, including spatiotemporal order (STO) and higher-order topological-insulating states.

\subsection{Hierarchical symmetry reduction in a \\ $\mathbb{Z}_4$ discrete time crystal}
\label{sec:4}

We begin with STO and consider a four-state clock model, whose kicked dynamics is generated by 
\begin{eqnarray}
\label{eq:clock_H}
        H_0 &=& \sum_{i=1}^L b_{i}(Z_i+Z_i^{\dagger})\nonumber\\
        H_1 &=& \sum_{\langle i,j\rangle}J_{ij}\left(Z_i^2Z_j^2-\eta (e^{i\phi}Z_i^{\dagger}Z_j+\text{h.c.})\right)\\
        & & + \sum_{i}h_{i}\left(Z_i^2-\frac{1}{2}\left(X_i+X_i^{\dagger}\right)\right) + \sum_{i}g_{i}X_i^2,\nonumber
\end{eqnarray}
where $J_{ij}$ is a nearest-neighbor interaction strength, $g_i$ an on-site $X^2$ {interaction}, and $b_{i}$ a parallel field; $h_{i}$ measures the strength of a combination of the on-site $Z^2$ interaction and the transverse field; $\eta$ and $\phi$ help stabilize the STO~\cite{surace2019floquet}.
We introduce randomness in the form of uniformly distributed spatial disorder in all couplings to reduce finite-size effects that manifest as temporal fluctuations in the dynamics of observables. 
In the $Z$ eigenbasis denoted by $|n\rangle$, $n=0,1,2,3$, 
\begin{equation}
Z=\begin{pmatrix}
  1&  0&  0&0 \\
  0&  -i&  0&0 \\
  0&  0&  -1&0 \\
  0&  0&  0&i
\end{pmatrix},\quad
X=\begin{pmatrix}
  0&  1&  0&0 \\
  0&  0&  1&0 \\
  0&  0&  0&1 \\
  1&  0&  0&0
\end{pmatrix}
\end{equation}
satisfy the commutation relations $Z_jX_k {=} e^{\frac{i\pi}{2}\delta_{jk}}X_kZ_j$ and $X_j^4 {=}1{=}Z_j^4$.
The $X$ operator shifts the population of the four $|n\rangle$ states cyclically. 
Hence, the internal level structure admits a $\mathbb{Z}_4$ symmetry. This symmetry is obeyed by the nearest-neighbor interaction and the $X^2$ terms and broken down to a $\mathbb{Z}_2$ subgroup by the $Z^2$ on-site term; the parallel field further reduces this remaining symmetry to the trivial group, $\mathbb{Z}_2\to E$.

When the kicked dynamics of the system, generated by $H_0$ and $H_1$, is interleaved with $T$-periodic $X$ kicks, $P_X = \prod_{i}X_i$,
\begin{eqnarray}
    U_F &=& e^{iH_0T/4}e^{-iH_1T/4}e^{-iH_0T/4}e^{-iH_1T/4}P_{X},
\label{eq:UF_DTC}
\end{eqnarray}
the $\mathbb{Z}_4$ symmetry can conspire with the discrete time-translation symmetry and induce spatiotemporal order, much like in a $\mathbb{Z}_4$-time crystal~\cite{surace2019floquet}.
The kick generators $H_0$, $H_1$ are designed to imprint \HS in the first few orders of the effective Hamiltonian associated with Eq.~\eqref{eq:UF_DTC}. 
In particular, using the relation $\sum_{q=0}^{3}[X^{ q}]^\dagger Z^{p}X^{q} {=} 0$ ($p{\in}\mathbb{N}$), one can show that the leading-order term $Q^{(0)}$ has a $\mathbb{Z}_4$ symmetry that is reduced to its $\mathbb{Z}_2$ subgroup by $Q^{(1)}$, while higher-order terms 
$Q^{(n\geq 2)}$ possess no symmetry; the explicit form of the effective Hamiltonian is given in Appendix~\ref{sec:sm3}.
As a result, the system exhibits two prethermal STO states within its time evolution (one for each emergent symmetry in the ladder), connected by a smooth crossover in time and stabilized by disorder.

\begin{figure}[t!]
    \centering
    \includegraphics[width= 1.0\linewidth]{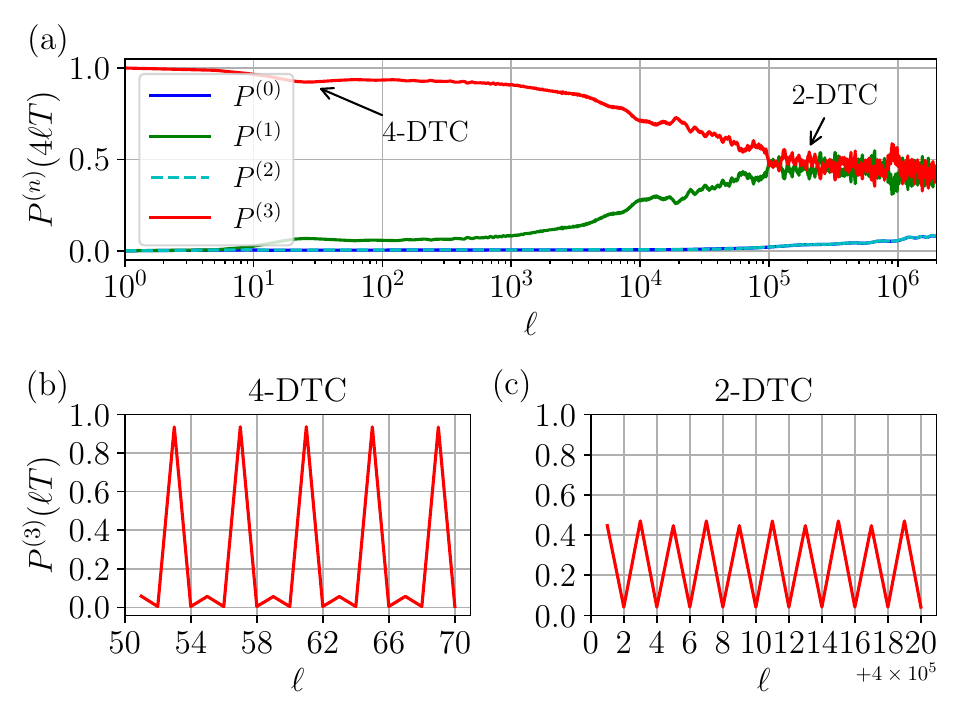}
    \caption{
    $\mathbb{Z}_4$ quantum clock model.
    Time evolution of average population $P^{(n)}(t_\ell) {=} L^{-1}\sum_{j=1}^L \langle \psi(t_\ell)|p_j^{(n)}|\psi(t_\ell)\rangle$ in the clock state $|n\rangle$ at $t_\ell = 4\ell T$, starting from the initial state $\left|\psi(0)\right\rangle {=} \left|3\right\rangle^{\otimes L}$ and averaged over 50 realizations. 
    (a) Population $P^{(n)}(t_\ell)$ at $t_\ell = 4\ell T$ and the dynamic crossover between two spatial-temporal orders.
    (b) $Z_4$ plateau. The system oscillates between these four local states and their superposition which is a 4-DTC behavior. (c) $\mathbb{Z}_2$ plateau. The odd and even local states merge in pairs (dashed line), and the system finally exhibits 2-DTC behavior.
    The parameters are $T=0.5$, $\eta {=} 0.35$, and $\phi {=} \pi/3$; and $J_{ij}\in(0.5,1.5)$, $g_{i}\in(0,0.3)$, $h_{i}\in(0,0.6)$, and $b_i\in(0,2.5)$ are drawn from a uniform distribution in the given interval. The system size is $L {=} 7.$
    }
    \label{fig:4}
\end{figure}

Figure~\ref{fig:4}(a) shows the dynamics of the population $p^{(n)}{=}|n\rangle\langle n|$ of each of the four clock states $|n\rangle$ at times $t_\ell {=} 4\ell T$, averaged over the lattice, starting from the initial product state $\left|\psi(0)\right\rangle {=} \left|3\right\rangle^{\otimes L}$.
Two prethermal plateaus, corresponding to $\mathbb{Z}_4$ and $\mathbb{Z}_2$ quasiconservation, are clearly visible in Fig.~\ref{fig:4}(a). Their lifetime can be increased by decreasing the drive period $T$.
In the $\mathbb{Z}_4$ plateau governed by $Q^{(0)}$, the population 
exhibits period-4 oscillations in time [Fig.~\ref{fig:4}(b)], characteristic of prethermal 4-DTC order. 
As time progresses, $Q^{(1)}$ asserts itself, and hence the dynamics crosses over to the $\mathbb{Z}_2$ plateau. The explicit breakdown of the original $\mathbb{Z}_4$ symmetry causes the population to redistribute, while subject to the surviving $\mathbb{Z}_2$ quasiconstraint. As a result, akin to prethermal 2-DTC order, the population keeps oscillating between two states [Fig.~\ref{fig:4}(c)], described by a statistical mixture of the bare even (odd) clock states that halves the oscillation amplitude. 

Hence, the manifestation of this spatiotemporal \HS ladder is reflected in the change of the characteristic periodic signature of observable expectations as a function of time.
Because of the ultimate breakdown of the $\mathbb{Z}_2$ symmetry by the higher-order effective Hamiltonians $Q^{(m\geq 2)}$, the population gradually spreads over all clock states, as evidenced by the rise of the blue and cyan curves in Fig.~\ref{fig:4}(a). 
Eventually, at even longer times, the final state of the system is evenly distributed among the four clock states, corresponding to a featureless infinite-temperature state (not shown).

\subsection{High-order topological insulators from \\ hierarchical symmetries}

Symmetries also play a fundamental role in topological quantum matter; this is perhaps most prominently encoded in the notion of symmetry-protected topological phases (SPTs), where topological stability is predicated on the presence of a particular symmetry \cite{pollmann2012symmetry,chen2013symmetry}. We now demonstrate how \HS can change the topological character of, say, electronic systems by altering the underlying symmetry. 

We start from a TI in a 2D lattice, protected by time-reversal symmetry $\mathcal{T}$ and crystalline inversion symmetry $\mathcal{I}$. 
Subject to open boundary conditions, the single-particle spectrum of the Hamiltonian $H_{\text{TI}}$ exhibits topological edge modes. In such materials, an initial state with significant support on the edge of a sample will keep this support during the time evolution generated by $H_{\text{TI}}.$ Perturbations that break time reversal $\mathcal{T}$ but preserve inversion symmetry $\mathcal{I}$ cause a topological phase transition from a TI to a HOTI ~\cite{ ahn2019failure, benalcazar2019quantization, khalaf2018symmetry, schindler2018higher, schindler2022topological}; see Appendix~\ref{sec:HOTI} for details. 
The characteristic feature of HOTI is the presence of corner states whose support is localized only at the corners of the sample, reflecting the reduced symmetry group.

A change of topology from a TI to a HOTI can naturally be exhibited by the transient dynamics of Floquet systems that realize the \HS ladder $\mathcal{T}{\times} \mathcal{I}{\rightarrow}\mathcal{I}{\rightarrow} E$. Intuitively, initial states with weight concentrated on the edge modes will remain stable over a controllable timescale before the leading-order symmetry-reducing term takes over; then, only modes supported on the corners survive, while other edge modes start delocalizing into the bulk as a result of broken time reversal. Eventually, if present, interactions will cause the system to heat up to an infinite-temperature state and lose all nontrivial topological properties. 

\begin{figure}[t!]
	\centering
	\includegraphics[width= 1.0\linewidth]{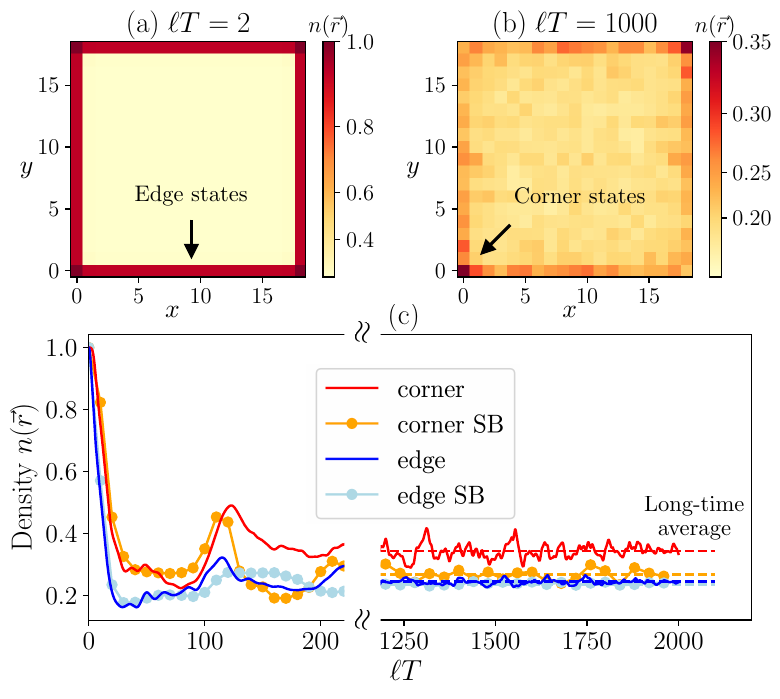}
    \caption{ 
    Topological corner states stabilized by \HS.
    Time evolution of the electron density on site $\vec{r}=(x,y)$, $n({\vec{r}}) {=}  \sum_{\sigma = \uparrow, \downarrow}\sum_{q = 0,1} \left\langle c_{\sigma,q}^{\dagger}(\vec{r})c_{\sigma,q}(\vec{r})\right\rangle$, starting from the initial state that fully covers the edge.
    (a),(b) Snapshots of the density distribution showing a dynamical change of topology, manifest in a transition from an edge state (TI) to a corner state (HOTI). 
    (c) Dynamics of the density at the corner (red line) $n_\text{corner} =(n{(0,0)}+n{(L,L)})/2$ and the edge (blue line) $n_\text{edge} = 2\sum_{i = L/4}^{3L/4}n{(i,0)}/L$ using a \HS ladder. If the first-order perturbation explicitly breaks $\mathcal{I}$ (labeled "corner SB" and "edge SB" for the corner and edge density respectively, where "SB" means symmetry breaking), support on the corner becomes less dominant (orange) while the edge maintains approximately the same density at long times (dashed lines).
    We use $M = 1.0, J=1.0, \Delta_0 = 1.0, \Delta_1 = 7.0, \Delta_2 = 12.0$ and $T = 0.2$ for numerical simulation. The linear size of the system is $L {=} 19$.
 }
\label{fig:6}
\end{figure}

To demonstrate this behavior explicitly, we consider the Floquet unitary 
\begin{eqnarray}
\label{eq:UF_HOTI}
    U_F &\!\!\!=\!\!\!& U\left(H_0,\frac{H_1}{2},\frac{H_1'}{2},-H_0,\frac{H_1}{2},\frac{H_1'}{2}, H_2\bigg| \frac{T}{10}\right){\times} \\
    && U\left(-\frac{H_1}{2},-\frac{H_1'}{2},H_0,-\frac{H_1}{2},-\frac{H_1'}{2}, -H_0, H_2\bigg| \frac{T}{10}\right)\nonumber
\end{eqnarray}
for a set of four four-band Hamiltonians $H_{0,1,2}^{(')}$ on a 2D square lattice, involving two orbital angular momentum and two spin degrees of freedom, described by the Pauli matrices $\tau$ and $\sigma$, respectively. Going to momentum space gives $H_j {=} \sum_{\vec{k}} \psi_{\vec{k}}^\dagger H_j(\vec{k}) \psi_{\vec{k}}$ with $\psi_{\vec{k}}{=}[c_{\uparrow,0}(\vec{k}), c_{\downarrow,0}(\vec{k}), c_{\uparrow,1}(\vec{k}),c_{\downarrow,1}(\vec{k})]$, where the $\uparrow$ and $\downarrow$ subscripts correspond to the fermion spin, while the 0 and 1 subscripts denote the orbital angular momentum of fermions. We have 
\begin{eqnarray}
\label{eq:H_HOTI}
     H_2(\vec{k}) &=&\left[M+J\sum_{i = \{x,y\}}\cos(k_i)\right]\; \tau_z\sigma_0 + \nonumber\\
     &&\Delta_0 \sum_{i=\{x,y\}} \sin(k_i) \tau_x \sigma_i\nonumber\\
     H_1(\vec{k}) &=& \Delta_1\tau_z(\sigma_x+\sigma_y),\nonumber\\
     H_1'(\vec{k}) &=& \Delta_1\tau_z\sigma_z\nonumber,\\
     H_0(\vec{k}) &=& \Delta_2\tau_x\sigma_y.
\end{eqnarray}
The representation of the Hamiltonian in real space is shown in Appendix~\ref{sec:HOTI}.
Compared to Eq.~\eqref{eq:general_seq}, Eq.~\eqref{eq:UF_HOTI} additionally features $H_1'$, which introduces an on-site $\mathcal{T}$-breaking term in the effective Hamiltonian that opens up the energy gap at the band touching point of $H_2$ to produce the HOTI.

The time-reversal symmetry can be represented as $\mathcal{T} {=} i\tau_0\sigma_y K$, with $K$ the complex conjugation, and acts on the Hamiltonian~\eqref{eq:H_HOTI} by $\vec{k}{\to}{-}\vec{k}$ and $\mathcal{T}\vec{\sigma}\mathcal{T}^{-1} {=} {-}\vec{\sigma}$; inversion symmetry $\mathcal{I} {=} \tau_z\sigma_0$ transforms $\vec{k}{\to}{-}\vec{k}$ and $\mathcal{I}\tau_{x,y}\mathcal{I}^{-1} {=} {-}\tau_{x,y}$. Therefore, $H_2$ is invariant under both $\mathcal{I}$ and $\mathcal{T}$, and it hosts the well-known $\mathbb{Z}_2$-TI for $|M|{<}2|J|$. As a consequence, the leading-order effective Hamiltonian $Q^{(0)}{\propto} H_2$ inherits the topological property of $H_2$. 
Moreover, note that the terms proportional to $\Delta_1$ break $\mathcal{T}$, while the $\Delta_2$ term breaks both $\mathcal{T}$ and $\mathcal{I}$; therefore, $Q^{(1)}{\propto}\tau_0(\sigma_x-\sigma_y)$ breaks $\mathcal{T}$ but preserves $\mathcal{I}$. This result allows the protocol in Eq.~\eqref{eq:UF_HOTI} to induce a dynamical crossover in topological order from TI to HOTI. 
The complete effective Hamiltonian and its eigenvalue spectrum can be found in Appendix~\ref{sec:HOTI}.

To exhibit this topological \HS ladder, we prepare an initial product state in the real-space Fock basis that fully covers the edge of the 2D lattice; on each edge site, the same internal degree of freedom $(\downarrow,1)$ is occupied, i.e., $\left|\psi_0\right\rangle {=} \prod_{\vec{r}\in \text{edge}} {c}^{\dagger}_{\downarrow,1}(\vec{r})\left|0\right\rangle$. Since this initial state has a large overlap with the localized edge state of $Q^{(0)}$, the initial configuration remains almost unchanged at the early evolution times, as shown by the real-space density $n(\vec{r})$ at time $\ell T=2$ 
in Fig.~\ref{fig:6}(a). 
The persistence of the edge density can be more pronounced if we initialize the system in one of the localized eigenstates of $Q^{(0)}$, cf. Appendix~\ref{sec:HOTI}.

At later times, the system starts delocalizing into the bulk, but the occupation around the corners persists, cf.~Fig.~\ref{fig:6}(b), due to the surviving inversion symmetry in $Q^{(1)}$, which is required for the HOTI (in our model, the zero-energy corner state occupies only two of the four corners). We also depict the density at the corner (red line) and the edge (blue line) in Fig.~\ref{fig:6}(c); clearly, the corner density has larger support, especially at long times (dashed lines).

To highlight the importance of using \HS in stabilizing the corner state, we also plot the dynamics (dotted lines) with different Hamiltonians $H_1 {=}\Delta_1 \tau_x(\sigma_x+\sigma_y)$, $H_1' {=} \Delta_1\tau_0\sigma_z$, such that the first-order IFE correction explicitly breaks all symmetries at once. As shown in Fig.~\ref{fig:6}(c), in this case, the support on the corner state (dotted orange) stabilizes at twice the smaller value compared to the \HS case~\footnote{Note that, since the system is non-interacting, heating to infinite temperature does not occur even at infinite times.}. By contrast, the edge density (light blue) remains approximately unchanged at long times. 

\section{Towards a Possible Experimental Implementation}
\label{sec:Exp}

As we have demonstrated, imprinting \HS in the dynamics of physical systems allows us to engineer a large variety of interesting phenomena, including the physics related to the breaking of both Abelian and non-Abelian symmetries, crossovers between time-crystalline-ordered states and topological-insulating states. We now turn our attention to some of the specifics of realizing \HS in present-day quantum simulators.

In many experimental settings, directly accessing a perfect SU(2) symmetric Hamiltonian can be difficult or even impossible in practice. For instance, in some superconducting qubit systems, couplings between neighboring qubits take the form of an $XY$ interaction, 
\begin{equation}  H_0{=}J\sum_{i}\sigma_i^x\sigma_{i+1}^x+\sigma_i^y\sigma_{i+1}^y,
\end{equation} 
which is only U(1) symmetric.
Nevertheless, a \HS structure can still be imprinted dynamically, even if the constituent Hamiltonian generators do not obey the underlying symmetry. 
In this section, we demonstrate how to design a driving sequence that involves single-qubit driving pulses, in addition to $H_0$, to realize the $\mathrm{SU(2)}\to\mathrm{U(1)}\to \mathbb{Z}_2\to E$ ladder as discussed in Sec.~\ref{sec:SU2HSB}. This construction is inspired by recent control protocols, e.g., Refs.~\cite{geier2021floquet, scholl2022microwave}, that have already been used to engineer SU(2) symmetric Hamiltonians in experiments. 

We first briefly review how to engineer the SU(2) symmetric Hamiltonian. Applying two $\pi/4$ single-site gates $P_x = \exp\left(-i\frac{\pi}{4}\sum_i\sigma_i^x\right)$ generated from a rapid and strong field in a given direction, we can rotate the U(1) axis of the Hamiltonian $H_0$:
\begin{equation}
\label{eq:trottererrorSU2}
    P_x\left(\sum_{i,j} \sigma_i^x\sigma_j^x+\sigma_i^y\sigma_j^y \right)  P^{\dagger}_x = \sum_{i,j}\sigma_i^x\sigma_j^x + \sigma_i^z\sigma_i^z .
\end{equation}
Repeating this process three times with $\pi/4$ gates along different axes effectively generates the Heisenberg model that is SU(2) symmetric. 

\begin{figure}[t!]
	\centering
	\includegraphics[width= 1.0\linewidth]{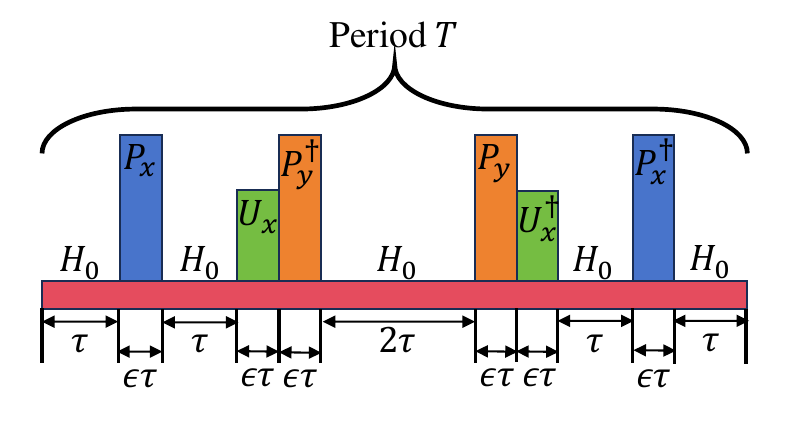}
    \caption{ 
    Schematic diagram of driving protocol in one period. $P_{x,y} and P_{x,y}^{\dagger}$ are $\pi/4$ single-site gates to effectively generate the Heisenberg model, and $U_x$ corresponds to the external pulse that reduces the symmetry group, $\mathrm{SU(2)}\to\mathrm{U(1)}$. 
    The underlying Hamiltonian $H_0$ of the system still affects the dynamics during the action of all pulses, reflecting the setup in experiments.
    }
\label{fig:driv_pro}
\end{figure}

Note that this Trotterized protocol assumes that one can suddenly switch off the coupling $H_0$ and instantaneously apply the single-site gate. Yet, in practice, there is normally a short timescale (quantified by a small {dimensionless} parameter $\epsilon$) where both $H_0$ and the strong $\pi/4$ pulses coexist, yielding errors proportional to $\epsilon$, in addition to the Trotter errors. 
Here, we design a protocol that never switches off the coupling $H_0$, and only applies extra stepwise single-body fields to the system, which are normally highly controllable in practice. We investigate the dynamics generated by continuous Gaussian pulses in Appendix~\ref{sec:Ap_exp}. Therefore, we significantly reduce the operational complexity and make the realization of \HS experimentally feasible.

Concretely, consider the Floquet protocol shown in Fig.~\ref{fig:driv_pro}, with the corresponding Floquet operator
\begin{eqnarray}
\begin{aligned}
\label{eq:exp_SU2}
    U_F =& e^{-iH_0 \tau} e^{-i(\frac{\pi}{4}X+H_0\epsilon \tau)} e^{-iH_0 \tau} 
    e^{-i(pH_x+H_0)\epsilon \tau}\\
    &e^{-i(-\frac{\pi}{4}Y+H_0\epsilon \tau)}e^{-iH_02\tau}
    e^{-i(\frac{\pi}{4}Y+H_0\epsilon \tau)}\\
    &e^{-i(-pH_x+H_0)\epsilon \tau}
    e^{-iH_0 \tau}
    e^{-i(-\frac{\pi}{4}X+H_0\epsilon \tau)}e^{-iH_0 \tau},
\end{aligned}
\end{eqnarray}
where we define $\tau = T/(6(1+\epsilon))$, with $T$ being the total driving period. Here, $X = \sum_i\sigma_i^x$ and $Y = \sum_i \sigma_i^y$ denote the total polarization in the $x$ and $y$ directions, respectively; by applying a strong field of strength $h_{x/y}$, they can generate the $\pi/4$ gate in a short time $\epsilon \tau$ ($\epsilon\ll 1$), where they coexist with the $XY$ term $H_0$. We also use a staggered field of (dimensionless) strength $p$, $H_x = p\sum_i(-1)^i\sigma_i^x$, to generate a nonvanishing correction $Q^{(1)}$ that preserves a U(1) symmetry along the $x$ axis (see below), as required for \HS.

We consider the regime $p\epsilon{\sim}\mathcal{O}(1)$ 
such that the strength of $H_x$ is effectively independent of $\epsilon$ and is comparable to the magnitude of the coupling strength $J$. 
We obtain the first two orders (in $T$) of the effective Hamiltonian as
\begin{eqnarray}
\begin{aligned}
\label{eq:epsilon_error}
        Q^{(0)} &{=} \frac{1}{2}\sum_{i}J(\sigma_i^x\sigma_{i+1}^x+\sigma_i^y\sigma_{i+1}^y+\sigma_i^z\sigma_{i+1}^z) + \mathcal{O}(\epsilon),\\
    Q^{(1)} &{=} T\frac{p\epsilon}{6}\sum_iJ(-1)^i(\sigma_i^y\sigma_{i+1}^z-\sigma_i^z\sigma_{i+1}^y)+\mathcal{O}(\epsilon T).
\end{aligned}
\end{eqnarray}
It becomes clear from these expressions that taking the limit $\epsilon{\to} 0$ while keeping $p\epsilon$ finite, $Q^{(0)}$ is the Heisenberg model preserving the $\mathrm{SU(2)}$ symmetry, while $Q^{(1)}$ preserves a $\mathrm{U(1)}$ symmetry along the $x$ axis and $Q^{(2)}$ preserves a $\mathbb{Z}_2$ symmetry generated by the parity operator $P_x = \prod_i\sigma_i^x$.

Finite $\epsilon$ corrections appear in $Q^{(0)}$, which may potentially speed up the breaking of the SU(2) and U(1) symmetries. However, notice that the corresponding timescale, as stipulated by FGR, is of order $\mathcal{O}(\epsilon^{-2})$, and these terms do not affect the dynamics at earlier times. Beyond times set by $\mathcal{O}(\epsilon^{-2})$, these corrections may indeed have an effect, but 
in the following we will show that a clear symmetry hierarchy can still be observed before this timescale. 
When the drive period is reduced, the $\mathcal{O}(\epsilon)$ terms will dominate, and one may not see clearly separated plateaus.

Depending on the details of the experimental settings, the specific Hamiltonian parameters for the protocol above will differ; in addition, experiments come with different coherence times when exposed to external drives. While there is no one-size-fits-all recipe, one can identify two competing effects when choosing a suitable value for the drive period $T$: 
(i) In the high-frequency regime $JT\ll 1$, one needs a very strong pulse to generate the single-site $\pi/4$ gates, which may not be feasible in practice; for weak pulses, $\epsilon$ is a large number, and it amplifies the symmetry-breaking processes in $Q^{(0)}$, cf.~Eq.~\eqref{eq:epsilon_error}, spoiling the symmetry ladder.
On the other hand, (ii) when $JT\sim 1$, the perturbative expansion may break down; additionally, the coherence time of the quantum simulator limits the total number of drive cycles that can be reached in practice. 
Hence, in practice, one should look for a sweet spot in $T$ to realize \HS.

To illustrate such a possibility, we consider a set of parameters for superconducting-qubit platforms, but a similar discussion can be applied to other quantum simulator platforms~\cite{geier2021floquet,mi2022time,scholl2022microwave}.
According to Ref.~\cite{xu2018emulating}, a typical value for the coupling strength $J$ is around  $2\pi\times$1-3 MHz, and a strong field strength $h_{x/y}$ can reach around $2\pi\times$50-100 MHz. 
Unlike previous sections, where we mostly focused on the high-frequency regime, here we consider a moderate driving frequency, e.g., $JT{\approx}1$; thus, a single drive cycle takes about $0.05-0.2\mu$s. This case suggests that before the qubits decohere (around $T_2=5\mu$s), one can perform a few tens to a hundred Floquet cycles. In addition, from the relation $\epsilon \tau h_{x/y}=\pi/4$, we estimate a value of $\epsilon \sim 0.05-0.3$. 

\begin{figure}[t!]
	\centering
	\includegraphics[width=\linewidth]{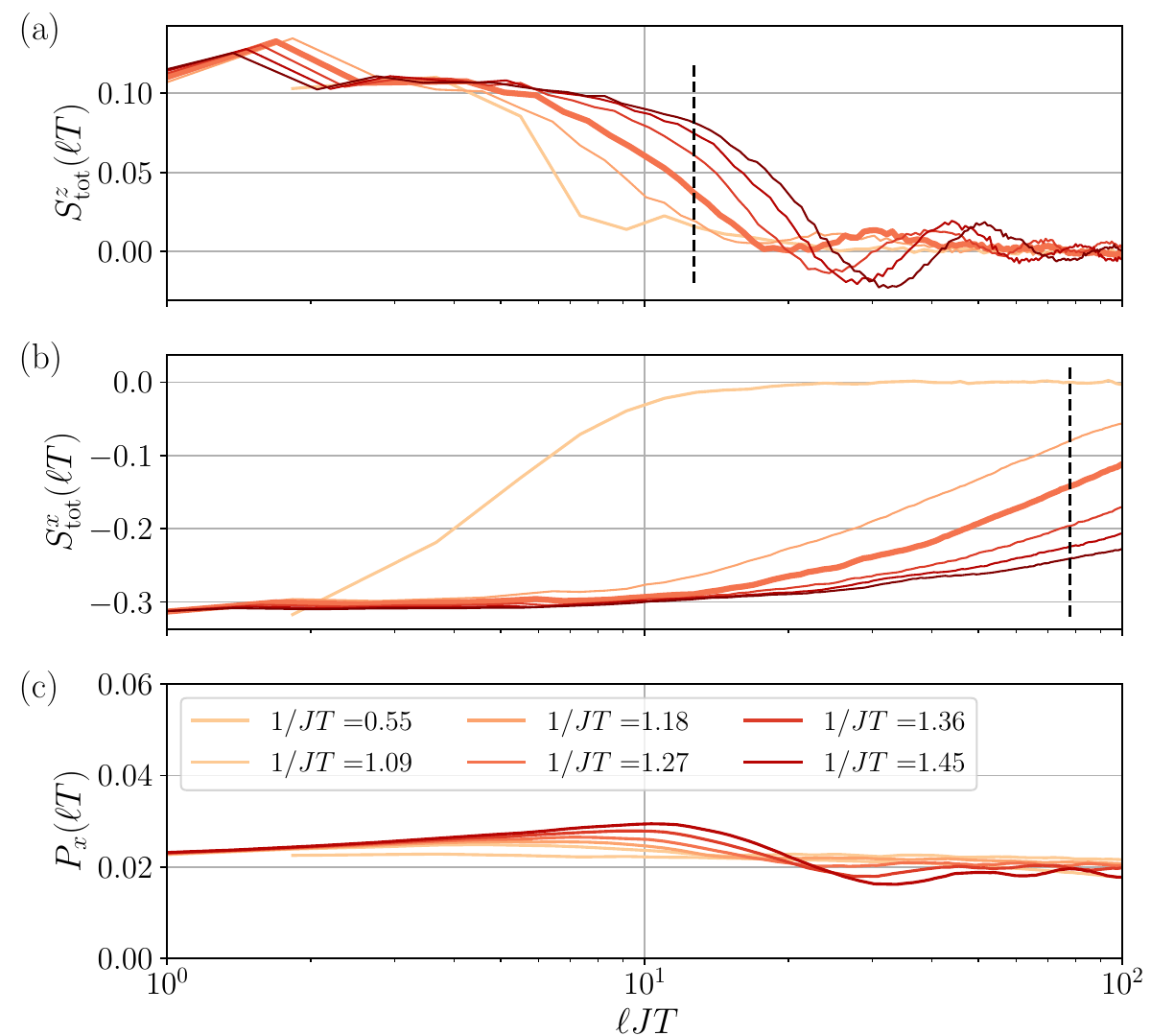}
	\caption{
Dynamical detection of the $\text{SU(2)}{\to}\text{U(1)} {\to} \mathbb{Z}_2$ symmetry ladder, with protocol from Eq.~\eqref{eq:exp_SU2}. A finite pulse width is included to mimic the experimental imperfection.
Dynamics of order parameters for the hierarchical quasiconservation laws--$S^z_\text{tot}$ for SU(2), $S^x_\text{tot}$ for U(1) and $P_x$ for $\mathbb{Z}_2$--are depicted in panels (a)-(c), respectively.
For low frequencies, e.g., $1/T {=} 0.55$, the order parameters of $\mathrm{SU(2)}$ and $\mathrm{U(1)}$ symmetry approximately have the same lifetime, around $lJT\approx 10$. A hierarchy in lifetimes is clearly visible for moderate driving frequencies, e.g., $1/T{=}1.18$ (marked with a thick line). In panels (a) and (b), the vertical dashed line corresponds to the evolution times to reach $e^{-1}S^z_{\mathrm{tot}}(0)$ and $e^{-0.8}S^x_{\mathrm{tot}}(0)$, respectively, with driving frequency $1/JT=1.18$, which are the lifetimes of the SU(2) and U(1) plateaus.
We use $p {=} 25$ as the strength of the staggered magnetic field, and the ratio of the pulse width to the drive period is $\epsilon {=} 0.1$. The system size is $L {=} 18$.
}
\label{fig:exp_fig}
\end{figure}

We now numerically show that a desired symmetry ladder can very well be observed in the above parameter regime.
The initial state is prepared as $|\psi(0)\rangle = |\downarrow \uparrow\uparrow\downarrow\uparrow\uparrow\downarrow\uparrow\uparrow\downarrow\uparrow\uparrow\downarrow\uparrow\uparrow\downarrow\rangle$ in the $z$-magnetization sector containing $N_{\downarrow}{=}6$ down spins out of all $L{=}16$ sites; it is then rotated around the $y$ axis by
{$\prod_{i}e^{-i\pi/5\sigma_i^y}$}, resulting in an ordered state with nonzero initial magnetization along the $z$ and $x$ directions (rotation angle is chosen arbitrarily).
During the evolution, we measure the magnetization density along the $z$ and $x$ axes, $S^{z/x}_\mathrm{tot}(\ell T)/N = \frac{1}{N}\langle \psi(\ell T)|\sum_i\sigma_i^{z/x}|\psi(\ell T)\rangle$, as the order parameters for $\mathrm{SU(2)}$ and $\mathrm{U(1)}$ symmetry, respectively. The parity operator $P_x = \langle \psi(\ell T)|\prod_i\sigma_i^x|\psi(\ell T)\rangle$ is used for detecting the $\mathbb{Z}_2$ symmetry. In practice, instead of measuring the nonlocal parity operator, one can preform projective measurements in the $x$ basis and simply count the particle number and check its even or odd parity.

In Fig.~\ref{fig:exp_fig}, we show the dynamics of the three order parameters for different drive frequencies. At low frequencies, e.g., $1/{JT} {=} 0.55$ (light orange), the order parameters of $\mathrm{SU(2)}$ and $\mathrm{U(1)}$ symmetry approximately have the same lifetime, around $lJT\approx 10$, indicating the breakdown of the inverse-frequency expansion. Their lifetimes can be clearly separated by increasing the driving frequency, and a hierarchy already becomes visible for a moderate frequency, e.g., $1/{JT}{=}1.18$ (thick line): The signal of $S^z_\text{tot}$ vanishes around $lJT{\approx} 20$ $[$vertical dashed line in panel (a)$]$ while $S^x_\text{tot}$ almost remains unchanged. Notable decay in $S^x_\text{tot}$ becomes visible after a few more tens of cycles $[$vertical dashed line in panel (b)$]$, which confirms the possibility in detecting \HS within the coherence time.  

Finally, let us note that the optimal parameter regimes may change for different platforms, and one should also tailor the parameters of the protocol accordingly. Yet, a few tens of Floquet cycles are already within reach of various state-of-the-art quantum simulator platforms. With the steady increase in coherence times achieved over the years, we believe that all quantum simulator platforms will be able to observe \HS in the near future.

\section{Discussion and Outlook}
\label{sec:outlook}

We present a constructive framework to engineer symmetry reduction hierarchically in driven many-body systems via a recursive time-dependent
ansatz. This permits us to impose hierarchical quasiconservation laws and to realize various kinds of order in nonequilibrium matter. The lifetime of such ordered states accessible via \HS is parametrically controllable and can be obtained using Fermi's golden rule, see Appendix~\ref{sec:FGR}. 

We demonstrate \HS in systems with different global symmetries, inducing dynamical crossovers between both equilibrium and nonequilibrium ordered states, including topological states.
A recent work generalized the phenomenon of order by disorder to Floquet systems~\cite{jin2024floquet}, where \HS also plays a crucial role.
The generalization to local gauge symmetries presents an interesting open direction~\cite{halimeh2023cold}; e.g., in the Floquet-engineered Kitaev honeycomb model~\cite{kalinowski2023non,jin2023fractionalized,sun2023engineering}, \HS can play an important role in stabilizing exotic fractionalized phases of matter.
\HS also have the potential to significantly suppress errors in the quantum simulation of time-dependent systems, thereby improving the robustness of quantum algorithms whenever symmetry plays a central role,  as in, e.g., quantum error correction~\cite{faist2020continuous,lieu2020symmetry,zhu2023nishimori}.

\HS\ constructions apply equally to fermionic and bosonic, interacting and noninteracting quantum models that exhibit hierarchical symmetries and are, therefore, widely accessible in various experimental platforms. For example, the Heisenberg model with tunable anisotropy in Eq.~\eqref{eq.su2drive} has already been realized in cold-atom systems and superconducting qubits~\cite{sun2021realization,jepsen2021transverse,wei2022quantum,keenan2023evidence}. As we have shown, there are parameter regimes in which \HS can be implemented in the lab; moreover, the underlying echo-out mechanism generalizes state-of-the-art dynamical decoupling techniques~\cite{choi2020robust} and can also be used to achieve enhanced control over higher-order corrections. 
Particularly intriguing is the possibility of experimentally studying the properties of distinct ordered states within different prethermal stages of the same time evolution.

On a fundamental level, SSB of \textit{approximate} (i.e., weakly explicitly broken) continuous symmetries gives rise to weakly \textit{gapped} Goldstone modes~\cite{watanabe2013massive}. These gapped excitations will naturally appear during the hierarchical symmetry breaking of $\text{SU(2)}{\rightarrow}\text{U(1)} {\rightarrow} \mathbb{Z}_2{\rightarrow} E$, provided that the system is initialized at low-enough temperatures~\cite{hou2024floquet}. Thus, hierarchical symmetry breaking opens up new avenues for stabilizing and manipulating quasiparticles via Floquet engineering. 

We emphasize that discussions regarding engineered symmetry breaking are not limited to quantum or Hamiltonian systems. It is intriguing to generalize the analysis of the symmetry structure to Liouvillians~\cite{albert2014symmetries,mori2018floquet,mori2023floquet,sieberer2023universality}, and we anticipate applications of \HS in time-dependent open quantum systems and classical many-body setups. In particular, numerical simulation of classical systems is not limited to small system sizes, and hence novel prethermal phases in higher dimensions induced by \HS can be explored~\cite{howell2019asymptotic,pizzi2021classical,ye2021floquet,yue2023prethermal,yan2023prethermalization}. Understanding the richer hierarchical structure of weak and strong symmetries in open systems is an open avenue for future studies~\cite{buvca2012note,lieu2020symmetry}. Moreover, generalizing the current framework to engineer integrability breaking ~\cite{surace2023weak} and to control classical chaos would be worthwhile to pursue.

Finally, our work also raises the intriguing question of whether or not one can design protocols that systematically enlarge a given symmetry group~\cite{geier2021floquet,scholl2022microwave}.

\textit{Acknowledgments.---}We thank A.~Dymarsky, B. Huang, M.~Heyl, Y.~Hou, F.~Pollmann, F.~Schindler, and K.~Xu 
for enlightening discussions. This work is supported by the Fundamental Research Funds for the Central Universities, Peking University, and by the High-performance Computing Platform of Peking University. This work is, in part, supported by the Deutsche Forschungsgemeinschaft  under  cluster of excellence ct.qmat (EXC 2147, Project No. 390858490),
funded by the European Union (ERC, QuSimCtrl, 101113633). Views and opinions expressed are, however those of the authors only and do not necessarily reflect those of the European Union or the European Research Council Executive Agency. Neither the European Union nor the granting authority can be held responsible for these views or opinions. 
This research was supported in part by the International Centre for Theoretical Sciences (ICTS) for participating in the program Stability of Quantum Matter in and out of Equilibrium at Various Scales ( ICTS/SQMVS2024/01).

\bibliography{Reference}

\begin{thebibliography}{98}%
\makeatletter
\providecommand \@ifxundefined [1]{%
 \@ifx{#1\undefined}
}%
\providecommand \@ifnum [1]{%
 \ifnum #1\expandafter \@firstoftwo
 \else \expandafter \@secondoftwo
 \fi
}%
\providecommand \@ifx [1]{%
 \ifx #1\expandafter \@firstoftwo
 \else \expandafter \@secondoftwo
 \fi
}%
\providecommand \natexlab [1]{#1}%
\providecommand \enquote  [1]{``#1''}%
\providecommand \bibnamefont  [1]{#1}%
\providecommand \bibfnamefont [1]{#1}%
\providecommand \citenamefont [1]{#1}%
\providecommand \href@noop [0]{\@secondoftwo}%
\providecommand \href [0]{\begingroup \@sanitize@url \@href}%
\providecommand \@href[1]{\@@startlink{#1}\@@href}%
\providecommand \@@href[1]{\endgroup#1\@@endlink}%
\providecommand \@sanitize@url [0]{\catcode `\\12\catcode `\$12\catcode `\&12\catcode `\#12\catcode `\^12\catcode `\_12\catcode `\%12\relax}%
\providecommand \@@startlink[1]{}%
\providecommand \@@endlink[0]{}%
\providecommand \url  [0]{\begingroup\@sanitize@url \@url }%
\providecommand \@url [1]{\endgroup\@href {#1}{\urlprefix }}%
\providecommand \urlprefix  [0]{URL }%
\providecommand \Eprint [0]{\href }%
\providecommand \doibase [0]{https://doi.org/}%
\providecommand \selectlanguage [0]{\@gobble}%
\providecommand \bibinfo  [0]{\@secondoftwo}%
\providecommand \bibfield  [0]{\@secondoftwo}%
\providecommand \translation [1]{[#1]}%
\providecommand \BibitemOpen [0]{}%
\providecommand \bibitemStop [0]{}%
\providecommand \bibitemNoStop [0]{.\EOS\space}%
\providecommand \EOS [0]{\spacefactor3000\relax}%
\providecommand \BibitemShut  [1]{\csname bibitem#1\endcsname}%
\let\auto@bib@innerbib\@empty
\bibitem [{\citenamefont {Landau}\ and\ \citenamefont {Lifshitz}(2013)}]{landau2013statistical}%
  \BibitemOpen
  \bibfield  {author} {\bibinfo {author} {\bibfnamefont {L.~D.}\ \bibnamefont {Landau}}\ and\ \bibinfo {author} {\bibfnamefont {E.~M.}\ \bibnamefont {Lifshitz}},\ }\href {https://doi.org/10.1016/B978-0-08-023039-9.50007-X} {\emph {\bibinfo {title} {Statistical Physics: Volume 5}}},\ Vol.~\bibinfo {volume} {5}\ (\bibinfo  {publisher} {Elsevier},\ \bibinfo {year} {2013})\BibitemShut {NoStop}%
\bibitem [{\citenamefont {Chaikin}\ \emph {et~al.}(1995)\citenamefont {Chaikin}, \citenamefont {Lubensky},\ and\ \citenamefont {Witten}}]{chaikin1995principles}%
  \BibitemOpen
  \bibfield  {author} {\bibinfo {author} {\bibfnamefont {P.~M.}\ \bibnamefont {Chaikin}}, \bibinfo {author} {\bibfnamefont {T.~C.}\ \bibnamefont {Lubensky}},\ and\ \bibinfo {author} {\bibfnamefont {T.~A.}\ \bibnamefont {Witten}},\ }\href {https://doi.org/https://doi.org/10.1017/CBO9780511813467} {\emph {\bibinfo {title} {Principles of condensed matter physics}}},\ Vol.~\bibinfo {volume} {10}\ (\bibinfo  {publisher} {Cambridge university press Cambridge},\ \bibinfo {year} {1995})\BibitemShut {NoStop}%
\bibitem [{\citenamefont {Pollmann}\ \emph {et~al.}(2012)\citenamefont {Pollmann}, \citenamefont {Berg}, \citenamefont {Turner},\ and\ \citenamefont {Oshikawa}}]{pollmann2012symmetry}%
  \BibitemOpen
  \bibfield  {author} {\bibinfo {author} {\bibfnamefont {F.}~\bibnamefont {Pollmann}}, \bibinfo {author} {\bibfnamefont {E.}~\bibnamefont {Berg}}, \bibinfo {author} {\bibfnamefont {A.~M.}\ \bibnamefont {Turner}},\ and\ \bibinfo {author} {\bibfnamefont {M.}~\bibnamefont {Oshikawa}},\ }\bibfield  {title} {\bibinfo {title} {Symmetry protection of topological phases in one-dimensional quantum spin systems},\ }\href {https://doi.org/10.1103/PhysRevB.85.075125} {\bibfield  {journal} {\bibinfo  {journal} {Physical review b}\ }\textbf {\bibinfo {volume} {85}},\ \bibinfo {pages} {075125} (\bibinfo {year} {2012})}\BibitemShut {NoStop}%
\bibitem [{\citenamefont {Chen}\ \emph {et~al.}(2013)\citenamefont {Chen}, \citenamefont {Gu}, \citenamefont {Liu},\ and\ \citenamefont {Wen}}]{chen2013symmetry}%
  \BibitemOpen
  \bibfield  {author} {\bibinfo {author} {\bibfnamefont {X.}~\bibnamefont {Chen}}, \bibinfo {author} {\bibfnamefont {Z.-C.}\ \bibnamefont {Gu}}, \bibinfo {author} {\bibfnamefont {Z.-X.}\ \bibnamefont {Liu}},\ and\ \bibinfo {author} {\bibfnamefont {X.-G.}\ \bibnamefont {Wen}},\ }\bibfield  {title} {\bibinfo {title} {Symmetry protected topological orders and the group cohomology of their symmetry group},\ }\href {https://doi.org/10.1103/PhysRevB.87.155114} {\bibfield  {journal} {\bibinfo  {journal} {Physical Review B}\ }\textbf {\bibinfo {volume} {87}},\ \bibinfo {pages} {155114} (\bibinfo {year} {2013})}\BibitemShut {NoStop}%
\bibitem [{\citenamefont {Anderson}(1952)}]{anderson1952approximate}%
  \BibitemOpen
  \bibfield  {author} {\bibinfo {author} {\bibfnamefont {P.~W.}\ \bibnamefont {Anderson}},\ }\bibfield  {title} {\bibinfo {title} {An approximate quantum theory of the antiferromagnetic ground state},\ }\href {https://doi.org/10.1103/PhysRev.86.694} {\bibfield  {journal} {\bibinfo  {journal} {Physical Review}\ }\textbf {\bibinfo {volume} {86}},\ \bibinfo {pages} {694} (\bibinfo {year} {1952})}\BibitemShut {NoStop}%
\bibitem [{\citenamefont {Dine}\ \emph {et~al.}(1981)\citenamefont {Dine}, \citenamefont {Fischler},\ and\ \citenamefont {Srednicki}}]{dine1981simple}%
  \BibitemOpen
  \bibfield  {author} {\bibinfo {author} {\bibfnamefont {M.}~\bibnamefont {Dine}}, \bibinfo {author} {\bibfnamefont {W.}~\bibnamefont {Fischler}},\ and\ \bibinfo {author} {\bibfnamefont {M.}~\bibnamefont {Srednicki}},\ }\bibfield  {title} {\bibinfo {title} {A simple solution to the strong cp problem with a harmless axion},\ }\href {https://doi.org/10.1016/0370-2693(81)90590-6} {\bibfield  {journal} {\bibinfo  {journal} {Physics letters B}\ }\textbf {\bibinfo {volume} {104}},\ \bibinfo {pages} {199} (\bibinfo {year} {1981})}\BibitemShut {NoStop}%
\bibitem [{\citenamefont {Brauner}(2010)}]{brauner2010spontaneous}%
  \BibitemOpen
  \bibfield  {author} {\bibinfo {author} {\bibfnamefont {T.}~\bibnamefont {Brauner}},\ }\bibfield  {title} {\bibinfo {title} {Spontaneous symmetry breaking and nambu--goldstone bosons in quantum many-body systems},\ }\href {https://doi.org/10.3390/sym2020609} {\bibfield  {journal} {\bibinfo  {journal} {Symmetry}\ }\textbf {\bibinfo {volume} {2}},\ \bibinfo {pages} {609} (\bibinfo {year} {2010})}\BibitemShut {NoStop}%
\bibitem [{\citenamefont {Castelnovo}\ \emph {et~al.}(2012)\citenamefont {Castelnovo}, \citenamefont {Moessner},\ and\ \citenamefont {Sondhi}}]{castelnovo2012spin}%
  \BibitemOpen
  \bibfield  {author} {\bibinfo {author} {\bibfnamefont {C.}~\bibnamefont {Castelnovo}}, \bibinfo {author} {\bibfnamefont {R.}~\bibnamefont {Moessner}},\ and\ \bibinfo {author} {\bibfnamefont {S.~L.}\ \bibnamefont {Sondhi}},\ }\bibfield  {title} {\bibinfo {title} {Spin ice, fractionalization, and topological order},\ }\href {https://doi.org/10.1146/annurev-conmatphys-020911-125058} {\bibfield  {journal} {\bibinfo  {journal} {Annu. Rev. Condens. Matter Phys.}\ }\textbf {\bibinfo {volume} {3}},\ \bibinfo {pages} {35} (\bibinfo {year} {2012})}\BibitemShut {NoStop}%
\bibitem [{\citenamefont {Abanin}\ \emph {et~al.}(2015)\citenamefont {Abanin}, \citenamefont {De~Roeck},\ and\ \citenamefont {Huveneers}}]{abanin2015exponentially}%
  \BibitemOpen
  \bibfield  {author} {\bibinfo {author} {\bibfnamefont {D.~A.}\ \bibnamefont {Abanin}}, \bibinfo {author} {\bibfnamefont {W.}~\bibnamefont {De~Roeck}},\ and\ \bibinfo {author} {\bibfnamefont {F.}~\bibnamefont {Huveneers}},\ }\bibfield  {title} {\bibinfo {title} {Exponentially slow heating in periodically driven many-body systems},\ }\href {https://doi.org/10.1103/PhysRevLett.115.256803} {\bibfield  {journal} {\bibinfo  {journal} {Physical review letters}\ }\textbf {\bibinfo {volume} {115}},\ \bibinfo {pages} {256803} (\bibinfo {year} {2015})}\BibitemShut {NoStop}%
\bibitem [{\citenamefont {Kuwahara}\ \emph {et~al.}(2016)\citenamefont {Kuwahara}, \citenamefont {Mori},\ and\ \citenamefont {Saito}}]{kuwahara2016floquet}%
  \BibitemOpen
  \bibfield  {author} {\bibinfo {author} {\bibfnamefont {T.}~\bibnamefont {Kuwahara}}, \bibinfo {author} {\bibfnamefont {T.}~\bibnamefont {Mori}},\ and\ \bibinfo {author} {\bibfnamefont {K.}~\bibnamefont {Saito}},\ }\bibfield  {title} {\bibinfo {title} {Floquet--magnus theory and generic transient dynamics in periodically driven many-body quantum systems},\ }\href {https://doi.org/10.1016/j.aop.2016.01.012} {\bibfield  {journal} {\bibinfo  {journal} {Annals of Physics}\ }\textbf {\bibinfo {volume} {367}},\ \bibinfo {pages} {96} (\bibinfo {year} {2016})}\BibitemShut {NoStop}%
\bibitem [{\citenamefont {Bertini}\ \emph {et~al.}(2021)\citenamefont {Bertini}, \citenamefont {Heidrich-Meisner}, \citenamefont {Karrasch}, \citenamefont {Prosen}, \citenamefont {Steinigeweg},\ and\ \citenamefont {{\v{Z}}nidari{\v{c}}}}]{bertini2021finite}%
  \BibitemOpen
  \bibfield  {author} {\bibinfo {author} {\bibfnamefont {B.}~\bibnamefont {Bertini}}, \bibinfo {author} {\bibfnamefont {F.}~\bibnamefont {Heidrich-Meisner}}, \bibinfo {author} {\bibfnamefont {C.}~\bibnamefont {Karrasch}}, \bibinfo {author} {\bibfnamefont {T.}~\bibnamefont {Prosen}}, \bibinfo {author} {\bibfnamefont {R.}~\bibnamefont {Steinigeweg}},\ and\ \bibinfo {author} {\bibfnamefont {M.}~\bibnamefont {{\v{Z}}nidari{\v{c}}}},\ }\bibfield  {title} {\bibinfo {title} {Finite-temperature transport in one-dimensional quantum lattice models},\ }\href {https://doi.org/10.1103/RevModPhys.93.025003} {\bibfield  {journal} {\bibinfo  {journal} {Reviews of Modern Physics}\ }\textbf {\bibinfo {volume} {93}},\ \bibinfo {pages} {025003} (\bibinfo {year} {2021})}\BibitemShut {NoStop}%
\bibitem [{\citenamefont {Martinez}\ \emph {et~al.}(2016)\citenamefont {Martinez}, \citenamefont {Muschik}, \citenamefont {Schindler}, \citenamefont {Nigg}, \citenamefont {Erhard}, \citenamefont {Heyl}, \citenamefont {Hauke}, \citenamefont {Dalmonte}, \citenamefont {Monz}, \citenamefont {Zoller} \emph {et~al.}}]{martinez2016real}%
  \BibitemOpen
  \bibfield  {author} {\bibinfo {author} {\bibfnamefont {E.~A.}\ \bibnamefont {Martinez}}, \bibinfo {author} {\bibfnamefont {C.~A.}\ \bibnamefont {Muschik}}, \bibinfo {author} {\bibfnamefont {P.}~\bibnamefont {Schindler}}, \bibinfo {author} {\bibfnamefont {D.}~\bibnamefont {Nigg}}, \bibinfo {author} {\bibfnamefont {A.}~\bibnamefont {Erhard}}, \bibinfo {author} {\bibfnamefont {M.}~\bibnamefont {Heyl}}, \bibinfo {author} {\bibfnamefont {P.}~\bibnamefont {Hauke}}, \bibinfo {author} {\bibfnamefont {M.}~\bibnamefont {Dalmonte}}, \bibinfo {author} {\bibfnamefont {T.}~\bibnamefont {Monz}}, \bibinfo {author} {\bibfnamefont {P.}~\bibnamefont {Zoller}}, \emph {et~al.},\ }\bibfield  {title} {\bibinfo {title} {Real-time dynamics of lattice gauge theories with a few-qubit quantum computer},\ }\href {https://doi.org/10.1038/nature18318} {\bibfield  {journal} {\bibinfo  {journal} {Nature}\ }\textbf {\bibinfo {volume} {534}},\ \bibinfo {pages} {516} (\bibinfo {year} {2016})}\BibitemShut {NoStop}%
\bibitem [{\citenamefont {Ji}\ \emph {et~al.}(2020)\citenamefont {Ji}, \citenamefont {Zhang}, \citenamefont {Wang}, \citenamefont {Zhang}, \citenamefont {Guo}, \citenamefont {Chai}, \citenamefont {Rong}, \citenamefont {Shi}, \citenamefont {Liu}, \citenamefont {Wang} \emph {et~al.}}]{ji2020quantum}%
  \BibitemOpen
  \bibfield  {author} {\bibinfo {author} {\bibfnamefont {W.}~\bibnamefont {Ji}}, \bibinfo {author} {\bibfnamefont {L.}~\bibnamefont {Zhang}}, \bibinfo {author} {\bibfnamefont {M.}~\bibnamefont {Wang}}, \bibinfo {author} {\bibfnamefont {L.}~\bibnamefont {Zhang}}, \bibinfo {author} {\bibfnamefont {Y.}~\bibnamefont {Guo}}, \bibinfo {author} {\bibfnamefont {Z.}~\bibnamefont {Chai}}, \bibinfo {author} {\bibfnamefont {X.}~\bibnamefont {Rong}}, \bibinfo {author} {\bibfnamefont {F.}~\bibnamefont {Shi}}, \bibinfo {author} {\bibfnamefont {X.-J.}\ \bibnamefont {Liu}}, \bibinfo {author} {\bibfnamefont {Y.}~\bibnamefont {Wang}}, \emph {et~al.},\ }\bibfield  {title} {\bibinfo {title} {Quantum simulation for three-dimensional chiral topological insulator},\ }\href {https://doi.org/10.1103/PhysRevLett.125.020504} {\bibfield  {journal} {\bibinfo  {journal} {Physical Review Letters}\ }\textbf {\bibinfo {volume} {125}},\ \bibinfo {pages} {020504} (\bibinfo {year} {2020})}\BibitemShut {NoStop}%
\bibitem [{\citenamefont {Jepsen}\ \emph {et~al.}(2021)\citenamefont {Jepsen}, \citenamefont {Ho}, \citenamefont {Amato-Grill}, \citenamefont {Dimitrova}, \citenamefont {Demler},\ and\ \citenamefont {Ketterle}}]{jepsen2021transverse}%
  \BibitemOpen
  \bibfield  {author} {\bibinfo {author} {\bibfnamefont {P.~N.}\ \bibnamefont {Jepsen}}, \bibinfo {author} {\bibfnamefont {W.~W.}\ \bibnamefont {Ho}}, \bibinfo {author} {\bibfnamefont {J.}~\bibnamefont {Amato-Grill}}, \bibinfo {author} {\bibfnamefont {I.}~\bibnamefont {Dimitrova}}, \bibinfo {author} {\bibfnamefont {E.}~\bibnamefont {Demler}},\ and\ \bibinfo {author} {\bibfnamefont {W.}~\bibnamefont {Ketterle}},\ }\bibfield  {title} {\bibinfo {title} {Transverse spin dynamics in the anisotropic heisenberg model realized with ultracold atoms},\ }\href {https://doi.org/10.1103/PhysRevX.11.041054} {\bibfield  {journal} {\bibinfo  {journal} {Physical Review X}\ }\textbf {\bibinfo {volume} {11}},\ \bibinfo {pages} {041054} (\bibinfo {year} {2021})}\BibitemShut {NoStop}%
\bibitem [{\citenamefont {Oka}\ and\ \citenamefont {Kitamura}(2019)}]{oka2019floquet}%
  \BibitemOpen
  \bibfield  {author} {\bibinfo {author} {\bibfnamefont {T.}~\bibnamefont {Oka}}\ and\ \bibinfo {author} {\bibfnamefont {S.}~\bibnamefont {Kitamura}},\ }\bibfield  {title} {\bibinfo {title} {Floquet engineering of quantum materials},\ }\href {https://doi.org/10.1146/annurev-conmatphys-031218-013423} {\bibfield  {journal} {\bibinfo  {journal} {Annual Review of Condensed Matter Physics}\ }\textbf {\bibinfo {volume} {10}},\ \bibinfo {pages} {387} (\bibinfo {year} {2019})}\BibitemShut {NoStop}%
\bibitem [{\citenamefont {Schweizer}\ \emph {et~al.}(2019)\citenamefont {Schweizer}, \citenamefont {Grusdt}, \citenamefont {Berngruber}, \citenamefont {Barbiero}, \citenamefont {Demler}, \citenamefont {Goldman}, \citenamefont {Bloch},\ and\ \citenamefont {Aidelsburger}}]{schweizer2019floquet}%
  \BibitemOpen
  \bibfield  {author} {\bibinfo {author} {\bibfnamefont {C.}~\bibnamefont {Schweizer}}, \bibinfo {author} {\bibfnamefont {F.}~\bibnamefont {Grusdt}}, \bibinfo {author} {\bibfnamefont {M.}~\bibnamefont {Berngruber}}, \bibinfo {author} {\bibfnamefont {L.}~\bibnamefont {Barbiero}}, \bibinfo {author} {\bibfnamefont {E.}~\bibnamefont {Demler}}, \bibinfo {author} {\bibfnamefont {N.}~\bibnamefont {Goldman}}, \bibinfo {author} {\bibfnamefont {I.}~\bibnamefont {Bloch}},\ and\ \bibinfo {author} {\bibfnamefont {M.}~\bibnamefont {Aidelsburger}},\ }\bibfield  {title} {\bibinfo {title} {Floquet approach to z2 lattice gauge theories with ultracold atoms in optical lattices},\ }\href {https://doi.org/10.1038/s41567-019-0649-7} {\bibfield  {journal} {\bibinfo  {journal} {Nature Physics}\ }\textbf {\bibinfo {volume} {15}},\ \bibinfo {pages} {1168} (\bibinfo {year} {2019})}\BibitemShut {NoStop}%
\bibitem [{\citenamefont {Geier}\ \emph {et~al.}(2021)\citenamefont {Geier}, \citenamefont {Thaicharoen}, \citenamefont {Hainaut}, \citenamefont {Franz}, \citenamefont {Salzinger}, \citenamefont {Tebben}, \citenamefont {Grimshandl}, \citenamefont {Z{\"u}rn},\ and\ \citenamefont {Weidem{\"u}ller}}]{geier2021floquet}%
  \BibitemOpen
  \bibfield  {author} {\bibinfo {author} {\bibfnamefont {S.}~\bibnamefont {Geier}}, \bibinfo {author} {\bibfnamefont {N.}~\bibnamefont {Thaicharoen}}, \bibinfo {author} {\bibfnamefont {C.}~\bibnamefont {Hainaut}}, \bibinfo {author} {\bibfnamefont {T.}~\bibnamefont {Franz}}, \bibinfo {author} {\bibfnamefont {A.}~\bibnamefont {Salzinger}}, \bibinfo {author} {\bibfnamefont {A.}~\bibnamefont {Tebben}}, \bibinfo {author} {\bibfnamefont {D.}~\bibnamefont {Grimshandl}}, \bibinfo {author} {\bibfnamefont {G.}~\bibnamefont {Z{\"u}rn}},\ and\ \bibinfo {author} {\bibfnamefont {M.}~\bibnamefont {Weidem{\"u}ller}},\ }\bibfield  {title} {\bibinfo {title} {Floquet hamiltonian engineering of an isolated many-body spin system},\ }\href {https://doi.org/10.1126/science.abd9547} {\bibfield  {journal} {\bibinfo  {journal} {Science}\ }\textbf {\bibinfo {volume} {374}},\ \bibinfo {pages} {1149} (\bibinfo {year} {2021})}\BibitemShut {NoStop}%
\bibitem [{\citenamefont {Petiziol}\ \emph {et~al.}(2022)\citenamefont {Petiziol}, \citenamefont {Wimberger}, \citenamefont {Eckardt},\ and\ \citenamefont {Mintert}}]{petiziol2022non}%
  \BibitemOpen
  \bibfield  {author} {\bibinfo {author} {\bibfnamefont {F.}~\bibnamefont {Petiziol}}, \bibinfo {author} {\bibfnamefont {S.}~\bibnamefont {Wimberger}}, \bibinfo {author} {\bibfnamefont {A.}~\bibnamefont {Eckardt}},\ and\ \bibinfo {author} {\bibfnamefont {F.}~\bibnamefont {Mintert}},\ }\bibfield  {title} {\bibinfo {title} {Non-perturbative floquet engineering of the toric-code hamiltonian and its ground state},\ }\href {https://doi.org/10.48550/arXiv.2211.09724} {\bibfield  {journal} {\bibinfo  {journal} {arXiv preprint arXiv:2211.09724}\ } (\bibinfo {year} {2022})}\BibitemShut {NoStop}%
\bibitem [{\citenamefont {Kalinowski}\ \emph {et~al.}(2023)\citenamefont {Kalinowski}, \citenamefont {Maskara},\ and\ \citenamefont {Lukin}}]{kalinowski2023non}%
  \BibitemOpen
  \bibfield  {author} {\bibinfo {author} {\bibfnamefont {M.}~\bibnamefont {Kalinowski}}, \bibinfo {author} {\bibfnamefont {N.}~\bibnamefont {Maskara}},\ and\ \bibinfo {author} {\bibfnamefont {M.~D.}\ \bibnamefont {Lukin}},\ }\bibfield  {title} {\bibinfo {title} {Non-abelian floquet spin liquids in a digital rydberg simulator},\ }\href {https://doi.org/10.1103/PhysRevX.13.031008} {\bibfield  {journal} {\bibinfo  {journal} {Physical Review X}\ }\textbf {\bibinfo {volume} {13}},\ \bibinfo {pages} {031008} (\bibinfo {year} {2023})}\BibitemShut {NoStop}%
\bibitem [{\citenamefont {Jin}\ \emph {et~al.}(2023)\citenamefont {Jin}, \citenamefont {Knolle},\ and\ \citenamefont {Knap}}]{jin2023fractionalized}%
  \BibitemOpen
  \bibfield  {author} {\bibinfo {author} {\bibfnamefont {H.-K.}\ \bibnamefont {Jin}}, \bibinfo {author} {\bibfnamefont {J.}~\bibnamefont {Knolle}},\ and\ \bibinfo {author} {\bibfnamefont {M.}~\bibnamefont {Knap}},\ }\bibfield  {title} {\bibinfo {title} {Fractionalized prethermalization in a driven quantum spin liquid},\ }\href {https://doi.org/10.1103/PhysRevLett.130.226701} {\bibfield  {journal} {\bibinfo  {journal} {Physical Review Letters}\ }\textbf {\bibinfo {volume} {130}},\ \bibinfo {pages} {226701} (\bibinfo {year} {2023})}\BibitemShut {NoStop}%
\bibitem [{\citenamefont {Sun}\ \emph {et~al.}(2023)\citenamefont {Sun}, \citenamefont {Goldman}, \citenamefont {Aidelsburger},\ and\ \citenamefont {Bukov}}]{sun2023engineering}%
  \BibitemOpen
  \bibfield  {author} {\bibinfo {author} {\bibfnamefont {B.-Y.}\ \bibnamefont {Sun}}, \bibinfo {author} {\bibfnamefont {N.}~\bibnamefont {Goldman}}, \bibinfo {author} {\bibfnamefont {M.}~\bibnamefont {Aidelsburger}},\ and\ \bibinfo {author} {\bibfnamefont {M.}~\bibnamefont {Bukov}},\ }\bibfield  {title} {\bibinfo {title} {Engineering and probing non-abelian chiral spin liquids using periodically driven ultracold atoms},\ }\href {https://doi.org/10.1103/PRXQuantum.4.020329} {\bibfield  {journal} {\bibinfo  {journal} {PRX Quantum}\ }\textbf {\bibinfo {volume} {4}},\ \bibinfo {pages} {020329} (\bibinfo {year} {2023})}\BibitemShut {NoStop}%
\bibitem [{\citenamefont {Kitagawa}\ \emph {et~al.}(2010)\citenamefont {Kitagawa}, \citenamefont {Berg}, \citenamefont {Rudner},\ and\ \citenamefont {Demler}}]{kitagawa2010topological}%
  \BibitemOpen
  \bibfield  {author} {\bibinfo {author} {\bibfnamefont {T.}~\bibnamefont {Kitagawa}}, \bibinfo {author} {\bibfnamefont {E.}~\bibnamefont {Berg}}, \bibinfo {author} {\bibfnamefont {M.}~\bibnamefont {Rudner}},\ and\ \bibinfo {author} {\bibfnamefont {E.}~\bibnamefont {Demler}},\ }\bibfield  {title} {\bibinfo {title} {Topological characterization of periodically driven quantum systems},\ }\href {https://doi.org/10.1103/PhysRevB.82.235114} {\bibfield  {journal} {\bibinfo  {journal} {Physical Review B}\ }\textbf {\bibinfo {volume} {82}},\ \bibinfo {pages} {235114} (\bibinfo {year} {2010})}\BibitemShut {NoStop}%
\bibitem [{\citenamefont {Potter}\ \emph {et~al.}(2016)\citenamefont {Potter}, \citenamefont {Morimoto},\ and\ \citenamefont {Vishwanath}}]{potter2016classification}%
  \BibitemOpen
  \bibfield  {author} {\bibinfo {author} {\bibfnamefont {A.~C.}\ \bibnamefont {Potter}}, \bibinfo {author} {\bibfnamefont {T.}~\bibnamefont {Morimoto}},\ and\ \bibinfo {author} {\bibfnamefont {A.}~\bibnamefont {Vishwanath}},\ }\bibfield  {title} {\bibinfo {title} {Classification of interacting topological floquet phases in one dimension},\ }\href {https://doi.org/10.1103/PhysRevX.6.041001} {\bibfield  {journal} {\bibinfo  {journal} {Physical Review X}\ }\textbf {\bibinfo {volume} {6}},\ \bibinfo {pages} {041001} (\bibinfo {year} {2016})}\BibitemShut {NoStop}%
\bibitem [{\citenamefont {Titum}\ \emph {et~al.}(2016)\citenamefont {Titum}, \citenamefont {Berg}, \citenamefont {Rudner}, \citenamefont {Refael},\ and\ \citenamefont {Lindner}}]{titum2016anomalous}%
  \BibitemOpen
  \bibfield  {author} {\bibinfo {author} {\bibfnamefont {P.}~\bibnamefont {Titum}}, \bibinfo {author} {\bibfnamefont {E.}~\bibnamefont {Berg}}, \bibinfo {author} {\bibfnamefont {M.~S.}\ \bibnamefont {Rudner}}, \bibinfo {author} {\bibfnamefont {G.}~\bibnamefont {Refael}},\ and\ \bibinfo {author} {\bibfnamefont {N.~H.}\ \bibnamefont {Lindner}},\ }\bibfield  {title} {\bibinfo {title} {Anomalous floquet-anderson insulator as a nonadiabatic quantized charge pump},\ }\href {https://doi.org/10.1103/PhysRevX.6.021013} {\bibfield  {journal} {\bibinfo  {journal} {Physical Review X}\ }\textbf {\bibinfo {volume} {6}},\ \bibinfo {pages} {021013} (\bibinfo {year} {2016})}\BibitemShut {NoStop}%
\bibitem [{\citenamefont {Khemani}\ \emph {et~al.}(2016)\citenamefont {Khemani}, \citenamefont {Lazarides}, \citenamefont {Moessner},\ and\ \citenamefont {Sondhi}}]{khemani2016phase}%
  \BibitemOpen
  \bibfield  {author} {\bibinfo {author} {\bibfnamefont {V.}~\bibnamefont {Khemani}}, \bibinfo {author} {\bibfnamefont {A.}~\bibnamefont {Lazarides}}, \bibinfo {author} {\bibfnamefont {R.}~\bibnamefont {Moessner}},\ and\ \bibinfo {author} {\bibfnamefont {S.~L.}\ \bibnamefont {Sondhi}},\ }\bibfield  {title} {\bibinfo {title} {Phase structure of driven quantum systems},\ }\href {https://doi.org/10.1103/PhysRevLett.116.250401} {\bibfield  {journal} {\bibinfo  {journal} {Physical review letters}\ }\textbf {\bibinfo {volume} {116}},\ \bibinfo {pages} {250401} (\bibinfo {year} {2016})}\BibitemShut {NoStop}%
\bibitem [{\citenamefont {Else}\ \emph {et~al.}(2016)\citenamefont {Else}, \citenamefont {Bauer},\ and\ \citenamefont {Nayak}}]{else2016floquet}%
  \BibitemOpen
  \bibfield  {author} {\bibinfo {author} {\bibfnamefont {D.~V.}\ \bibnamefont {Else}}, \bibinfo {author} {\bibfnamefont {B.}~\bibnamefont {Bauer}},\ and\ \bibinfo {author} {\bibfnamefont {C.}~\bibnamefont {Nayak}},\ }\bibfield  {title} {\bibinfo {title} {Floquet time crystals},\ }\href {https://doi.org/10.1103/PhysRevLett.117.090402} {\bibfield  {journal} {\bibinfo  {journal} {Physical review letters}\ }\textbf {\bibinfo {volume} {117}},\ \bibinfo {pages} {090402} (\bibinfo {year} {2016})}\BibitemShut {NoStop}%
\bibitem [{\citenamefont {Yao}\ \emph {et~al.}(2017)\citenamefont {Yao}, \citenamefont {Potter}, \citenamefont {Potirniche},\ and\ \citenamefont {Vishwanath}}]{yao2017discrete}%
  \BibitemOpen
  \bibfield  {author} {\bibinfo {author} {\bibfnamefont {N.~Y.}\ \bibnamefont {Yao}}, \bibinfo {author} {\bibfnamefont {A.~C.}\ \bibnamefont {Potter}}, \bibinfo {author} {\bibfnamefont {I.-D.}\ \bibnamefont {Potirniche}},\ and\ \bibinfo {author} {\bibfnamefont {A.}~\bibnamefont {Vishwanath}},\ }\bibfield  {title} {\bibinfo {title} {Discrete time crystals: Rigidity, criticality, and realizations},\ }\href {https://doi.org/10.1103/PhysRevLett.118.030401} {\bibfield  {journal} {\bibinfo  {journal} {Physical review letters}\ }\textbf {\bibinfo {volume} {118}},\ \bibinfo {pages} {030401} (\bibinfo {year} {2017})}\BibitemShut {NoStop}%
\bibitem [{\citenamefont {Else}\ \emph {et~al.}(2020)\citenamefont {Else}, \citenamefont {Ho},\ and\ \citenamefont {Dumitrescu}}]{else2020long}%
  \BibitemOpen
  \bibfield  {author} {\bibinfo {author} {\bibfnamefont {D.~V.}\ \bibnamefont {Else}}, \bibinfo {author} {\bibfnamefont {W.~W.}\ \bibnamefont {Ho}},\ and\ \bibinfo {author} {\bibfnamefont {P.~T.}\ \bibnamefont {Dumitrescu}},\ }\bibfield  {title} {\bibinfo {title} {Long-lived interacting phases of matter protected by multiple time-translation symmetries in quasiperiodically driven systems},\ }\href {https://doi.org/10.1103/PhysRevX.10.021032} {\bibfield  {journal} {\bibinfo  {journal} {Physical Review X}\ }\textbf {\bibinfo {volume} {10}},\ \bibinfo {pages} {021032} (\bibinfo {year} {2020})}\BibitemShut {NoStop}%
\bibitem [{\citenamefont {Dumitrescu}\ \emph {et~al.}(2022)\citenamefont {Dumitrescu}, \citenamefont {Bohnet}, \citenamefont {Gaebler}, \citenamefont {Hankin}, \citenamefont {Hayes}, \citenamefont {Kumar}, \citenamefont {Neyenhuis}, \citenamefont {Vasseur},\ and\ \citenamefont {Potter}}]{dumitrescu2022dynamical}%
  \BibitemOpen
  \bibfield  {author} {\bibinfo {author} {\bibfnamefont {P.~T.}\ \bibnamefont {Dumitrescu}}, \bibinfo {author} {\bibfnamefont {J.~G.}\ \bibnamefont {Bohnet}}, \bibinfo {author} {\bibfnamefont {J.~P.}\ \bibnamefont {Gaebler}}, \bibinfo {author} {\bibfnamefont {A.}~\bibnamefont {Hankin}}, \bibinfo {author} {\bibfnamefont {D.}~\bibnamefont {Hayes}}, \bibinfo {author} {\bibfnamefont {A.}~\bibnamefont {Kumar}}, \bibinfo {author} {\bibfnamefont {B.}~\bibnamefont {Neyenhuis}}, \bibinfo {author} {\bibfnamefont {R.}~\bibnamefont {Vasseur}},\ and\ \bibinfo {author} {\bibfnamefont {A.~C.}\ \bibnamefont {Potter}},\ }\bibfield  {title} {\bibinfo {title} {Dynamical topological phase realized in a trapped-ion quantum simulator},\ }\href {https://doi.org/10.1038/s41586-022-04853-4} {\bibfield  {journal} {\bibinfo  {journal} {Nature}\ }\textbf {\bibinfo {volume} {607}},\ \bibinfo {pages} {463} (\bibinfo {year} {2022})}\BibitemShut {NoStop}%
\bibitem [{\citenamefont {Zhang}\ \emph {et~al.}(2022)\citenamefont {Zhang}, \citenamefont {Jiang}, \citenamefont {Deng}, \citenamefont {Wang}, \citenamefont {Chen}, \citenamefont {Zhang}, \citenamefont {Ren}, \citenamefont {Dong}, \citenamefont {Xu}, \citenamefont {Gao} \emph {et~al.}}]{zhang2022digital}%
  \BibitemOpen
  \bibfield  {author} {\bibinfo {author} {\bibfnamefont {X.}~\bibnamefont {Zhang}}, \bibinfo {author} {\bibfnamefont {W.}~\bibnamefont {Jiang}}, \bibinfo {author} {\bibfnamefont {J.}~\bibnamefont {Deng}}, \bibinfo {author} {\bibfnamefont {K.}~\bibnamefont {Wang}}, \bibinfo {author} {\bibfnamefont {J.}~\bibnamefont {Chen}}, \bibinfo {author} {\bibfnamefont {P.}~\bibnamefont {Zhang}}, \bibinfo {author} {\bibfnamefont {W.}~\bibnamefont {Ren}}, \bibinfo {author} {\bibfnamefont {H.}~\bibnamefont {Dong}}, \bibinfo {author} {\bibfnamefont {S.}~\bibnamefont {Xu}}, \bibinfo {author} {\bibfnamefont {Y.}~\bibnamefont {Gao}}, \emph {et~al.},\ }\bibfield  {title} {\bibinfo {title} {Digital quantum simulation of floquet symmetry-protected topological phases},\ }\href {https://doi.org/10.1038/s41586-022-04854-3} {\bibfield  {journal} {\bibinfo  {journal} {Nature}\ }\textbf {\bibinfo {volume} {607}},\ \bibinfo {pages} {468} (\bibinfo {year} {2022})}\BibitemShut {NoStop}%
\bibitem [{\citenamefont {Mi}\ \emph {et~al.}(2022)\citenamefont {Mi}, \citenamefont {Ippoliti}, \citenamefont {Quintana}, \citenamefont {Greene}, \citenamefont {Chen}, \citenamefont {Gross}, \citenamefont {Arute}, \citenamefont {Arya}, \citenamefont {Atalaya}, \citenamefont {Babbush} \emph {et~al.}}]{mi2022time}%
  \BibitemOpen
  \bibfield  {author} {\bibinfo {author} {\bibfnamefont {X.}~\bibnamefont {Mi}}, \bibinfo {author} {\bibfnamefont {M.}~\bibnamefont {Ippoliti}}, \bibinfo {author} {\bibfnamefont {C.}~\bibnamefont {Quintana}}, \bibinfo {author} {\bibfnamefont {A.}~\bibnamefont {Greene}}, \bibinfo {author} {\bibfnamefont {Z.}~\bibnamefont {Chen}}, \bibinfo {author} {\bibfnamefont {J.}~\bibnamefont {Gross}}, \bibinfo {author} {\bibfnamefont {F.}~\bibnamefont {Arute}}, \bibinfo {author} {\bibfnamefont {K.}~\bibnamefont {Arya}}, \bibinfo {author} {\bibfnamefont {J.}~\bibnamefont {Atalaya}}, \bibinfo {author} {\bibfnamefont {R.}~\bibnamefont {Babbush}}, \emph {et~al.},\ }\bibfield  {title} {\bibinfo {title} {Time-crystalline eigenstate order on a quantum processor},\ }\href {https://doi.org/10.1038/s41586-021-04257-w} {\bibfield  {journal} {\bibinfo  {journal} {Nature}\ }\textbf {\bibinfo {volume} {601}},\ \bibinfo {pages} {531} (\bibinfo {year} {2022})}\BibitemShut {NoStop}%
\bibitem [{\citenamefont {D'Alessio}\ \emph {et~al.}(2016)\citenamefont {D'Alessio}, \citenamefont {Kafri}, \citenamefont {Polkovnikov},\ and\ \citenamefont {Rigol}}]{d2016quantum}%
  \BibitemOpen
  \bibfield  {author} {\bibinfo {author} {\bibfnamefont {L.}~\bibnamefont {D'Alessio}}, \bibinfo {author} {\bibfnamefont {Y.}~\bibnamefont {Kafri}}, \bibinfo {author} {\bibfnamefont {A.}~\bibnamefont {Polkovnikov}},\ and\ \bibinfo {author} {\bibfnamefont {M.}~\bibnamefont {Rigol}},\ }\bibfield  {title} {\bibinfo {title} {From quantum chaos and eigenstate thermalization to statistical mechanics and thermodynamics},\ }\href {https://doi.org/10.1080/00018732.2016.1198134} {\bibfield  {journal} {\bibinfo  {journal} {Advances in Physics}\ }\textbf {\bibinfo {volume} {65}},\ \bibinfo {pages} {239} (\bibinfo {year} {2016})}\BibitemShut {NoStop}%
\bibitem [{\citenamefont {Vidmar}\ and\ \citenamefont {Rigol}(2016)}]{vidmar2016generalized}%
  \BibitemOpen
  \bibfield  {author} {\bibinfo {author} {\bibfnamefont {L.}~\bibnamefont {Vidmar}}\ and\ \bibinfo {author} {\bibfnamefont {M.}~\bibnamefont {Rigol}},\ }\bibfield  {title} {\bibinfo {title} {Generalized gibbs ensemble in integrable lattice models},\ }\href {https://doi.org/10.1088/1742-5468/2016/06/064007} {\bibfield  {journal} {\bibinfo  {journal} {Journal of Statistical Mechanics: Theory and Experiment}\ }\textbf {\bibinfo {volume} {2016}},\ \bibinfo {pages} {064007} (\bibinfo {year} {2016})}\BibitemShut {NoStop}%
\bibitem [{\citenamefont {Hainaut}\ \emph {et~al.}(2018)\citenamefont {Hainaut}, \citenamefont {Manai}, \citenamefont {Cl{\'e}ment}, \citenamefont {Garreau}, \citenamefont {Szriftgiser}, \citenamefont {Lemari{\'e}}, \citenamefont {Cherroret}, \citenamefont {Delande},\ and\ \citenamefont {Chicireanu}}]{hainaut2018controlling}%
  \BibitemOpen
  \bibfield  {author} {\bibinfo {author} {\bibfnamefont {C.}~\bibnamefont {Hainaut}}, \bibinfo {author} {\bibfnamefont {I.}~\bibnamefont {Manai}}, \bibinfo {author} {\bibfnamefont {J.-F.}\ \bibnamefont {Cl{\'e}ment}}, \bibinfo {author} {\bibfnamefont {J.~C.}\ \bibnamefont {Garreau}}, \bibinfo {author} {\bibfnamefont {P.}~\bibnamefont {Szriftgiser}}, \bibinfo {author} {\bibfnamefont {G.}~\bibnamefont {Lemari{\'e}}}, \bibinfo {author} {\bibfnamefont {N.}~\bibnamefont {Cherroret}}, \bibinfo {author} {\bibfnamefont {D.}~\bibnamefont {Delande}},\ and\ \bibinfo {author} {\bibfnamefont {R.}~\bibnamefont {Chicireanu}},\ }\bibfield  {title} {\bibinfo {title} {Controlling symmetry and localization with an artificial gauge field in a disordered quantum system},\ }\href {https://doi.org/10.1038/s41467-018-03481-9} {\bibfield  {journal} {\bibinfo  {journal} {Nature communications}\ }\textbf {\bibinfo {volume} {9}},\ \bibinfo {pages} {1382} (\bibinfo {year} {2018})}\BibitemShut {NoStop}%
\bibitem [{\citenamefont {Haldar}\ \emph {et~al.}(2018)\citenamefont {Haldar}, \citenamefont {Moessner},\ and\ \citenamefont {Das}}]{haldar2018onset}%
  \BibitemOpen
  \bibfield  {author} {\bibinfo {author} {\bibfnamefont {A.}~\bibnamefont {Haldar}}, \bibinfo {author} {\bibfnamefont {R.}~\bibnamefont {Moessner}},\ and\ \bibinfo {author} {\bibfnamefont {A.}~\bibnamefont {Das}},\ }\bibfield  {title} {\bibinfo {title} {Onset of floquet thermalization},\ }\href {https://doi.org/10.1103/PhysRevB.97.245122} {\bibfield  {journal} {\bibinfo  {journal} {Physical Review B}\ }\textbf {\bibinfo {volume} {97}},\ \bibinfo {pages} {245122} (\bibinfo {year} {2018})}\BibitemShut {NoStop}%
\bibitem [{\citenamefont {Agrawal}\ \emph {et~al.}(2022)\citenamefont {Agrawal}, \citenamefont {Zabalo}, \citenamefont {Chen}, \citenamefont {Wilson}, \citenamefont {Potter}, \citenamefont {Pixley}, \citenamefont {Gopalakrishnan},\ and\ \citenamefont {Vasseur}}]{agrawal2022entanglement}%
  \BibitemOpen
  \bibfield  {author} {\bibinfo {author} {\bibfnamefont {U.}~\bibnamefont {Agrawal}}, \bibinfo {author} {\bibfnamefont {A.}~\bibnamefont {Zabalo}}, \bibinfo {author} {\bibfnamefont {K.}~\bibnamefont {Chen}}, \bibinfo {author} {\bibfnamefont {J.~H.}\ \bibnamefont {Wilson}}, \bibinfo {author} {\bibfnamefont {A.~C.}\ \bibnamefont {Potter}}, \bibinfo {author} {\bibfnamefont {J.}~\bibnamefont {Pixley}}, \bibinfo {author} {\bibfnamefont {S.}~\bibnamefont {Gopalakrishnan}},\ and\ \bibinfo {author} {\bibfnamefont {R.}~\bibnamefont {Vasseur}},\ }\bibfield  {title} {\bibinfo {title} {Entanglement and charge-sharpening transitions in u (1) symmetric monitored quantum circuits},\ }\href {https://doi.org/10.1103/PhysRevX.12.041002} {\bibfield  {journal} {\bibinfo  {journal} {Physical Review X}\ }\textbf {\bibinfo {volume} {12}},\ \bibinfo {pages} {041002} (\bibinfo {year} {2022})}\BibitemShut {NoStop}%
\bibitem [{\citenamefont {Murthy}\ \emph {et~al.}(2023)\citenamefont {Murthy}, \citenamefont {Babakhani}, \citenamefont {Iniguez}, \citenamefont {Srednicki},\ and\ \citenamefont {Halpern}}]{murthy2023non}%
  \BibitemOpen
  \bibfield  {author} {\bibinfo {author} {\bibfnamefont {C.}~\bibnamefont {Murthy}}, \bibinfo {author} {\bibfnamefont {A.}~\bibnamefont {Babakhani}}, \bibinfo {author} {\bibfnamefont {F.}~\bibnamefont {Iniguez}}, \bibinfo {author} {\bibfnamefont {M.}~\bibnamefont {Srednicki}},\ and\ \bibinfo {author} {\bibfnamefont {N.~Y.}\ \bibnamefont {Halpern}},\ }\bibfield  {title} {\bibinfo {title} {Non-abelian eigenstate thermalization hypothesis},\ }\href {https://doi.org/10.1103/PhysRevLett.130.140402} {\bibfield  {journal} {\bibinfo  {journal} {Physical Review Letters}\ }\textbf {\bibinfo {volume} {130}},\ \bibinfo {pages} {140402} (\bibinfo {year} {2023})}\BibitemShut {NoStop}%
\bibitem [{\citenamefont {Kranzl}\ \emph {et~al.}(2023)\citenamefont {Kranzl}, \citenamefont {Lasek}, \citenamefont {Joshi}, \citenamefont {Kalev}, \citenamefont {Blatt}, \citenamefont {Roos},\ and\ \citenamefont {Halpern}}]{kranzl2023experimental}%
  \BibitemOpen
  \bibfield  {author} {\bibinfo {author} {\bibfnamefont {F.}~\bibnamefont {Kranzl}}, \bibinfo {author} {\bibfnamefont {A.}~\bibnamefont {Lasek}}, \bibinfo {author} {\bibfnamefont {M.~K.}\ \bibnamefont {Joshi}}, \bibinfo {author} {\bibfnamefont {A.}~\bibnamefont {Kalev}}, \bibinfo {author} {\bibfnamefont {R.}~\bibnamefont {Blatt}}, \bibinfo {author} {\bibfnamefont {C.~F.}\ \bibnamefont {Roos}},\ and\ \bibinfo {author} {\bibfnamefont {N.~Y.}\ \bibnamefont {Halpern}},\ }\bibfield  {title} {\bibinfo {title} {Experimental observation of thermalization with noncommuting charges},\ }\href {https://doi.org/10.1103/PRXQuantum.4.020318} {\bibfield  {journal} {\bibinfo  {journal} {PRX Quantum}\ }\textbf {\bibinfo {volume} {4}},\ \bibinfo {pages} {020318} (\bibinfo {year} {2023})}\BibitemShut {NoStop}%
\bibitem [{\citenamefont {Luitz}\ \emph {et~al.}(2020)\citenamefont {Luitz}, \citenamefont {Moessner}, \citenamefont {Sondhi},\ and\ \citenamefont {Khemani}}]{luitz2020prethermalization}%
  \BibitemOpen
  \bibfield  {author} {\bibinfo {author} {\bibfnamefont {D.~J.}\ \bibnamefont {Luitz}}, \bibinfo {author} {\bibfnamefont {R.}~\bibnamefont {Moessner}}, \bibinfo {author} {\bibfnamefont {S.}~\bibnamefont {Sondhi}},\ and\ \bibinfo {author} {\bibfnamefont {V.}~\bibnamefont {Khemani}},\ }\bibfield  {title} {\bibinfo {title} {Prethermalization without temperature},\ }\href {https://doi.org/10.1103/PhysRevX.10.021046} {\bibfield  {journal} {\bibinfo  {journal} {Physical Review X}\ }\textbf {\bibinfo {volume} {10}},\ \bibinfo {pages} {021046} (\bibinfo {year} {2020})}\BibitemShut {NoStop}%
\bibitem [{\citenamefont {Beatrez}\ \emph {et~al.}(2023)\citenamefont {Beatrez}, \citenamefont {Fleckenstein}, \citenamefont {Pillai}, \citenamefont {de~Leon~Sanchez}, \citenamefont {Akkiraju}, \citenamefont {Diaz~Alcala}, \citenamefont {Conti}, \citenamefont {Reshetikhin}, \citenamefont {Druga}, \citenamefont {Bukov} \emph {et~al.}}]{beatrez2023critical}%
  \BibitemOpen
  \bibfield  {author} {\bibinfo {author} {\bibfnamefont {W.}~\bibnamefont {Beatrez}}, \bibinfo {author} {\bibfnamefont {C.}~\bibnamefont {Fleckenstein}}, \bibinfo {author} {\bibfnamefont {A.}~\bibnamefont {Pillai}}, \bibinfo {author} {\bibfnamefont {E.}~\bibnamefont {de~Leon~Sanchez}}, \bibinfo {author} {\bibfnamefont {A.}~\bibnamefont {Akkiraju}}, \bibinfo {author} {\bibfnamefont {J.}~\bibnamefont {Diaz~Alcala}}, \bibinfo {author} {\bibfnamefont {S.}~\bibnamefont {Conti}}, \bibinfo {author} {\bibfnamefont {P.}~\bibnamefont {Reshetikhin}}, \bibinfo {author} {\bibfnamefont {E.}~\bibnamefont {Druga}}, \bibinfo {author} {\bibfnamefont {M.}~\bibnamefont {Bukov}}, \emph {et~al.},\ }\bibfield  {title} {\bibinfo {title} {Critical prethermal discrete time crystal created by two-frequency driving},\ }\href {https://doi.org/10.1038/s41567-022-01891-7} {\bibfield  {journal} {\bibinfo  {journal} {Nature Physics}\ }\textbf {\bibinfo {volume} {19}},\ \bibinfo {pages} {407} (\bibinfo {year} {2023})}\BibitemShut {NoStop}%
\bibitem [{\citenamefont {Trenkwalder}\ \emph {et~al.}(2016)\citenamefont {Trenkwalder}, \citenamefont {Spagnolli}, \citenamefont {Semeghini}, \citenamefont {Coop}, \citenamefont {Landini}, \citenamefont {Castilho}, \citenamefont {Pezze}, \citenamefont {Modugno}, \citenamefont {Inguscio}, \citenamefont {Smerzi} \emph {et~al.}}]{trenkwalder2016quantum}%
  \BibitemOpen
  \bibfield  {author} {\bibinfo {author} {\bibfnamefont {A.}~\bibnamefont {Trenkwalder}}, \bibinfo {author} {\bibfnamefont {G.}~\bibnamefont {Spagnolli}}, \bibinfo {author} {\bibfnamefont {G.}~\bibnamefont {Semeghini}}, \bibinfo {author} {\bibfnamefont {S.}~\bibnamefont {Coop}}, \bibinfo {author} {\bibfnamefont {M.}~\bibnamefont {Landini}}, \bibinfo {author} {\bibfnamefont {P.}~\bibnamefont {Castilho}}, \bibinfo {author} {\bibfnamefont {L.}~\bibnamefont {Pezze}}, \bibinfo {author} {\bibfnamefont {G.}~\bibnamefont {Modugno}}, \bibinfo {author} {\bibfnamefont {M.}~\bibnamefont {Inguscio}}, \bibinfo {author} {\bibfnamefont {A.}~\bibnamefont {Smerzi}}, \emph {et~al.},\ }\bibfield  {title} {\bibinfo {title} {Quantum phase transitions with parity-symmetry breaking and hysteresis},\ }\href {https://doi.org/10.1038/nphys3743} {\bibfield  {journal} {\bibinfo  {journal} {Nature physics}\ }\textbf {\bibinfo {volume} {12}},\ \bibinfo {pages} {826} (\bibinfo {year} {2016})}\BibitemShut {NoStop}%
\bibitem [{\citenamefont {Garc{\'\i}a-Pintos}\ \emph {et~al.}(2019)\citenamefont {Garc{\'\i}a-Pintos}, \citenamefont {Tielas},\ and\ \citenamefont {Del~Campo}}]{garcia2019spontaneous}%
  \BibitemOpen
  \bibfield  {author} {\bibinfo {author} {\bibfnamefont {L.~P.}\ \bibnamefont {Garc{\'\i}a-Pintos}}, \bibinfo {author} {\bibfnamefont {D.}~\bibnamefont {Tielas}},\ and\ \bibinfo {author} {\bibfnamefont {A.}~\bibnamefont {Del~Campo}},\ }\bibfield  {title} {\bibinfo {title} {Spontaneous symmetry breaking induced by quantum monitoring},\ }\href {https://doi.org/10.1103/PhysRevLett.123.090403} {\bibfield  {journal} {\bibinfo  {journal} {Physical Review Letters}\ }\textbf {\bibinfo {volume} {123}},\ \bibinfo {pages} {090403} (\bibinfo {year} {2019})}\BibitemShut {NoStop}%
\bibitem [{\citenamefont {Kokail}\ \emph {et~al.}(2019)\citenamefont {Kokail}, \citenamefont {Maier}, \citenamefont {van Bijnen}, \citenamefont {Brydges}, \citenamefont {Joshi}, \citenamefont {Jurcevic}, \citenamefont {Muschik}, \citenamefont {Silvi}, \citenamefont {Blatt}, \citenamefont {Roos} \emph {et~al.}}]{kokail2019self}%
  \BibitemOpen
  \bibfield  {author} {\bibinfo {author} {\bibfnamefont {C.}~\bibnamefont {Kokail}}, \bibinfo {author} {\bibfnamefont {C.}~\bibnamefont {Maier}}, \bibinfo {author} {\bibfnamefont {R.}~\bibnamefont {van Bijnen}}, \bibinfo {author} {\bibfnamefont {T.}~\bibnamefont {Brydges}}, \bibinfo {author} {\bibfnamefont {M.~K.}\ \bibnamefont {Joshi}}, \bibinfo {author} {\bibfnamefont {P.}~\bibnamefont {Jurcevic}}, \bibinfo {author} {\bibfnamefont {C.~A.}\ \bibnamefont {Muschik}}, \bibinfo {author} {\bibfnamefont {P.}~\bibnamefont {Silvi}}, \bibinfo {author} {\bibfnamefont {R.}~\bibnamefont {Blatt}}, \bibinfo {author} {\bibfnamefont {C.~F.}\ \bibnamefont {Roos}}, \emph {et~al.},\ }\bibfield  {title} {\bibinfo {title} {Self-verifying variational quantum simulation of lattice models},\ }\href {https://doi.org/10.1038/s41586-019-1177-4} {\bibfield  {journal} {\bibinfo  {journal} {Nature}\ }\textbf {\bibinfo {volume} {569}},\ \bibinfo {pages} {355} (\bibinfo {year} {2019})}\BibitemShut {NoStop}%
\bibitem [{\citenamefont {Choi}\ \emph {et~al.}(2020)\citenamefont {Choi}, \citenamefont {Zhou}, \citenamefont {Knowles}, \citenamefont {Landig}, \citenamefont {Choi},\ and\ \citenamefont {Lukin}}]{choi2020robust}%
  \BibitemOpen
  \bibfield  {author} {\bibinfo {author} {\bibfnamefont {J.}~\bibnamefont {Choi}}, \bibinfo {author} {\bibfnamefont {H.}~\bibnamefont {Zhou}}, \bibinfo {author} {\bibfnamefont {H.~S.}\ \bibnamefont {Knowles}}, \bibinfo {author} {\bibfnamefont {R.}~\bibnamefont {Landig}}, \bibinfo {author} {\bibfnamefont {S.}~\bibnamefont {Choi}},\ and\ \bibinfo {author} {\bibfnamefont {M.~D.}\ \bibnamefont {Lukin}},\ }\bibfield  {title} {\bibinfo {title} {Robust dynamic hamiltonian engineering of many-body spin systems},\ }\href {https://doi.org/10.1103/PhysRevX.10.031002} {\bibfield  {journal} {\bibinfo  {journal} {Phys. Rev. X}\ }\textbf {\bibinfo {volume} {10}},\ \bibinfo {pages} {031002} (\bibinfo {year} {2020})}\BibitemShut {NoStop}%
\bibitem [{\citenamefont {Tran}\ \emph {et~al.}(2021)\citenamefont {Tran}, \citenamefont {Su}, \citenamefont {Carney},\ and\ \citenamefont {Taylor}}]{tran2021faster}%
  \BibitemOpen
  \bibfield  {author} {\bibinfo {author} {\bibfnamefont {M.~C.}\ \bibnamefont {Tran}}, \bibinfo {author} {\bibfnamefont {Y.}~\bibnamefont {Su}}, \bibinfo {author} {\bibfnamefont {D.}~\bibnamefont {Carney}},\ and\ \bibinfo {author} {\bibfnamefont {J.~M.}\ \bibnamefont {Taylor}},\ }\bibfield  {title} {\bibinfo {title} {Faster digital quantum simulation by symmetry protection},\ }\href {https://doi.org/10.1103/PRXQuantum.2.010323} {\bibfield  {journal} {\bibinfo  {journal} {PRX Quantum}\ }\textbf {\bibinfo {volume} {2}},\ \bibinfo {pages} {010323} (\bibinfo {year} {2021})}\BibitemShut {NoStop}%
\bibitem [{\citenamefont {Richter}\ and\ \citenamefont {Pal}(2021)}]{richter2021simulating}%
  \BibitemOpen
  \bibfield  {author} {\bibinfo {author} {\bibfnamefont {J.}~\bibnamefont {Richter}}\ and\ \bibinfo {author} {\bibfnamefont {A.}~\bibnamefont {Pal}},\ }\bibfield  {title} {\bibinfo {title} {Simulating hydrodynamics on noisy intermediate-scale quantum devices with random circuits},\ }\href {https://doi.org/10.1103/PhysRevLett.126.230501} {\bibfield  {journal} {\bibinfo  {journal} {Physical Review Letters}\ }\textbf {\bibinfo {volume} {126}},\ \bibinfo {pages} {230501} (\bibinfo {year} {2021})}\BibitemShut {NoStop}%
\bibitem [{\citenamefont {Keenan}\ \emph {et~al.}(2023)\citenamefont {Keenan}, \citenamefont {Robertson}, \citenamefont {Murphy}, \citenamefont {Zhuk},\ and\ \citenamefont {Goold}}]{keenan2023evidence}%
  \BibitemOpen
  \bibfield  {author} {\bibinfo {author} {\bibfnamefont {N.}~\bibnamefont {Keenan}}, \bibinfo {author} {\bibfnamefont {N.~F.}\ \bibnamefont {Robertson}}, \bibinfo {author} {\bibfnamefont {T.}~\bibnamefont {Murphy}}, \bibinfo {author} {\bibfnamefont {S.}~\bibnamefont {Zhuk}},\ and\ \bibinfo {author} {\bibfnamefont {J.}~\bibnamefont {Goold}},\ }\bibfield  {title} {\bibinfo {title} {Evidence of kardar-parisi-zhang scaling on a digital quantum simulator},\ }\href {https://doi.org/10.1038/s41534-023-00742-4} {\bibfield  {journal} {\bibinfo  {journal} {npj Quantum Information}\ }\textbf {\bibinfo {volume} {9}},\ \bibinfo {pages} {72} (\bibinfo {year} {2023})}\BibitemShut {NoStop}%
\bibitem [{\citenamefont {Zhao}\ \emph {et~al.}(2023)\citenamefont {Zhao}, \citenamefont {Bukov}, \citenamefont {Heyl},\ and\ \citenamefont {Moessner}}]{zhao2023making}%
  \BibitemOpen
  \bibfield  {author} {\bibinfo {author} {\bibfnamefont {H.}~\bibnamefont {Zhao}}, \bibinfo {author} {\bibfnamefont {M.}~\bibnamefont {Bukov}}, \bibinfo {author} {\bibfnamefont {M.}~\bibnamefont {Heyl}},\ and\ \bibinfo {author} {\bibfnamefont {R.}~\bibnamefont {Moessner}},\ }\bibfield  {title} {\bibinfo {title} {Making trotterization adaptive and energy-self-correcting for nisq devices and beyond},\ }\href {https://doi.org/10.1103/PRXQuantum.4.030319} {\bibfield  {journal} {\bibinfo  {journal} {PRX Quantum}\ }\textbf {\bibinfo {volume} {4}},\ \bibinfo {pages} {030319} (\bibinfo {year} {2023})}\BibitemShut {NoStop}%
\bibitem [{\citenamefont {Zhao}\ \emph {et~al.}(2019)\citenamefont {Zhao}, \citenamefont {Mintert},\ and\ \citenamefont {Knolle}}]{zhao2019floquet}%
  \BibitemOpen
  \bibfield  {author} {\bibinfo {author} {\bibfnamefont {H.}~\bibnamefont {Zhao}}, \bibinfo {author} {\bibfnamefont {F.}~\bibnamefont {Mintert}},\ and\ \bibinfo {author} {\bibfnamefont {J.}~\bibnamefont {Knolle}},\ }\bibfield  {title} {\bibinfo {title} {Floquet time spirals and stable discrete-time quasicrystals in quasiperiodically driven quantum many-body systems},\ }\href {https://doi.org/10.1103/PhysRevB.100.134302} {\bibfield  {journal} {\bibinfo  {journal} {Physical Review B}\ }\textbf {\bibinfo {volume} {100}},\ \bibinfo {pages} {134302} (\bibinfo {year} {2019})}\BibitemShut {NoStop}%
\bibitem [{\citenamefont {Lapierre}\ \emph {et~al.}(2020)\citenamefont {Lapierre}, \citenamefont {Choo}, \citenamefont {Tiwari}, \citenamefont {Tauber}, \citenamefont {Neupert},\ and\ \citenamefont {Chitra}}]{lapierre2020fine}%
  \BibitemOpen
  \bibfield  {author} {\bibinfo {author} {\bibfnamefont {B.}~\bibnamefont {Lapierre}}, \bibinfo {author} {\bibfnamefont {K.}~\bibnamefont {Choo}}, \bibinfo {author} {\bibfnamefont {A.}~\bibnamefont {Tiwari}}, \bibinfo {author} {\bibfnamefont {C.}~\bibnamefont {Tauber}}, \bibinfo {author} {\bibfnamefont {T.}~\bibnamefont {Neupert}},\ and\ \bibinfo {author} {\bibfnamefont {R.}~\bibnamefont {Chitra}},\ }\bibfield  {title} {\bibinfo {title} {Fine structure of heating in a quasiperiodically driven critical quantum system},\ }\href {https://doi.org/10.1103/PhysRevResearch.2.033461} {\bibfield  {journal} {\bibinfo  {journal} {Physical Review Research}\ }\textbf {\bibinfo {volume} {2}},\ \bibinfo {pages} {033461} (\bibinfo {year} {2020})}\BibitemShut {NoStop}%
\bibitem [{\citenamefont {Wen}\ \emph {et~al.}(2021)\citenamefont {Wen}, \citenamefont {Fan}, \citenamefont {Vishwanath},\ and\ \citenamefont {Gu}}]{wen2021periodically}%
  \BibitemOpen
  \bibfield  {author} {\bibinfo {author} {\bibfnamefont {X.}~\bibnamefont {Wen}}, \bibinfo {author} {\bibfnamefont {R.}~\bibnamefont {Fan}}, \bibinfo {author} {\bibfnamefont {A.}~\bibnamefont {Vishwanath}},\ and\ \bibinfo {author} {\bibfnamefont {Y.}~\bibnamefont {Gu}},\ }\bibfield  {title} {\bibinfo {title} {Periodically, quasiperiodically, and randomly driven conformal field theories},\ }\href {https://doi.org/10.1103/PhysRevResearch.3.023044} {\bibfield  {journal} {\bibinfo  {journal} {Physical Review Research}\ }\textbf {\bibinfo {volume} {3}},\ \bibinfo {pages} {023044} (\bibinfo {year} {2021})}\BibitemShut {NoStop}%
\bibitem [{\citenamefont {Mori}\ \emph {et~al.}(2021)\citenamefont {Mori}, \citenamefont {Zhao}, \citenamefont {Mintert}, \citenamefont {Knolle},\ and\ \citenamefont {Moessner}}]{mori2021rigorous}%
  \BibitemOpen
  \bibfield  {author} {\bibinfo {author} {\bibfnamefont {T.}~\bibnamefont {Mori}}, \bibinfo {author} {\bibfnamefont {H.}~\bibnamefont {Zhao}}, \bibinfo {author} {\bibfnamefont {F.}~\bibnamefont {Mintert}}, \bibinfo {author} {\bibfnamefont {J.}~\bibnamefont {Knolle}},\ and\ \bibinfo {author} {\bibfnamefont {R.}~\bibnamefont {Moessner}},\ }\bibfield  {title} {\bibinfo {title} {Rigorous bounds on the heating rate in thue-morse quasiperiodically and randomly driven quantum many-body systems},\ }\href {https://doi.org/10.1103/PhysRevLett.127.050602} {\bibfield  {journal} {\bibinfo  {journal} {Physical Review Letters}\ }\textbf {\bibinfo {volume} {127}},\ \bibinfo {pages} {050602} (\bibinfo {year} {2021})}\BibitemShut {NoStop}%
\bibitem [{\citenamefont {Long}\ \emph {et~al.}(2022)\citenamefont {Long}, \citenamefont {Crowley},\ and\ \citenamefont {Chandran}}]{long2022many}%
  \BibitemOpen
  \bibfield  {author} {\bibinfo {author} {\bibfnamefont {D.~M.}\ \bibnamefont {Long}}, \bibinfo {author} {\bibfnamefont {P.~J.}\ \bibnamefont {Crowley}},\ and\ \bibinfo {author} {\bibfnamefont {A.}~\bibnamefont {Chandran}},\ }\bibfield  {title} {\bibinfo {title} {Many-body localization with quasiperiodic driving},\ }\href {https://doi.org/10.1103/PhysRevB.105.144204} {\bibfield  {journal} {\bibinfo  {journal} {Physical Review B}\ }\textbf {\bibinfo {volume} {105}},\ \bibinfo {pages} {144204} (\bibinfo {year} {2022})}\BibitemShut {NoStop}%
\bibitem [{\citenamefont {Nathan}\ \emph {et~al.}(2022)\citenamefont {Nathan}, \citenamefont {Martin},\ and\ \citenamefont {Refael}}]{nathan2022topological}%
  \BibitemOpen
  \bibfield  {author} {\bibinfo {author} {\bibfnamefont {F.}~\bibnamefont {Nathan}}, \bibinfo {author} {\bibfnamefont {I.}~\bibnamefont {Martin}},\ and\ \bibinfo {author} {\bibfnamefont {G.}~\bibnamefont {Refael}},\ }\bibfield  {title} {\bibinfo {title} {Topological frequency conversion in weyl semimetals},\ }\href {https://doi.org/10.1103/PhysRevResearch.4.043060} {\bibfield  {journal} {\bibinfo  {journal} {Physical Review Research}\ }\textbf {\bibinfo {volume} {4}},\ \bibinfo {pages} {043060} (\bibinfo {year} {2022})}\BibitemShut {NoStop}%
\bibitem [{\citenamefont {He}\ \emph {et~al.}(2023)\citenamefont {He}, \citenamefont {Ye}, \citenamefont {Gong}, \citenamefont {Liu}, \citenamefont {Murch}, \citenamefont {Yao},\ and\ \citenamefont {Zu}}]{he2023quasi}%
  \BibitemOpen
  \bibfield  {author} {\bibinfo {author} {\bibfnamefont {G.}~\bibnamefont {He}}, \bibinfo {author} {\bibfnamefont {B.}~\bibnamefont {Ye}}, \bibinfo {author} {\bibfnamefont {R.}~\bibnamefont {Gong}}, \bibinfo {author} {\bibfnamefont {Z.}~\bibnamefont {Liu}}, \bibinfo {author} {\bibfnamefont {K.~W.}\ \bibnamefont {Murch}}, \bibinfo {author} {\bibfnamefont {N.~Y.}\ \bibnamefont {Yao}},\ and\ \bibinfo {author} {\bibfnamefont {C.}~\bibnamefont {Zu}},\ }\bibfield  {title} {\bibinfo {title} {Quasi-floquet prethermalization in a disordered dipolar spin ensemble in diamond},\ }\href {https://doi.org/10.1103/PhysRevLett.131.130401} {\bibfield  {journal} {\bibinfo  {journal} {Physical Review Letters}\ }\textbf {\bibinfo {volume} {131}},\ \bibinfo {pages} {130401} (\bibinfo {year} {2023})}\BibitemShut {NoStop}%
\bibitem [{\citenamefont {Zhao}\ \emph {et~al.}(2021)\citenamefont {Zhao}, \citenamefont {Mintert}, \citenamefont {Moessner},\ and\ \citenamefont {Knolle}}]{zhao2021random}%
  \BibitemOpen
  \bibfield  {author} {\bibinfo {author} {\bibfnamefont {H.}~\bibnamefont {Zhao}}, \bibinfo {author} {\bibfnamefont {F.}~\bibnamefont {Mintert}}, \bibinfo {author} {\bibfnamefont {R.}~\bibnamefont {Moessner}},\ and\ \bibinfo {author} {\bibfnamefont {J.}~\bibnamefont {Knolle}},\ }\bibfield  {title} {\bibinfo {title} {Random multipolar driving: Tunably slow heating through spectral engineering},\ }\href {https://doi.org/10.1103/PhysRevLett.126.040601} {\bibfield  {journal} {\bibinfo  {journal} {Physical Review Letters}\ }\textbf {\bibinfo {volume} {126}},\ \bibinfo {pages} {040601} (\bibinfo {year} {2021})}\BibitemShut {NoStop}%
\bibitem [{\citenamefont {Guarnieri}\ \emph {et~al.}(2022)\citenamefont {Guarnieri}, \citenamefont {Mitchison}, \citenamefont {Purkayastha}, \citenamefont {Jaksch}, \citenamefont {Bu{\v{c}}a},\ and\ \citenamefont {Goold}}]{guarnieri2022time}%
  \BibitemOpen
  \bibfield  {author} {\bibinfo {author} {\bibfnamefont {G.}~\bibnamefont {Guarnieri}}, \bibinfo {author} {\bibfnamefont {M.~T.}\ \bibnamefont {Mitchison}}, \bibinfo {author} {\bibfnamefont {A.}~\bibnamefont {Purkayastha}}, \bibinfo {author} {\bibfnamefont {D.}~\bibnamefont {Jaksch}}, \bibinfo {author} {\bibfnamefont {B.}~\bibnamefont {Bu{\v{c}}a}},\ and\ \bibinfo {author} {\bibfnamefont {J.}~\bibnamefont {Goold}},\ }\bibfield  {title} {\bibinfo {title} {Time periodicity from randomness in quantum systems},\ }\href {https://doi.org/10.1103/PhysRevA.106.022209} {\bibfield  {journal} {\bibinfo  {journal} {Physical Review A}\ }\textbf {\bibinfo {volume} {106}},\ \bibinfo {pages} {022209} (\bibinfo {year} {2022})}\BibitemShut {NoStop}%
\bibitem [{Note1()}]{Note1}%
  \BibitemOpen
  \bibinfo {note} {The proof in \ref {sec:sm1} relies on the Baker-Campbell-Hausdorff (BCH) formula which is generalized to continuous drives via the Floquet-Magnus expansion. Since in the latter, the commutator structure decouples from the time-ordered integrals, HS can also be engineered for continuous drives.}\BibitemShut {Stop}%
\bibitem [{\citenamefont {Bukov}\ \emph {et~al.}(2015)\citenamefont {Bukov}, \citenamefont {D'Alessio},\ and\ \citenamefont {Polkovnikov}}]{bukov2015universal}%
  \BibitemOpen
  \bibfield  {author} {\bibinfo {author} {\bibfnamefont {M.}~\bibnamefont {Bukov}}, \bibinfo {author} {\bibfnamefont {L.}~\bibnamefont {D'Alessio}},\ and\ \bibinfo {author} {\bibfnamefont {A.}~\bibnamefont {Polkovnikov}},\ }\bibfield  {title} {\bibinfo {title} {Universal high-frequency behavior of periodically driven systems: from dynamical stabilization to floquet engineering},\ }\href {https://doi.org/10.1080/00018732.2015.1055918} {\bibfield  {journal} {\bibinfo  {journal} {Advances in Physics}\ }\textbf {\bibinfo {volume} {64}},\ \bibinfo {pages} {139} (\bibinfo {year} {2015})}\BibitemShut {NoStop}%
\bibitem [{\citenamefont {Mori}\ \emph {et~al.}(2016)\citenamefont {Mori}, \citenamefont {Kuwahara},\ and\ \citenamefont {Saito}}]{mori2016rigorous}%
  \BibitemOpen
  \bibfield  {author} {\bibinfo {author} {\bibfnamefont {T.}~\bibnamefont {Mori}}, \bibinfo {author} {\bibfnamefont {T.}~\bibnamefont {Kuwahara}},\ and\ \bibinfo {author} {\bibfnamefont {K.}~\bibnamefont {Saito}},\ }\bibfield  {title} {\bibinfo {title} {Rigorous bound on energy absorption and generic relaxation in periodically driven quantum systems},\ }\href {https://doi.org/10.1103/PhysRevLett.116.120401} {\bibfield  {journal} {\bibinfo  {journal} {Physical review letters}\ }\textbf {\bibinfo {volume} {116}},\ \bibinfo {pages} {120401} (\bibinfo {year} {2016})}\BibitemShut {NoStop}%
\bibitem [{\citenamefont {Abanin}\ \emph {et~al.}(2017)\citenamefont {Abanin}, \citenamefont {De~Roeck}, \citenamefont {Ho},\ and\ \citenamefont {Huveneers}}]{abanin2017effective}%
  \BibitemOpen
  \bibfield  {author} {\bibinfo {author} {\bibfnamefont {D.~A.}\ \bibnamefont {Abanin}}, \bibinfo {author} {\bibfnamefont {W.}~\bibnamefont {De~Roeck}}, \bibinfo {author} {\bibfnamefont {W.~W.}\ \bibnamefont {Ho}},\ and\ \bibinfo {author} {\bibfnamefont {F.}~\bibnamefont {Huveneers}},\ }\bibfield  {title} {\bibinfo {title} {Effective hamiltonians, prethermalization, and slow energy absorption in periodically driven many-body systems},\ }\href {https://doi.org/10.1103/PhysRevB.95.014112} {\bibfield  {journal} {\bibinfo  {journal} {Physical Review B}\ }\textbf {\bibinfo {volume} {95}},\ \bibinfo {pages} {014112} (\bibinfo {year} {2017})}\BibitemShut {NoStop}%
\bibitem [{Note2()}]{Note2}%
  \BibitemOpen
  \bibinfo {note} {Note the difference in notation between $Q_{[M]}$ which denotes the effective IFE Hamiltonian truncated to order $M$, and $Q_n$ which conserves the symmetry group $G_n$.}\BibitemShut {Stop}%
\bibitem [{\citenamefont {Xu}\ \emph {et~al.}(2018)\citenamefont {Xu}, \citenamefont {Chen}, \citenamefont {Zeng}, \citenamefont {Zhang}, \citenamefont {Song}, \citenamefont {Liu}, \citenamefont {Guo}, \citenamefont {Zhang}, \citenamefont {Xu}, \citenamefont {Deng} \emph {et~al.}}]{xu2018emulating}%
  \BibitemOpen
  \bibfield  {author} {\bibinfo {author} {\bibfnamefont {K.}~\bibnamefont {Xu}}, \bibinfo {author} {\bibfnamefont {J.-J.}\ \bibnamefont {Chen}}, \bibinfo {author} {\bibfnamefont {Y.}~\bibnamefont {Zeng}}, \bibinfo {author} {\bibfnamefont {Y.-R.}\ \bibnamefont {Zhang}}, \bibinfo {author} {\bibfnamefont {C.}~\bibnamefont {Song}}, \bibinfo {author} {\bibfnamefont {W.}~\bibnamefont {Liu}}, \bibinfo {author} {\bibfnamefont {Q.}~\bibnamefont {Guo}}, \bibinfo {author} {\bibfnamefont {P.}~\bibnamefont {Zhang}}, \bibinfo {author} {\bibfnamefont {D.}~\bibnamefont {Xu}}, \bibinfo {author} {\bibfnamefont {H.}~\bibnamefont {Deng}}, \emph {et~al.},\ }\bibfield  {title} {\bibinfo {title} {Emulating many-body localization with a superconducting quantum processor},\ }\href {https://link.aps.org/doi/10.1103/PhysRevLett.120.050507} {\bibfield  {journal} {\bibinfo  {journal} {Physical review letters}\ }\textbf {\bibinfo {volume} {120}},\ \bibinfo {pages} {050507} (\bibinfo {year} {2018})}\BibitemShut {NoStop}%
\bibitem [{\citenamefont {Scholl}\ \emph {et~al.}(2022)\citenamefont {Scholl}, \citenamefont {Williams}, \citenamefont {Bornet}, \citenamefont {Wallner}, \citenamefont {Barredo}, \citenamefont {Henriet}, \citenamefont {Signoles}, \citenamefont {Hainaut}, \citenamefont {Franz}, \citenamefont {Geier} \emph {et~al.}}]{scholl2022microwave}%
  \BibitemOpen
  \bibfield  {author} {\bibinfo {author} {\bibfnamefont {P.}~\bibnamefont {Scholl}}, \bibinfo {author} {\bibfnamefont {H.~J.}\ \bibnamefont {Williams}}, \bibinfo {author} {\bibfnamefont {G.}~\bibnamefont {Bornet}}, \bibinfo {author} {\bibfnamefont {F.}~\bibnamefont {Wallner}}, \bibinfo {author} {\bibfnamefont {D.}~\bibnamefont {Barredo}}, \bibinfo {author} {\bibfnamefont {L.}~\bibnamefont {Henriet}}, \bibinfo {author} {\bibfnamefont {A.}~\bibnamefont {Signoles}}, \bibinfo {author} {\bibfnamefont {C.}~\bibnamefont {Hainaut}}, \bibinfo {author} {\bibfnamefont {T.}~\bibnamefont {Franz}}, \bibinfo {author} {\bibfnamefont {S.}~\bibnamefont {Geier}}, \emph {et~al.},\ }\bibfield  {title} {\bibinfo {title} {Microwave engineering of programmable xxz hamiltonians in arrays of rydberg atoms},\ }\href {https://link.aps.org/doi/10.1103/PRXQuantum.3.020303} {\bibfield  {journal} {\bibinfo  {journal} {PRX Quantum}\ }\textbf {\bibinfo {volume} {3}},\ \bibinfo {pages} {020303} (\bibinfo {year} {2022})}\BibitemShut {NoStop}%
\bibitem [{Note3()}]{Note3}%
  \BibitemOpen
  \bibinfo {note} {Note that there is another $\protect \mathbb {Z}_2$ generated by $P_x=\DOTSB \prod@ \slimits@ _{i}\sigma ^x_i$ which will not be broken by the drive. However, since this is not a subgroup of the $\protect \mathbb {Z}_2$ generated by $P_z$, we do not consider it in the present HS\protect \xspace example.}\BibitemShut {Stop}%
\bibitem [{Note4()}]{Note4}%
  \BibitemOpen
  \bibinfo {note} {Such an initial state is a high-temperature state with respect to $Q_{\pm }^{(0)}$ as verified by the energy density which is close to zero.}\BibitemShut {Stop}%
\bibitem [{Note5()}]{Note5}%
  \BibitemOpen
  \bibinfo {note} {These threshold values are chosen for numerical simplicity and the scaling exponent of the prethermal lifetime in the high-frequency regime does not depend on these specific choices.}\BibitemShut {Stop}%
\bibitem [{\citenamefont {Hou}\ \emph {et~al.}(2024)\citenamefont {Hou}, \citenamefont {Fu}, \citenamefont {Moessner}, \citenamefont {Bukov},\ and\ \citenamefont {Zhao}}]{hou2024floquet}%
  \BibitemOpen
  \bibfield  {author} {\bibinfo {author} {\bibfnamefont {Y.}~\bibnamefont {Hou}}, \bibinfo {author} {\bibfnamefont {Z.}~\bibnamefont {Fu}}, \bibinfo {author} {\bibfnamefont {R.}~\bibnamefont {Moessner}}, \bibinfo {author} {\bibfnamefont {M.}~\bibnamefont {Bukov}},\ and\ \bibinfo {author} {\bibfnamefont {H.}~\bibnamefont {Zhao}},\ }\bibfield  {title} {\bibinfo {title} {Floquet-engineered emergent massive nambu-goldstone modes},\ }\href {https://doi.org/10.48550/arXiv.2409.01902} {\bibfield  {journal} {\bibinfo  {journal} {arXiv preprint arXiv:2409.01902}\ } (\bibinfo {year} {2024})}\BibitemShut {NoStop}%
\bibitem [{\citenamefont {Surace}\ \emph {et~al.}(2019)\citenamefont {Surace}, \citenamefont {Russomanno}, \citenamefont {Dalmonte}, \citenamefont {Silva}, \citenamefont {Fazio},\ and\ \citenamefont {Iemini}}]{surace2019floquet}%
  \BibitemOpen
  \bibfield  {author} {\bibinfo {author} {\bibfnamefont {F.~M.}\ \bibnamefont {Surace}}, \bibinfo {author} {\bibfnamefont {A.}~\bibnamefont {Russomanno}}, \bibinfo {author} {\bibfnamefont {M.}~\bibnamefont {Dalmonte}}, \bibinfo {author} {\bibfnamefont {A.}~\bibnamefont {Silva}}, \bibinfo {author} {\bibfnamefont {R.}~\bibnamefont {Fazio}},\ and\ \bibinfo {author} {\bibfnamefont {F.}~\bibnamefont {Iemini}},\ }\bibfield  {title} {\bibinfo {title} {Floquet time crystals in clock models},\ }\href {https://doi.org/10.1103/PhysRevB.99.104303} {\bibfield  {journal} {\bibinfo  {journal} {Physical Review B}\ }\textbf {\bibinfo {volume} {99}},\ \bibinfo {pages} {104303} (\bibinfo {year} {2019})}\BibitemShut {NoStop}%
\bibitem [{\citenamefont {Ahn}\ \emph {et~al.}(2019)\citenamefont {Ahn}, \citenamefont {Park},\ and\ \citenamefont {Yang}}]{ahn2019failure}%
  \BibitemOpen
  \bibfield  {author} {\bibinfo {author} {\bibfnamefont {J.}~\bibnamefont {Ahn}}, \bibinfo {author} {\bibfnamefont {S.}~\bibnamefont {Park}},\ and\ \bibinfo {author} {\bibfnamefont {B.-J.}\ \bibnamefont {Yang}},\ }\bibfield  {title} {\bibinfo {title} {Failure of nielsen-ninomiya theorem and fragile topology in two-dimensional systems with space-time inversion symmetry: application to twisted bilayer graphene at magic angle},\ }\href {https://link.aps.org/doi/10.1103/PhysRevX.9.021013} {\bibfield  {journal} {\bibinfo  {journal} {Physical Review X}\ }\textbf {\bibinfo {volume} {9}},\ \bibinfo {pages} {021013} (\bibinfo {year} {2019})}\BibitemShut {NoStop}%
\bibitem [{\citenamefont {Benalcazar}\ \emph {et~al.}(2019)\citenamefont {Benalcazar}, \citenamefont {Li},\ and\ \citenamefont {Hughes}}]{benalcazar2019quantization}%
  \BibitemOpen
  \bibfield  {author} {\bibinfo {author} {\bibfnamefont {W.~A.}\ \bibnamefont {Benalcazar}}, \bibinfo {author} {\bibfnamefont {T.}~\bibnamefont {Li}},\ and\ \bibinfo {author} {\bibfnamefont {T.~L.}\ \bibnamefont {Hughes}},\ }\bibfield  {title} {\bibinfo {title} {Quantization of fractional corner charge in c n-symmetric higher-order topological crystalline insulators},\ }\href {https://link.aps.org/doi/10.1103/PhysRevB.99.245151} {\bibfield  {journal} {\bibinfo  {journal} {Physical Review B}\ }\textbf {\bibinfo {volume} {99}},\ \bibinfo {pages} {245151} (\bibinfo {year} {2019})}\BibitemShut {NoStop}%
\bibitem [{\citenamefont {Khalaf}\ \emph {et~al.}(2018)\citenamefont {Khalaf}, \citenamefont {Po}, \citenamefont {Vishwanath},\ and\ \citenamefont {Watanabe}}]{khalaf2018symmetry}%
  \BibitemOpen
  \bibfield  {author} {\bibinfo {author} {\bibfnamefont {E.}~\bibnamefont {Khalaf}}, \bibinfo {author} {\bibfnamefont {H.~C.}\ \bibnamefont {Po}}, \bibinfo {author} {\bibfnamefont {A.}~\bibnamefont {Vishwanath}},\ and\ \bibinfo {author} {\bibfnamefont {H.}~\bibnamefont {Watanabe}},\ }\bibfield  {title} {\bibinfo {title} {Symmetry indicators and anomalous surface states of topological crystalline insulators},\ }\href {https://link.aps.org/doi/10.1103/PhysRevX.8.031070} {\bibfield  {journal} {\bibinfo  {journal} {Physical Review X}\ }\textbf {\bibinfo {volume} {8}},\ \bibinfo {pages} {031070} (\bibinfo {year} {2018})}\BibitemShut {NoStop}%
\bibitem [{\citenamefont {Schindler}\ \emph {et~al.}(2018)\citenamefont {Schindler}, \citenamefont {Cook}, \citenamefont {Vergniory}, \citenamefont {Wang}, \citenamefont {Parkin}, \citenamefont {Bernevig},\ and\ \citenamefont {Neupert}}]{schindler2018higher}%
  \BibitemOpen
  \bibfield  {author} {\bibinfo {author} {\bibfnamefont {F.}~\bibnamefont {Schindler}}, \bibinfo {author} {\bibfnamefont {A.~M.}\ \bibnamefont {Cook}}, \bibinfo {author} {\bibfnamefont {M.~G.}\ \bibnamefont {Vergniory}}, \bibinfo {author} {\bibfnamefont {Z.}~\bibnamefont {Wang}}, \bibinfo {author} {\bibfnamefont {S.~S.}\ \bibnamefont {Parkin}}, \bibinfo {author} {\bibfnamefont {B.~A.}\ \bibnamefont {Bernevig}},\ and\ \bibinfo {author} {\bibfnamefont {T.}~\bibnamefont {Neupert}},\ }\bibfield  {title} {\bibinfo {title} {Higher-order topological insulators},\ }\href {https://doi.org/10.1126/sciadv.aat0346} {\bibfield  {journal} {\bibinfo  {journal} {Science advances}\ }\textbf {\bibinfo {volume} {4}},\ \bibinfo {pages} {eaat0346} (\bibinfo {year} {2018})}\BibitemShut {NoStop}%
\bibitem [{\citenamefont {Schindler}\ \emph {et~al.}(2022)\citenamefont {Schindler}, \citenamefont {Tsirkin}, \citenamefont {Neupert}, \citenamefont {Andrei~Bernevig},\ and\ \citenamefont {Wieder}}]{schindler2022topological}%
  \BibitemOpen
  \bibfield  {author} {\bibinfo {author} {\bibfnamefont {F.}~\bibnamefont {Schindler}}, \bibinfo {author} {\bibfnamefont {S.~S.}\ \bibnamefont {Tsirkin}}, \bibinfo {author} {\bibfnamefont {T.}~\bibnamefont {Neupert}}, \bibinfo {author} {\bibfnamefont {B.}~\bibnamefont {Andrei~Bernevig}},\ and\ \bibinfo {author} {\bibfnamefont {B.~J.}\ \bibnamefont {Wieder}},\ }\bibfield  {title} {\bibinfo {title} {Topological zero-dimensional defect and flux states in three-dimensional insulators},\ }\href {https://doi.org/10.1038/s41467-022-33471-x} {\bibfield  {journal} {\bibinfo  {journal} {Nature communications}\ }\textbf {\bibinfo {volume} {13}},\ \bibinfo {pages} {5791} (\bibinfo {year} {2022})}\BibitemShut {NoStop}%
\bibitem [{Note6()}]{Note6}%
  \BibitemOpen
  \bibinfo {note} {Note that, since the system is non-interacting, heating to infinite temperature does not occur even at infinite times.}\BibitemShut {Stop}%
\bibitem [{\citenamefont {Jin}\ and\ \citenamefont {Knolle}(2024)}]{jin2024floquet}%
  \BibitemOpen
  \bibfield  {author} {\bibinfo {author} {\bibfnamefont {H.-K.}\ \bibnamefont {Jin}}\ and\ \bibinfo {author} {\bibfnamefont {J.}~\bibnamefont {Knolle}},\ }\bibfield  {title} {\bibinfo {title} {Floquet prethermal order by disorder},\ }\href {https://arxiv.org/abs/2403.17118} {\bibfield  {journal} {\bibinfo  {journal} {arXiv preprint arXiv:2403.17118}\ } (\bibinfo {year} {2024})}\BibitemShut {NoStop}%
\bibitem [{\citenamefont {Halimeh}\ \emph {et~al.}(2023)\citenamefont {Halimeh}, \citenamefont {Aidelsburger}, \citenamefont {Grusdt}, \citenamefont {Hauke},\ and\ \citenamefont {Yang}}]{halimeh2023cold}%
  \BibitemOpen
  \bibfield  {author} {\bibinfo {author} {\bibfnamefont {J.~C.}\ \bibnamefont {Halimeh}}, \bibinfo {author} {\bibfnamefont {M.}~\bibnamefont {Aidelsburger}}, \bibinfo {author} {\bibfnamefont {F.}~\bibnamefont {Grusdt}}, \bibinfo {author} {\bibfnamefont {P.}~\bibnamefont {Hauke}},\ and\ \bibinfo {author} {\bibfnamefont {B.}~\bibnamefont {Yang}},\ }\bibfield  {title} {\bibinfo {title} {Cold-atom quantum simulators of gauge theories},\ }\href {https://doi.org/10.48550/arXiv.2310.12201} {\bibfield  {journal} {\bibinfo  {journal} {arXiv preprint arXiv:2310.12201}\ } (\bibinfo {year} {2023})}\BibitemShut {NoStop}%
\bibitem [{\citenamefont {Faist}\ \emph {et~al.}(2020)\citenamefont {Faist}, \citenamefont {Nezami}, \citenamefont {Albert}, \citenamefont {Salton}, \citenamefont {Pastawski}, \citenamefont {Hayden},\ and\ \citenamefont {Preskill}}]{faist2020continuous}%
  \BibitemOpen
  \bibfield  {author} {\bibinfo {author} {\bibfnamefont {P.}~\bibnamefont {Faist}}, \bibinfo {author} {\bibfnamefont {S.}~\bibnamefont {Nezami}}, \bibinfo {author} {\bibfnamefont {V.~V.}\ \bibnamefont {Albert}}, \bibinfo {author} {\bibfnamefont {G.}~\bibnamefont {Salton}}, \bibinfo {author} {\bibfnamefont {F.}~\bibnamefont {Pastawski}}, \bibinfo {author} {\bibfnamefont {P.}~\bibnamefont {Hayden}},\ and\ \bibinfo {author} {\bibfnamefont {J.}~\bibnamefont {Preskill}},\ }\bibfield  {title} {\bibinfo {title} {Continuous symmetries and approximate quantum error correction},\ }\href {https://doi.org/10.1103/PhysRevX.10.041018} {\bibfield  {journal} {\bibinfo  {journal} {Physical Review X}\ }\textbf {\bibinfo {volume} {10}},\ \bibinfo {pages} {041018} (\bibinfo {year} {2020})}\BibitemShut {NoStop}%
\bibitem [{\citenamefont {Lieu}\ \emph {et~al.}(2020)\citenamefont {Lieu}, \citenamefont {Belyansky}, \citenamefont {Young}, \citenamefont {Lundgren}, \citenamefont {Albert},\ and\ \citenamefont {Gorshkov}}]{lieu2020symmetry}%
  \BibitemOpen
  \bibfield  {author} {\bibinfo {author} {\bibfnamefont {S.}~\bibnamefont {Lieu}}, \bibinfo {author} {\bibfnamefont {R.}~\bibnamefont {Belyansky}}, \bibinfo {author} {\bibfnamefont {J.~T.}\ \bibnamefont {Young}}, \bibinfo {author} {\bibfnamefont {R.}~\bibnamefont {Lundgren}}, \bibinfo {author} {\bibfnamefont {V.~V.}\ \bibnamefont {Albert}},\ and\ \bibinfo {author} {\bibfnamefont {A.~V.}\ \bibnamefont {Gorshkov}},\ }\bibfield  {title} {\bibinfo {title} {Symmetry breaking and error correction in open quantum systems},\ }\href {https://doi.org/10.1103/PhysRevLett.125.240405} {\bibfield  {journal} {\bibinfo  {journal} {Physical Review Letters}\ }\textbf {\bibinfo {volume} {125}},\ \bibinfo {pages} {240405} (\bibinfo {year} {2020})}\BibitemShut {NoStop}%
\bibitem [{\citenamefont {Zhu}\ \emph {et~al.}(2023)\citenamefont {Zhu}, \citenamefont {Tantivasadakarn}, \citenamefont {Vishwanath}, \citenamefont {Trebst},\ and\ \citenamefont {Verresen}}]{zhu2023nishimori}%
  \BibitemOpen
  \bibfield  {author} {\bibinfo {author} {\bibfnamefont {G.-Y.}\ \bibnamefont {Zhu}}, \bibinfo {author} {\bibfnamefont {N.}~\bibnamefont {Tantivasadakarn}}, \bibinfo {author} {\bibfnamefont {A.}~\bibnamefont {Vishwanath}}, \bibinfo {author} {\bibfnamefont {S.}~\bibnamefont {Trebst}},\ and\ \bibinfo {author} {\bibfnamefont {R.}~\bibnamefont {Verresen}},\ }\bibfield  {title} {\bibinfo {title} {Nishimori’s cat: stable long-range entanglement from finite-depth unitaries and weak measurements},\ }\href {https://doi.org/10.1103/PhysRevLett.131.200201} {\bibfield  {journal} {\bibinfo  {journal} {Physical Review Letters}\ }\textbf {\bibinfo {volume} {131}},\ \bibinfo {pages} {200201} (\bibinfo {year} {2023})}\BibitemShut {NoStop}%
\bibitem [{\citenamefont {Sun}\ \emph {et~al.}(2021)\citenamefont {Sun}, \citenamefont {Yang}, \citenamefont {Wang}, \citenamefont {Zhou}, \citenamefont {Su}, \citenamefont {Dai}, \citenamefont {Yuan},\ and\ \citenamefont {Pan}}]{sun2021realization}%
  \BibitemOpen
  \bibfield  {author} {\bibinfo {author} {\bibfnamefont {H.}~\bibnamefont {Sun}}, \bibinfo {author} {\bibfnamefont {B.}~\bibnamefont {Yang}}, \bibinfo {author} {\bibfnamefont {H.-Y.}\ \bibnamefont {Wang}}, \bibinfo {author} {\bibfnamefont {Z.-Y.}\ \bibnamefont {Zhou}}, \bibinfo {author} {\bibfnamefont {G.-X.}\ \bibnamefont {Su}}, \bibinfo {author} {\bibfnamefont {H.-N.}\ \bibnamefont {Dai}}, \bibinfo {author} {\bibfnamefont {Z.-S.}\ \bibnamefont {Yuan}},\ and\ \bibinfo {author} {\bibfnamefont {J.-W.}\ \bibnamefont {Pan}},\ }\bibfield  {title} {\bibinfo {title} {Realization of a bosonic antiferromagnet},\ }\href {https://doi.org/10.1038/s41567-021-01277-1} {\bibfield  {journal} {\bibinfo  {journal} {Nature Physics}\ }\textbf {\bibinfo {volume} {17}},\ \bibinfo {pages} {990} (\bibinfo {year} {2021})}\BibitemShut {NoStop}%
\bibitem [{\citenamefont {Wei}\ \emph {et~al.}(2022)\citenamefont {Wei}, \citenamefont {Rubio-Abadal}, \citenamefont {Ye}, \citenamefont {Machado}, \citenamefont {Kemp}, \citenamefont {Srakaew}, \citenamefont {Hollerith}, \citenamefont {Rui}, \citenamefont {Gopalakrishnan}, \citenamefont {Yao} \emph {et~al.}}]{wei2022quantum}%
  \BibitemOpen
  \bibfield  {author} {\bibinfo {author} {\bibfnamefont {D.}~\bibnamefont {Wei}}, \bibinfo {author} {\bibfnamefont {A.}~\bibnamefont {Rubio-Abadal}}, \bibinfo {author} {\bibfnamefont {B.}~\bibnamefont {Ye}}, \bibinfo {author} {\bibfnamefont {F.}~\bibnamefont {Machado}}, \bibinfo {author} {\bibfnamefont {J.}~\bibnamefont {Kemp}}, \bibinfo {author} {\bibfnamefont {K.}~\bibnamefont {Srakaew}}, \bibinfo {author} {\bibfnamefont {S.}~\bibnamefont {Hollerith}}, \bibinfo {author} {\bibfnamefont {J.}~\bibnamefont {Rui}}, \bibinfo {author} {\bibfnamefont {S.}~\bibnamefont {Gopalakrishnan}}, \bibinfo {author} {\bibfnamefont {N.~Y.}\ \bibnamefont {Yao}}, \emph {et~al.},\ }\bibfield  {title} {\bibinfo {title} {Quantum gas microscopy of kardar-parisi-zhang superdiffusion},\ }\href {https://doi.org/10.1126/science.abk2397} {\bibfield  {journal} {\bibinfo  {journal} {Science}\ }\textbf {\bibinfo {volume} {376}},\ \bibinfo {pages} {716} (\bibinfo {year} {2022})}\BibitemShut {NoStop}%
\bibitem [{\citenamefont {Watanabe}\ \emph {et~al.}(2013)\citenamefont {Watanabe}, \citenamefont {Brauner},\ and\ \citenamefont {Murayama}}]{watanabe2013massive}%
  \BibitemOpen
  \bibfield  {author} {\bibinfo {author} {\bibfnamefont {H.}~\bibnamefont {Watanabe}}, \bibinfo {author} {\bibfnamefont {T.}~\bibnamefont {Brauner}},\ and\ \bibinfo {author} {\bibfnamefont {H.}~\bibnamefont {Murayama}},\ }\bibfield  {title} {\bibinfo {title} {Massive nambu-goldstone bosons},\ }\href {https://doi.org/10.1103/PhysRevLett.111.021601} {\bibfield  {journal} {\bibinfo  {journal} {Physical Review Letters}\ }\textbf {\bibinfo {volume} {111}},\ \bibinfo {pages} {021601} (\bibinfo {year} {2013})}\BibitemShut {NoStop}%
\bibitem [{\citenamefont {Albert}\ and\ \citenamefont {Jiang}(2014)}]{albert2014symmetries}%
  \BibitemOpen
  \bibfield  {author} {\bibinfo {author} {\bibfnamefont {V.~V.}\ \bibnamefont {Albert}}\ and\ \bibinfo {author} {\bibfnamefont {L.}~\bibnamefont {Jiang}},\ }\bibfield  {title} {\bibinfo {title} {Symmetries and conserved quantities in lindblad master equations},\ }\href {https://doi.org/10.1103/PhysRevA.89.022118} {\bibfield  {journal} {\bibinfo  {journal} {Physical Review A}\ }\textbf {\bibinfo {volume} {89}},\ \bibinfo {pages} {022118} (\bibinfo {year} {2014})}\BibitemShut {NoStop}%
\bibitem [{\citenamefont {Mori}(2018)}]{mori2018floquet}%
  \BibitemOpen
  \bibfield  {author} {\bibinfo {author} {\bibfnamefont {T.}~\bibnamefont {Mori}},\ }\bibfield  {title} {\bibinfo {title} {Floquet prethermalization in periodically driven classical spin systems},\ }\href {https://doi.org/10.1103/PhysRevB.98.104303} {\bibfield  {journal} {\bibinfo  {journal} {Physical Review B}\ }\textbf {\bibinfo {volume} {98}},\ \bibinfo {pages} {104303} (\bibinfo {year} {2018})}\BibitemShut {NoStop}%
\bibitem [{\citenamefont {Mori}(2023)}]{mori2023floquet}%
  \BibitemOpen
  \bibfield  {author} {\bibinfo {author} {\bibfnamefont {T.}~\bibnamefont {Mori}},\ }\bibfield  {title} {\bibinfo {title} {Floquet states in open quantum systems},\ }\href {https://doi.org/10.1146/annurev-conmatphys-040721-015537} {\bibfield  {journal} {\bibinfo  {journal} {Annual Review of Condensed Matter Physics}\ }\textbf {\bibinfo {volume} {14}},\ \bibinfo {pages} {35} (\bibinfo {year} {2023})}\BibitemShut {NoStop}%
\bibitem [{\citenamefont {Sieberer}\ \emph {et~al.}(2023)\citenamefont {Sieberer}, \citenamefont {Buchhold}, \citenamefont {Marino},\ and\ \citenamefont {Diehl}}]{sieberer2023universality}%
  \BibitemOpen
  \bibfield  {author} {\bibinfo {author} {\bibfnamefont {L.~M.}\ \bibnamefont {Sieberer}}, \bibinfo {author} {\bibfnamefont {M.}~\bibnamefont {Buchhold}}, \bibinfo {author} {\bibfnamefont {J.}~\bibnamefont {Marino}},\ and\ \bibinfo {author} {\bibfnamefont {S.}~\bibnamefont {Diehl}},\ }\bibfield  {title} {\bibinfo {title} {Universality in driven open quantum matter},\ }\href {https://doi.org/10.48550/arXiv.2312.03073} {\bibfield  {journal} {\bibinfo  {journal} {arXiv preprint arXiv:2312.03073}\ } (\bibinfo {year} {2023})}\BibitemShut {NoStop}%
\bibitem [{\citenamefont {Howell}\ \emph {et~al.}(2019)\citenamefont {Howell}, \citenamefont {Weinberg}, \citenamefont {Sels}, \citenamefont {Polkovnikov},\ and\ \citenamefont {Bukov}}]{howell2019asymptotic}%
  \BibitemOpen
  \bibfield  {author} {\bibinfo {author} {\bibfnamefont {O.}~\bibnamefont {Howell}}, \bibinfo {author} {\bibfnamefont {P.}~\bibnamefont {Weinberg}}, \bibinfo {author} {\bibfnamefont {D.}~\bibnamefont {Sels}}, \bibinfo {author} {\bibfnamefont {A.}~\bibnamefont {Polkovnikov}},\ and\ \bibinfo {author} {\bibfnamefont {M.}~\bibnamefont {Bukov}},\ }\bibfield  {title} {\bibinfo {title} {Asymptotic prethermalization in periodically driven classical spin chains},\ }\href {https://doi.org/10.1103/PhysRevLett.122.010602} {\bibfield  {journal} {\bibinfo  {journal} {Physical review letters}\ }\textbf {\bibinfo {volume} {122}},\ \bibinfo {pages} {010602} (\bibinfo {year} {2019})}\BibitemShut {NoStop}%
\bibitem [{\citenamefont {Pizzi}\ \emph {et~al.}(2021)\citenamefont {Pizzi}, \citenamefont {Nunnenkamp},\ and\ \citenamefont {Knolle}}]{pizzi2021classical}%
  \BibitemOpen
  \bibfield  {author} {\bibinfo {author} {\bibfnamefont {A.}~\bibnamefont {Pizzi}}, \bibinfo {author} {\bibfnamefont {A.}~\bibnamefont {Nunnenkamp}},\ and\ \bibinfo {author} {\bibfnamefont {J.}~\bibnamefont {Knolle}},\ }\bibfield  {title} {\bibinfo {title} {Classical prethermal phases of matter},\ }\href {https://doi.org/10.1103/PhysRevLett.127.140602} {\bibfield  {journal} {\bibinfo  {journal} {Physical Review Letters}\ }\textbf {\bibinfo {volume} {127}},\ \bibinfo {pages} {140602} (\bibinfo {year} {2021})}\BibitemShut {NoStop}%
\bibitem [{\citenamefont {Ye}\ \emph {et~al.}(2021)\citenamefont {Ye}, \citenamefont {Machado},\ and\ \citenamefont {Yao}}]{ye2021floquet}%
  \BibitemOpen
  \bibfield  {author} {\bibinfo {author} {\bibfnamefont {B.}~\bibnamefont {Ye}}, \bibinfo {author} {\bibfnamefont {F.}~\bibnamefont {Machado}},\ and\ \bibinfo {author} {\bibfnamefont {N.~Y.}\ \bibnamefont {Yao}},\ }\bibfield  {title} {\bibinfo {title} {Floquet phases of matter via classical prethermalization},\ }\href {https://doi.org/10.1103/PhysRevLett.127.140603} {\bibfield  {journal} {\bibinfo  {journal} {Physical Review Letters}\ }\textbf {\bibinfo {volume} {127}},\ \bibinfo {pages} {140603} (\bibinfo {year} {2021})}\BibitemShut {NoStop}%
\bibitem [{\citenamefont {Yue}\ and\ \citenamefont {Cai}(2023)}]{yue2023prethermal}%
  \BibitemOpen
  \bibfield  {author} {\bibinfo {author} {\bibfnamefont {M.}~\bibnamefont {Yue}}\ and\ \bibinfo {author} {\bibfnamefont {Z.}~\bibnamefont {Cai}},\ }\bibfield  {title} {\bibinfo {title} {Prethermal time-crystalline spin ice and monopole confinement in a driven magnet},\ }\href {https://doi.org/10.1103/PhysRevLett.131.056502} {\bibfield  {journal} {\bibinfo  {journal} {Physical Review Letters}\ }\textbf {\bibinfo {volume} {131}},\ \bibinfo {pages} {056502} (\bibinfo {year} {2023})}\BibitemShut {NoStop}%
\bibitem [{\citenamefont {Yan}\ \emph {et~al.}(2023)\citenamefont {Yan}, \citenamefont {Moessner},\ and\ \citenamefont {Zhao}}]{yan2023prethermalization}%
  \BibitemOpen
  \bibfield  {author} {\bibinfo {author} {\bibfnamefont {J.}~\bibnamefont {Yan}}, \bibinfo {author} {\bibfnamefont {R.}~\bibnamefont {Moessner}},\ and\ \bibinfo {author} {\bibfnamefont {H.}~\bibnamefont {Zhao}},\ }\bibfield  {title} {\bibinfo {title} {Prethermalization in aperiodically kicked many-body dynamics},\ }\href {https://doi.org/10.48550/arXiv.2306.16144} {\bibfield  {journal} {\bibinfo  {journal} {arXiv preprint arXiv:2306.16144}\ } (\bibinfo {year} {2023})}\BibitemShut {NoStop}%
\bibitem [{\citenamefont {Bu{\v{c}}a}\ and\ \citenamefont {Prosen}(2012)}]{buvca2012note}%
  \BibitemOpen
  \bibfield  {author} {\bibinfo {author} {\bibfnamefont {B.}~\bibnamefont {Bu{\v{c}}a}}\ and\ \bibinfo {author} {\bibfnamefont {T.}~\bibnamefont {Prosen}},\ }\bibfield  {title} {\bibinfo {title} {A note on symmetry reductions of the lindblad equation: transport in constrained open spin chains},\ }\href {https://doi.org/10.1088/1367-2630/14/7/073007} {\bibfield  {journal} {\bibinfo  {journal} {New Journal of Physics}\ }\textbf {\bibinfo {volume} {14}},\ \bibinfo {pages} {073007} (\bibinfo {year} {2012})}\BibitemShut {NoStop}%
\bibitem [{\citenamefont {Surace}\ and\ \citenamefont {Motrunich}(2023)}]{surace2023weak}%
  \BibitemOpen
  \bibfield  {author} {\bibinfo {author} {\bibfnamefont {F.~M.}\ \bibnamefont {Surace}}\ and\ \bibinfo {author} {\bibfnamefont {O.}~\bibnamefont {Motrunich}},\ }\bibfield  {title} {\bibinfo {title} {Weak integrability breaking perturbations of integrable models},\ }\href {https://doi.org/10.1103/PhysRevResearch.5.043019} {\bibfield  {journal} {\bibinfo  {journal} {Phys. Rev. Res.}\ }\textbf {\bibinfo {volume} {5}},\ \bibinfo {pages} {043019} (\bibinfo {year} {2023})}\BibitemShut {NoStop}%
\bibitem [{\citenamefont {Ikeda}\ and\ \citenamefont {Polkovnikov}(2021)}]{ikeda2021fermi}%
  \BibitemOpen
  \bibfield  {author} {\bibinfo {author} {\bibfnamefont {T.~N.}\ \bibnamefont {Ikeda}}\ and\ \bibinfo {author} {\bibfnamefont {A.}~\bibnamefont {Polkovnikov}},\ }\bibfield  {title} {\bibinfo {title} {Fermi's golden rule for heating in strongly driven floquet systems},\ }\href {https://doi.org/10.1103/PhysRevB.104.134308} {\bibfield  {journal} {\bibinfo  {journal} {Physical Review B}\ }\textbf {\bibinfo {volume} {104}},\ \bibinfo {pages} {134308} (\bibinfo {year} {2021})}\BibitemShut {NoStop}%
\bibitem [{\citenamefont {Yeh}\ \emph {et~al.}(2023)\citenamefont {Yeh}, \citenamefont {Rosch},\ and\ \citenamefont {Mitra}}]{yeh2023decay}%
  \BibitemOpen
  \bibfield  {author} {\bibinfo {author} {\bibfnamefont {H.-C.}\ \bibnamefont {Yeh}}, \bibinfo {author} {\bibfnamefont {A.}~\bibnamefont {Rosch}},\ and\ \bibinfo {author} {\bibfnamefont {A.}~\bibnamefont {Mitra}},\ }\bibfield  {title} {\bibinfo {title} {Decay rates of almost strong modes in floquet spin chains beyond fermi's golden rule},\ }\href {https://doi.org/10.1103/PhysRevB.108.075112} {\bibfield  {journal} {\bibinfo  {journal} {Phys. Rev. B}\ }\textbf {\bibinfo {volume} {108}},\ \bibinfo {pages} {075112} (\bibinfo {year} {2023})}\BibitemShut {NoStop}%
\bibitem [{\citenamefont {Fleckenstein}\ and\ \citenamefont {Bukov}(2021)}]{fleckenstein2021thermalization}%
  \BibitemOpen
  \bibfield  {author} {\bibinfo {author} {\bibfnamefont {C.}~\bibnamefont {Fleckenstein}}\ and\ \bibinfo {author} {\bibfnamefont {M.}~\bibnamefont {Bukov}},\ }\bibfield  {title} {\bibinfo {title} {Thermalization and prethermalization in periodically kicked quantum spin chains},\ }\href {https://doi.org/10.1103/PhysRevB.103.144307} {\bibfield  {journal} {\bibinfo  {journal} {Physical Review B}\ }\textbf {\bibinfo {volume} {103}},\ \bibinfo {pages} {144307} (\bibinfo {year} {2021})}\BibitemShut {NoStop}%
\bibitem [{\citenamefont {Benalcazar}\ \emph {et~al.}(2017)\citenamefont {Benalcazar}, \citenamefont {Bernevig},\ and\ \citenamefont {Hughes}}]{benalcazar2017quantized}%
  \BibitemOpen
  \bibfield  {author} {\bibinfo {author} {\bibfnamefont {W.~A.}\ \bibnamefont {Benalcazar}}, \bibinfo {author} {\bibfnamefont {B.~A.}\ \bibnamefont {Bernevig}},\ and\ \bibinfo {author} {\bibfnamefont {T.~L.}\ \bibnamefont {Hughes}},\ }\bibfield  {title} {\bibinfo {title} {Quantized electric multipole insulators},\ }\href {https://www.science.org/doi/abs/10.1126/science.aah6442} {\bibfield  {journal} {\bibinfo  {journal} {Science}\ }\textbf {\bibinfo {volume} {357}},\ \bibinfo {pages} {61} (\bibinfo {year} {2017})}\BibitemShut {NoStop}%
\end{thebibliography}%

\clearpage

\appendix

\let\addcontentsline\oldaddcontentsline
\cleardoublepage
\onecolumngrid

	

\tableofcontents
\label{SM}

\section{GENERAL PROTOCOL FOR ARBITRARY ORDER OF \HS}
\label{sec:sm1}

 \subsection{General driving protocol: Iterative proof}
\label{sec.generalproof}

Here, we prove that the recursive time-dependent construction in Eq.~\eqref{eq:general_seq}, introduced in the main text, can hierarchically realize the symmetries of the symmetry ladder 
\begin{equation}
    G_{n}\supset G_{n-1} \supset G_{n-2} \supset \cdots \supset G_1 \supset G_0.
\label{equ1}
\end{equation} 
We illustrate the basic idea with a simple and concrete example before discussing the general proof. 

Consider a Hamiltonian $H_1$ that preserves the symmetry $G_1$ and a Hamiltonian $H_0$ that only preserves a subgroup $G_0$ of $G_1$, i.e., $G_1 \supset G_0$. We construct the following Floquet operator:
\begin{equation}
\label{eq.U1construction}
    U_1 \equiv e^{-iQ_1l_1T} = e^{-iH_0T}e^{-iH_1T}e^{iH_0T}e^{-iH_1T},
\end{equation}
where $l_{n} \equiv -2+3\times2^{n}$. 
The effective Hamiltonian reads $Q_1 = Q_1^{(0)}+Q_1^{(1)}+\mathcal{O}(T^2)$, with $Q_1^{(0)} \sim H_1$ preserving $G_1$ and $Q_1^{(1)} \sim T[H_1, H_0]$ reducing $G_1$ to $G_0$. Notice that in Eq.~\eqref{eq.U1construction} the prefactors in front of the two $H_{0}$ operators differ by a sign, ensuring the exact cancellation of $H_1$ symmetry-reducing terms in the leading-order (time-average) $Q_1^{(0)}$. This mechanism is the key ingredient for the proposed recursive construction.

Now, we go one step further by introducing yet a new Hamiltonian $H_2$ that preserves the symmetry $G_2$, such that $G_2\supset G_1 \supset G_0$. Therefore, $H_2$ also preserves $G_1$ and $G_0$ by construction. The new time- evolution operator is 
\begin{equation}
    U_2 \equiv e^{-iQ_2l_2T} = e^{-iQ_1l_1T}e^{-iH_2T}e^{iQ_1l_1T}e^{-iH_2T},
\end{equation}
where $Q_2 = Q_2^{(0)} + Q_2^{(1)} + Q_2^{(2)} + \mathcal{O}(T^3)$. Similar to the case above, the prefactors in front of $Q_1$ differ by a sign, and one can verify that $Q_2^{(0)} \sim H_2$ preserves $G_2$, and $Q_2^{(1)} \sim T[H_2, Q_1^{(0)}]\sim T[H_2, H_1]$ reduces $G_2$ to $G_1$. Finally, $Q_2^{(2)} \sim T^2(\alpha_1[H_2, Q_1^{(1)}]+\alpha_2[H_2,[H_2,Q_1^{(0)}]+\alpha_3[Q_1^{(0)}, [Q_1^{(0)},H_2]])$ for some coefficients $\alpha_i$, and, importantly, the term with prefactor $\alpha_1$ reduces $G_1$ to $G_0$ since $Q_1^{(1)}$ does so alone. In general, it follows that the $m$th order term $Q_2^{(m)}$ contains at most the $(m-1)$th order term  $Q_1^{(m-1)}$ from the IFE expansion of $Q_1$.

These basic observations suggest that if we iteratively construct a new evolution operator $U_n$, the previous \HS structure can be embedded into the effective Hamiltonian of $U_n$, while the zeroth order of the effective Hamiltonian is proportional to the newly added Hamiltonian which obeys the highest symmetry group.
In addition, the structure of the level $(n-1)$ \HS ladder is embedded in the level $n$ \HS ladder.
In practice, $e^{iQ_{1}l_{1}T}$ can be implemented by reversing the order of the temporal sequence of Eq.~\eqref{eq.U1construction} and conjugating each individual driving element (i.e., going backward in time).

We can now give a more general construction and its proof based on induction. Consider a set of Hamiltonians $\{H_r\}$ with a symmetry ladder $\{G_{r}\}$ and corresponding symmetry generators $\{S_{r}\}$:
\begin{equation}
    \begin{aligned}
        &H_{n-1}, H_{n-2}, \cdots, H_1, H_0; \\
        &G_{n-1}\supset G_{n-2} \supset \cdots \supset G_1 \supset G_0; \\
        &[H_{p}, S_{q}] = 0, \forall q\le p.
    \end{aligned}
\end{equation}
Let us assume that the $(n-1)$th order time-evolution operator $U_{n-1}= e^{-iQ_{n-1}l_{n-1}T}$ has the following property
\begin{equation}
    \begin{aligned}
    \label{Eq.induction}
        &Q_{n-1} = \sum_{m}Q_{n-1}^{(m)},\\
        &[Q_{n-1}^{(m)}, S_q] = 0, \forall q<n-m-1, \qquad [Q_{n-1}^{(m)}, S_q] \ne 0, \forall q\ge n-m-1,\\
    \end{aligned}
\end{equation}
such that $U_{n-1}$ already implements  the above level-$(n-1)$ \HS. 

Next we add a new Hamiltonian $H_n$ which obeys a higher symmetry $G_n$,
\begin{equation}
  [H_n,S_n] = 0, G_n\supset G_{n-1},    
\end{equation}
and extend the drive protocol to
\begin{equation}
\label{eq.protocol}
    U_n  =  e^{-iQ_{n-1} l_{n-1} T} e^{-iH_n T} e^{iQ_{n-1} l_{n-1} T} e^{-iH_n T}\equiv e^{-iQ_n l_{n} T}.
\end{equation}
Again, the prefactors in front of two $Q_{n-1}$ differ by a sign. One can check that, for $n=1$, we recover Eq.~\eqref{eq.U1construction}.

Using the Baker-Campbell-Hausdorff (BCH) expansion, we get the perturbative expansion of $Q_n$ as a power series of $T$
\begin{equation}
\begin{aligned}
    &Q_n = \frac{2}{l_n} H_n + \lambda_1T[H_n,Q_{n-1}] + \lambda_2T^2([H_n,[H_n,Q_{n-1}]] + l_{n-1}[Q_{n-1}, [ Q_{n-1}, H_n]]) + \cdots,\\
    &\lambda_1 = i\frac{l_{n-1}}{l_n}, \lambda_2 = -\frac{l_{n-1}}{2l_n}.
\end{aligned}
\end{equation}
Observe that the leading order term preserves $G_n$; the strength of the $G_{n-1}$-reducing term is renormalized by an extra factor of $T$; the strength of the $G_{n-2}$-reducing term is renormalized by the extra factor $T^2$, etc. 
More precisely,
the $\mathcal{O}(T^m)$ term in BCH expansion of $Q_{n}$ takes the following form
\begin{equation}
    Q_{n}^{(m)} = \sum_{p=1}^{m}f_p(Q_{n-1}^{(m-p), (m-p-1), \cdots, 0}),
\end{equation}
where $f_p(Q_{n-1}^{(m-p), (m-p-1), \cdots, 0})$ involves nested commutators of $H_n$ and $Q_{n-1}^{(m-p)}, Q_{n-1}^{(m-p-1)}, \cdots, Q_{n-1}^{0}$; for example, $f_1(Q_{n-1}^{(m-1)}) = \lambda_1T[H_n, Q_{n-1}^{(m-1)}]$. Since the Hamiltonian with the highest order of $T$ in $f_p(Q_{n-1}^{(m-p), (m-p-1), \cdots, 0})$ is $Q_{n-1}^{(m-p)}$, which explicitly breaks the symmetry group $G_{n-1-(m-p)}$ but preserves all lower-order symmetries $G_q$, we have
\begin{equation}
    [f_p(Q_{n-1}^{(m-p),\cdots}), S_q] = 0, \qquad \forall q<n-1-m+p.
\end{equation}
Therefore, for $Q^{(m)}_{n}$, the first symmetry group along the ladder which is explicitly broken is $G_{n-m}$, i.e.,
\begin{equation}
    [Q_{n}^{(l)}, S_q] = 0,\qquad  \forall q<n-m.
\end{equation} 
Note that the conditions in Eq.~\eqref{Eq.induction} are now generalized to $Q_n$, which completes the induction step. The new symmetry ladder is Eq.~\eqref{equ1}, as desired.

\subsection{Implementing symmetry breaking at arbitrary order of the inverse-frequency expansion}

The \HS structure is not necessary to have the perturbations aligned order by order in the effective Hamiltonian. We can make symmetry-breaking terms appear in an arbitrarily high-order term using our proposed protocol. The proof is very intuitive: Assuming that we already have the evolution operator $U_{n-1}\equiv e^{-i\ell_{n-1}Q_{n-1}T} = e^{-i\ell_{n-2}Q_{n-2}T}e^{-iH_{n-1}T}e^{i\ell_{n-2}Q_{n-2}T}e^{-iH_{n-1}T}$ with \HS structure, we know that 
 normally, $Q_{n-1}^{(1)}$ breaks $G_{n-1}$. However, if we impose that the newly introduced Hamiltonian $H_n$ in the protocol preserves the symmetry $G_{n-1}$ rather than $G_n$, it is clear that $Q_{n}^{(0)}$ and $Q_{n}^{(1)}$ will both preserve $G_{n-1}$, while the $G_{n-1}$-breaking term will become a higher-order perturbation. 

One of the simplest ways to engineer this is to choose $H_n = H_{n-1}$; the evolution operator in this case is
\begin{equation}
U_n \equiv  e^{-i\ell_{n}Q_{n}T} = (e^{-i\ell_{n-2}Q_{n-2}T}e^{-iH_{n-1}T}e^{i\ell_{n-2}Q_{n-2}T})e^{-iH_{n-1}T} (e^{-i\ell_{n-2}Q_{n-2}T}e^{iH_{n-1}T}e^{i\ell_{n-2}Q_{n-2}T})e^{-iH_{n-1}T}.
\end{equation}
Repeating this operation we can push the symmetry-breaking term to an arbitrarily high order. Moreover, this observation holds for any symmetry in the symmetry ladder.

\subsection{Shortening the \HS drive sequence}
The general construction shown in Appendix.~\ref{sec.generalproof} is exponentially long in $n$. Here, we illustrate the possibility of using additional properties of the generating Hamiltonians to shorten the drive sequence. Consider the example in Eq.~\eqref{equ:evo} discussed in the main text,
\begin{equation}
    U_2 = (e^{-i H_0T}e^{-i H_1T}e^{-i H_2T})(e^{iH_0T}e^{iH_1T}e^{-iH_2T}).
\end{equation}
Using the property $[H_0, H_1+H_2]=0$ we can achieve level-2 \HS with a shorter driving sequence (compared to the general construction presented in Appendix.~\ref{sec:sm1}). 
For level-3 \HS, an option for reducing the drive sequence is by using $U_2$ above and defining $U_3 = U_2e^{-iH_3T}U_2^{\dagger}e^{-iH_3T}$ without imposing any extra limitations on $H_3$, as in the SU(2) case in the main text(length of the driving sequence is $\ell=14$).

Another way is to impose a condition on the structure of the newly added Hamiltonian $H_3$: Consider the evolution operator
\begin{equation}
    U_3 \equiv e^{-i8Q_3T}= (e^{-i H_0T}e^{-i H_1T}e^{-i H_2T}e^{-i H_3T})(e^{iH_1T}e^{iH_2T}e^{iH_0T}e^{-iH_3T}).
\end{equation}
The first three orders of $Q_3$ are
\begin{eqnarray}
     Q_3^{(0)} &\sim& H_3\nonumber\\
     Q_3^{(1)} &\sim& iT([H_3, H_2]+[H_3+H_2,H_1]+[H_3,H_0])\nonumber\\
     Q_3^{(2)} &\sim& T^2(Q_{1,2,3}+[H_0,[H_0,H_3]]-[H_3,[H_0,H_3]]+2[H_0, ([H_1,H_2+H_3]+[H_2,H_3])]),
\end{eqnarray}
where $Q_{1,2,3}$ is a short-hand notation for the effective Hamiltonian that contains only commutators of $H_1$, $H_2$, and $H_3$. In order to achieve \HS structure, the terms containing $H_1$ and $H_0$ should vanish in $Q_3^{(1)}$ and terms containing $H_0$ should vanish in $Q_3^{(2)}$. Therefore, one of the simplest conditions one can impose is
\begin{equation}
    [H_0, H_3] = 0,\qquad [H_0, H_2] = 0,\qquad [H_1, H_2+H_3] = 0.
\end{equation}

These special cases can be used as building blocks for longer symmetry ladders. For even higher-level \HS, a driving sequence can be constructed based on the above two special cases by the previous iterative method to reduce the sequence length. Whether there exists further shortening of the pair of driving sequences with pragmatically meaningful conditions on the Hamiltonians for higher-level \HS, remains an open question.

\subsection{Generalizations of the hierarchical symmetry protocols}
\label{sec.manipulate}
In this section, we will discuss two more generalizations of the protocol we proposed before, which enable us to manipulate the effective Hamiltonian in a different way. 
The first one generalizes the single Hamiltonian $H_n$ to an effective Hamiltonian $P_n$ from the symmetry-preserving driving sequence. This has important implications for the realization of HS on real quantum devices because sometimes, instead of directly simulating $H_n$, one may approximate its time evolution via Trotterization. One example is  the 1D Heisenberg model, whose time evolution can be Trotterized into even and odd-bond updates, both of which satisfy the desired SU(2) symmetry~\cite{richter2021simulating}. The second one allows us to embed specific terms in the effective Hamiltonian to tailor its properties. As we will discuss in the following, we use this method to open the energy gap in the topological example, which generates a corner state. 

\textbf{Generalization 1.} Considering the following evolution operator
\begin{eqnarray}
    &&U_{n}\equiv e^{-iQ_nl_nT} = e^{-iQ_{n-1}l_{n-1}T}e^{-iP_nT}e^{iQ_{n-1}l_{n-1}T}e^{-iP_n'T},
    \label{eq:sm1seq1}\\
    &&e^{-iP_nT}\equiv \prod_{i=1}^{p_n}e^{-iH_{n,i}T/p_n}, \qquad e^{-iP_n'T}\equiv \prod_{i=1}^{p_n'}e^{-iH_{n,i}'T/p_n'},
    \label{eq:sm1seq2}
\end{eqnarray}
where $P_n$ and $P_n'$ are defined as the effective Hamiltonian of the driving sequence defined in Eq.~\eqref{eq:sm1seq2} and $H_{n,i}$, $H_{n,i}'$ all preserve the symmetry $G_n$. The effective Hamiltonian reads
\begin{eqnarray}
    Q_{n} &=& \frac{1}{\ell_n}(P_n+P_n')+\frac{iT}{\ell_n}\left(\ell_{n-1}[P_n,Q_{n-1}]+\frac{1}{2}[P_n', P_n]\right)\nonumber\\
    &&-\frac{T^2}{12\ell_n}\left(6\ell_{n-1}^2[Q_{n-1},[Q_{n-1},P_n]]+6\ell_{n-1}[P_n',[P_n,Q_{n-1}]]+[P_n-P_n',[P_n,P_n']]\right)+\cdots
\end{eqnarray}
Similar to the proof in the previous section, we can see that for $G_q$-breaking perturbation, the leading order term will be $\frac{i\ell_{n-1}T}{\ell_n}[P_n^{(0)}, Q_{n-1}^{(n-q)}]$, so that the \HS structure still holds.

\textbf{Generalization 2.} We begin with the following two evolution operators
\begin{eqnarray}
    U_{n-1} &\equiv& e^{-i\ell_{n-1}Q_{n-1}T} = e^{-iQ_{n-2}\ell_{n-2}T}e^{-iTR_{n-1}}e^{iQ_{n-2}\ell_{n-2}T}e^{-iTR_{n-1}},\nonumber\\
    U_{n-1}' &\equiv& e^{-i\ell_{n-1}Q_{n-1}'T} = e^{-iQ_{n-2}T}e^{-iTR_{n-1}'}e^{iQ_{n-2}T}e^{-iTR_{n-1}'},
\end{eqnarray}
where $R_{n-1}$ and $R_{n-1}'$ are $G_{n-1}$-preserving effective Hamiltonians. From Generalization 1 above with $P_{n-1}=R_{n-1}$ and $P_{n-1}'=R_{n-1}'$ in Eq.~\eqref{eq:sm1seq1}, it follows that both $U_{n-1}$ and $U_{n-1}'$ have a \HS structure. Let us now consider the following evolution operator
\begin{eqnarray}
    U_n & = U_{n-1}e^{-iTR_n}U_{n-1}'^{\dagger}e^{-iTR_n}\equiv& e^{-i\ell_n Q_n T}.\nonumber\\
    \label{eq:sm1U}
\end{eqnarray}
The effective Hamiltonian $Q_n$ is
\begin{eqnarray}
    Q_n &=& \frac{1}{\ell_n}\left(2R_n+\ell_{n-1}(Q_{n-1}-Q_{n-1}')\right) + \frac{iT}{\ell_n}\left(\ell_{n-1}[R_n, Q_{n-1}]+\frac{\ell_{n-1}^2}{2}[Q_{n-1}', Q_{n-1}]\right)\nonumber\\
    &&-\frac{T^2}{6\ell_n}\bigg(\ell_{n-1}^2[Q_{n-1},[Q_{n-1}, R_n]]+\ell_{n-1}^2[Q_{n-1}',[Q_{n-1}', R_n]]+\ell_{n-1}^2[Q_{n-1}',[Q_{n-1}, R_n]]\nonumber\\
    &&\phantom{---}+\ell_{n-1}^2[R_n,[Q_{n-1}, Q_{n-1}']]+\ell_{n-1}[R_n,[R_n,2Q_{n-1}+Q_{n-1}']]+\frac{\ell_{n-1}^3}{2}[Q_{n-1}+Q_{n-1}',[Q_{n-1}', Q_{n-1}]]\bigg)\nonumber\\
    &&+ \cdots
    \label{eq:sm1eff}
\end{eqnarray}
Similarly, for a $G_q$-breaking perturbation with $q<n$ (i.e., terms containing $Q_{n-2}^{(n-q-1)}$), the leading- and subleading- order term are found to be
\begin{eqnarray}
    \mathcal{O}(T^{n-q})&:& \frac{\ell_{n-1}}{\ell_n}\left(Q_{n-1}^{(n-q)}-Q_{n-1}'^{(n-q)}\right)
    = \frac{i\ell_{n-2}T}{\ell_n}[R_{n-1}^{(0)}-R_{n-1}'^{(0)}, Q_{n-2}^{(n-q-1)}] + G_q\;\text{-preserving terms},
    \label{eq:sm1led}\\
    \mathcal{O}(T^{n-q+1})&:& \frac{iT}{\ell_n}\left(\ell_{n-1}[R_n, Q_{n-1}^{(n-q)}]+\frac{\ell_{n-1}^2}{2}[Q_{n-1}'^{(0)}, Q_{n-1}^{(n-q)}]+\frac{\ell_{n-1}^2}{2}[Q_{n-1}'^{(n-q)}, Q_{n-1}^{(0)}]\right).
    \label{eq:sm1subled}
\end{eqnarray}
Now, if we impose that $R_{n-1}^{(0)} = R_{n-1}'^{(0)}$, the leading order term in Eq.~\eqref{eq:sm1led} will be echoed out, and the regular $\mathcal{O}(T^{n-q+1})$ terms will be retained. 
For a $G_n$-breaking perturbation, in addition to the common term $[R_n,Q_{n-1}^{(0)}]$ we have inserted a term $Q_{n-1}^{(1)}-Q_{n-1}'^{(1)}\sim R_{n-1}^{(1)}-R_{n-1}'^{(1)}$ by means of the definition of $U_n$. Therefore $Q_n$ preserves the \HS structure.
The inserted term consists only of $G_{n-1}$-preserving Hamiltonian and it may help us to achieve some specific HS. For example, in the HOTI case (c.f. Appendix ~\ref{sec:HOTI}), without this generalization, we can only have $i[H_2, H_1]$ as the $\mathcal{T}$-breaking perturbation, which cannot open an energy gap to  a obtain corner state. 

\section{FERMI GOLDEN RULE AND DYNAMIC TRANSITIONS BETWEEN PRETHERMAL SUBPLATEAUS}
\label{sec:FGR}

The constructions presented in the last section realize a temporal sequence of \HS. As shown in Fig.~\ref{fig:SU2} in the main text, \HS can modify significantly the thermalization pathways before the ultimate heat death. In particular, in Fig.~\ref{fig:SU2} (b), we numerically show that the prethermal lifetime of the conservation laws follows the algebraic scaling $T^{\alpha}$ where $\alpha\approx 2$ and $\alpha\approx4$, depending on the strength of the term that reduces the corresponding symmetry group. Here we justify this scaling law using a Fermi's Golden rule type argument. We first consider the dynamics of the conservation laws (or, more precisely, the order parameters) associated with the corresponding symmetry. Then we analyze the dynamics of auto-correlation functions. Both of them lead to the same scaling exponent, which matches well our numerical observations in the high-frequency regime.

\subsection{Decay rate of order parameter}

We first analyze the dynamics of the expectation value of the symmetry generators $S_{q,r}$ where $q\in\{n,n-1,\dots 0\}$ associated with the relevant symmetry ladder $G_q$ and $r$ labels non-commuting generators of $G_q$. 
For example, in the case of $\text{SU(2)}{\rightarrow}\text{U(1)} {\rightarrow} \mathbb{Z}_2{\rightarrow} E$, we have $n=3$ and the order parameters can be chosen as $S_{3,(x,y,z)} = {S}_{(x,y,z)}$ for SU(2) and $S_2 = {S}_z$ for U(1); $S_1$ is the parity order parameter (see main text). At short times and in the high-frequency regime, the highest symmetry is approximately preserved and hence $\langle S_{n,r}\rangle$ remains almost constant in time. The system prethermalizes in the Hilbert space restricted by the highest symmetry. This soft constraint becomes less stringent at late times when the effect of the next-order perturbation, which reduces the symmetry group, becomes sizeable. Therefore, in the high-frequency limit, there should appear $n$-step relaxation processes, and here we want to understand the transition rate between them associated with different symmetry sectors. Technically, we follow Ref.~\cite{ikeda2021fermi} and generalize their results to cases when an approximate symmetry is present.

In the $(n-q)$-th relaxation step, the effective Hamiltonian reads as
\begin{equation}
    Q_{n,[n-q]} = \sum_{i = 0}^{n-q} Q_n^{(i)},
\end{equation}
where $[n-q]$ denotes that the perturbative BCH expansion is truncated at the order $n-q$. Since $[Q_{n,[n-q]}, S_{q,r}] = 0 $, there is a set of eigenstates shared by $Q_{n,[n-q]}$ and $S_{q,r}$: $\{\left|\ell_r\right\rangle\} = \{\left|\ell_{r,s}, \ell_{r,e}, \cdots\right\rangle\}$. Where $\ell_{r,s}$ and $\ell_{r,e}$ are the quantum number of $S_{q,r}$ and $Q_{n,[n-q]}$ respectively (for U(1) there is only one quantum number, but for SU(2) there are two). The corresponding eigenvalues are given by:
\begin{equation}
    Q_{n,[n-q]}\left|\ell_r\right\rangle = E_{q,\ell_{r}} \left|\ell_r\right\rangle, \qquad 
    S_{q,r}\left|\ell_r\right\rangle = N_{q,\ell_{r}} \left|\ell_r\right\rangle.
\end{equation}
In a fixed prethermal plateau, the state of the system is approximately described by the generalized Gibbs ensemble \cite{vidmar2016generalized} (for plateaus of non-Abelian symmetries, the corresponding system state is called the non-Abelian thermal state \cite{murthy2023non}).
We assume that the different prethermal plateaus are well separated in time; for a fixed prethermal plateau corresponding to $G_q$, this timescale separation ensures that only those Lagrange multipliers that reflect the associated quasi-conservation law are taken into account. Therefore the generalized ensemble which characterizes the local properties of the system in the $q$-th prethermal state reads as 
\begin{eqnarray}
    & &\rho_t = \frac{e^{-\beta(t)Q_{n,[n-q]}-\sum_{r}\mu_{q,r}(t)S_{q,r}}}{Z_t}\nonumber\\
    & & Z_t = \text{Tr}(e^{-\beta(t)Q_{n,[n-q]}-\sum_{r}\mu_{q,r}(t)S_{q,r}}).
\end{eqnarray}
Here $\mu_{q,r}$ and $\beta$ are time-dependent Lagrange multipliers. They can be determined from the expectation values of the quasi-conserved quantities, $\langle S_{q,r}\rangle$, $\langle Q_{n,[n-q]}\rangle$ and their evolution equations are
\begin{eqnarray}
    \label{equ:FGR_1}
    \frac{d\langle S_{q,r}\rangle}{dt} &=& \text{Tr}(S_{q,r}\frac{d}{dt}\rho_t)\nonumber\\
    \frac{d\langle Q_{n,[n-q]}\rangle}{dt} &=& \text{Tr}(Q_{n,[n-q]}\frac{d}{dt}\rho_t)
\end{eqnarray}
However, the Lagrange multipliers $\beta$ and $\mu_q$ are still undetermined. To find them we define the probability distribution function as
\begin{equation}
    P_{\ell_r}^{\mu_q(t)} = \langle\ell_r|\rho_t|\ell_r\rangle,
    \langle S_{q,r}\rangle = \sum_{\ell_r}N_{q,\ell_r}P_{\ell_r}^{\mu_q(t)},
\end{equation}
where $\mu_q(t) = \{\beta(t), \mu_{q,0}(t), \mu_{q,1}(t), \cdots\}$. We consider the master equation for $P_{\ell_r}^{\mu_q(t)}$:
\begin{equation}
    \frac{dP_{\ell_r}^{\mu_q(t)}}{dt} = \sum_{m_r} \left(w_{m_r\rightarrow\ell_r}^{(q)}P_{m_r}^{\mu_q(t)} - w_{\ell_r\rightarrow m_r}^{(q)}P_{\ell_r}^{\mu_q(t)}\right),
\end{equation}
where $w_{m_r\rightarrow\ell_r}^{(q)}$ is the transition rate of $S_{q,r}$ corresponding to the $G_q$-reducing perturbation derived from Fermi-Golden rule (FGR). Therefore we obtain a second set of evolution equations:
\begin{eqnarray}
    \frac{d\langle S_{q,r}\rangle}{dt} &=& \sum_{\ell_r}N_{q,\ell_r}\frac{dP_{\ell_r}^{\mu_q(t)}}{dt} = \sum_{\ell_r, m_r}(N_{q,\ell_r}-N_{q,m_r})P_{m_r}^{\mu_q(t)}w_{m_r\rightarrow\ell_r}^{(q)},\nonumber\\
    \frac{d\langle Q_{n,[n-q]}\rangle}{dt} &=& \sum_{\ell_r}E_{\ell_r}\frac{dP_{\ell_r}^{\mu_q(t)}}{dt} = \sum_{\ell_r, m_r}(E_{\ell_r}-E_{m_r})P_{m_r}^{\mu_q(t)}w_{m_r\rightarrow\ell_r}^{(q)}.
    \label{eq:sm2N}
\end{eqnarray}
Equations~\eqref{equ:FGR_1} and~\eqref{eq:sm2N}
form a set of self-consistent evolution equations for the variables $\langle S_{q,r}\rangle$, $\langle Q_{n,[n-q]}\rangle$, $\mu_{q,r}(t)$, $\beta(t)$ describing the approximate evolution of the ensemble.

For a general time-dependent Hamiltonian $H(t) = H_n+g(t)V$, $g_{f} = \int_{0}^{T}\frac{dt}{T}g(t)e^{if\omega t}$ and $\omega=2\pi/T$, the FGR transition rate is given by the expression
\begin{equation}
    w^\text{FGR}_{m_r\rightarrow \ell_r} = 2\pi\sum_{f\in\mathbb{Z}}|{_0\langle} \ell_r|V|m_r\rangle_0|^2|g_{f}|^2\delta(E_{\ell}-E_m - f\omega),
\end{equation}
where $\left|m_r\right\rangle_0$ and $\left|\ell_r\right\rangle_0$ are the eigenstates of $H_n$ and $S_{q,r}$. 

To obtain an estimate of the transition rates $w^\text{FGR}_{m_r\rightarrow \ell_r }$, we first reorganize the terms in our driving protocol in Eq.~\eqref{eq:general_seq} by moving the right-most unitary to the right-hand side, resulting in:
\begin{equation}
\label{eq:Q'}
    e^{-iQ_{n-1} l_{n-1} T} e^{-iH_n T} e^{iQ_{n-1} l_{n-1} T} = e^{-iQ_n l_{n} T} e^{iH_nT} \equiv e^{-iQ'_{n-1} (2l_{n-1}+1)T}.
\end{equation}
Since $Q_n$ has a level-$n$ \HS structure and $H_n$ preserves the highest symmetry, $Q'_{n-1}$ also inherits the same \HS structure; hence, $Q_{n-1}^{'(0)} = H_nT/(2l_{n-1}+1)$. Thus the Floquet operator of our original protocol can be equivalently written as
\begin{equation}
    U_n = e^{-iQ'_{n-1} (2l_{n-1}+1)T} e^{-iH_nT} = e^{-i(H_n+V_n)T}e^{-iH_nT},
\end{equation}
where $V_n = (2l_{n-1}+1)(Q_n^{'(1)}+Q_n^{'(2)}+Q_n^{'(3)}+\cdots)$ can be regarded as the time-dependent perturbation on top of the static part $H_n$ in the context of applying FGR. 

The time-dependent function is $g(t+2T)=g(t)$ is given by
\begin{equation}
    g(t) = 
     \begin{cases}
       1, &\quad\text{for}\quad 0<t \text{ mod } T\le T\\
       0, &\quad\text{for}\quad T<t \text{ mod } T\le 2T\\ 
     \end{cases}
\end{equation}
Fourier transforming, we obtain $|g_f|^2 = \frac{\sin^2(fT/2)}{T^2f^2}$. When the driving frequency $\omega = 2\pi/T$ is much larger than the norm of the local effective Hamiltonian, the system can only absorb a finite number of energy quanta (i.e., $f$ is a finite integer). Hence in the high-frequency limit $|g_f|^2\approx \frac{1}{4}$.  

We now focus on the breaking of the symmetry $G_q$ $(0<q\le n)$. Note that Eq.~\eqref{eq:sm2N} is written in the eigenstates $|\ell_r\rangle$ and $|m_r\rangle$ of $Q_{n,[n-q]}$ and $S_{q,r}$, but the matrix element in FGR's heating rate corresponds to the eigenstates of the unperturbed Hamiltonian $H_n$ and $S_{q,r}$. Therefore, we need to show that the matrix elements in the heating rate under these two sets of eigenbases are equal in leading order. 
The perturbation $Q_n^{'(n-q+1)}\sim\mathcal{O}(T^{n-q+1})$ will break symmetry $G_{q}$ and preserve symmetry $G_{p}, \forall p<q$. 
Since $Q_{n,[n-q]} = Q_n^{(0)} + \mathcal{O}(T) = \frac{1}{l_{n-1}+1}H_n + \mathcal{O}(T)$, it follows that $\left|\ell_r\right\rangle = \left|\ell_r\right\rangle_0 + \mathcal{O}(T)$. Therefore $\langle \ell_r|Q_n'^{(n-q+1)}| m_r\rangle = 
{_0\langle} \ell_r|Q_n'^{(n-q+1)}| m_r\rangle_0 + \mathcal{O}(T^{n-q+2})$, which means that the leading order contribution to the heating rate can be captured by the perturbed eigenstates of $H_n$. Now when we consider the contributions to the FGR rate we also need to take into account contributions from cross-terms in the calculation of the FGR rate $w^{(q)}_{m_r\rightarrow\ell_r}\propto |{\langle} \ell_r|V|m_r\rangle|^2$. 
However, since $[Q_n^{'(n-p+1)}, S_{q,r}] = 0$ for all $q<p$ and $r$, the matrix element $\langle \ell_r|Q_n'^{(n-p+1)}| m_r\rangle$ is nonzero only when $N_{q,\ell_r} = N_{q,m_r}$, for all $q<p$ and $r$. 
Notice that $\langle \ell_r|Q_n'^{(n-q+1)}| m_r\rangle \langle m_r|Q_n'^{(n-p+1)}| \ell_r\rangle$ ($q<p$) does not contribute to the heating rate associated with the order parameter $\langle S_{q,r}\rangle$ of $G_q$ because in Eq.~\eqref{eq:sm2N} $N_{q,\ell_r}=N_{q,m_r}$. Therefore, the leading order contribution is
\begin{equation}
\label{eq.FGR_rate}
    w^{(q)}_{m_r\rightarrow\ell_r} \sim |\langle \ell_r|Q_n'^{(n-q+1)}| m_r\rangle|^2\sim \mathcal{O}(T^{2(n-q+1)}).
\end{equation}

\subsection{Decay rate of infinite-temperature auto-correlation function}

Here we supply another analytical approach to analyse the stability of the emergent symmetries. This is achieved by studying the time evolution of the symmetry generators $S_{n,r}(t)$, and the auto-correlation function 
\begin{equation}
\label{eq:auto_corr}
    C_{n,r}(t)=\frac{1}{2^L}\text{Tr}(S_{n,r}(t)S_{n,r}(0))
\end{equation}
at stroboscopic times $t=\ell T$ for a system of size $L$. If the corresponding symmetry $G_n$ is well preserved,  $C_{n,r}$ is a constant. Whereas if the symmetry is perturbed, $C_{n,r}$ generally decays and in the following, we explicitly calculate the leading order contribution to the decay rate. The result has the same scaling dependence on $T$ as in Eq.~\eqref{eq.FGR_rate}.

As we proposed, the evolution operator for $n$-th order \HS is
\begin{eqnarray}
    U_n &\equiv& e^{-iQ_nl_nT} = e^{-iQ_{n-1}l_{n-1}T}e^{-iH_nT}e^{iQ_{n-1}l_{n-1}T}e^{-iH_nT} \nonumber\\
    &=& e^{-iH_nT}\left(\underbrace{e^{iH_nT}e^{-iQ_{n-1}l_{n-1}T}e^{-iH_nT}e^{iQ_{n-1}l_{n-1}T}}_{U_n'}e^{-2iH_nT}\right)e^{iH_nT}.
\end{eqnarray}
Similarly, $U_n'$ corresponding to Eq.~\eqref{eq:Q'} also has the $n$-th order \HS structure, which means that if we define the effective Hamiltonian $Q_n'$ via the relation $U_n' \equiv e^{-iQ_n'l_nT}$, it will preserve $G_n$ and break $G_{n-1}$. Note that, since $Q_n'^{(0)} = 0$, it can be perturbatively constructed as $Q_n' = Q_n'^{(1)} + Q_n'^{(2)} + \cdots$.
We can also expand the operator $U_n'$ accordingly as
\begin{equation}
    U_n' = 1-iT\left(Q_n'^{(1)} + Q_n'^{(2)} + \cdots\right)-\frac{T^2}{2}\left(Q_n'^{(1)} + Q_n'^{(2)} + \cdots)(Q_n'^{(1)} + Q_n'^{(2)} + \cdots\right) + \cdots.
\end{equation}

To calculate the auto-correlation function~\eqref{eq:auto_corr}, we first need to derive the time-evolved operator $S_{q,r}(T) = U_n^{\dagger}S_{q,r}U_n$. For simplicity, here we only consider the first two orders in the perturbation $V_1 = Q_n'^{(1)}/T$ and $V_2 = Q_n'^{(2)}/T^2$ and calculate the auto-correlation function of $S_{n,r}$ as an example. In the high-frequency regime,  we can expand it as a power series in $T$
\begin{equation}
\label{eq.sn}
    S_{n,r}(\ell T) {\approx}  e^{i\mathcal{L}_nT}\left(e^{-2i\mathcal{L}_nT}\left(1+iT^2\mathcal{L}_{V_1}+iT^3\mathcal{L}_{V_2}+T^4\mathcal{G}_{V_1}+T^6\mathcal{G}_{V_2}-\frac{T^4}{2}\mathcal{F}_{V_1}-\frac{T^6}{2}\mathcal{F}_{V_2}+T^5\mathcal{G}_{V_{1,2}}-\frac{T^5}{2}\mathcal{F}_{V_{1,2}}\right)\right)^\ell S_{n,r}(0),
\end{equation}
where $\mathcal{L}_nS_{n,r} = [H_n,S_{n,r}]$, $\mathcal{L}_{V_i}S_{n,r} = [V_i,S_{n,r}]$, $\mathcal{G}_{V_i}S_{n,r} = V_iS_{n,r}V_i$, $\mathcal{F}_{V_i}S_{n,r} = \{V_i^2, S_{n,r}\}$,
$\mathcal{G}_{V_{i,j}}S_{n,r} = V_iS_{n,r}V_j+V_jS_{n,r}V_i$,
$\mathcal{F}_{V_{i,j}}S_{n,r} = \{\{V_i,V_j\},S_{n,r}\}$ with $\{\cdot,\cdot\}$ the anticommutator, and the perturbation $V_p = Q_n'^{(p)}/T^p$ which is indeed $T$-independent. We also used the property that $H_n$ preserves all the symmetries in the symmetry ladder, i.e., $e^{iH_nT}S_{q,r}e^{-iH_nT} = S_{q,r}$ for $0 \le q \le n$ and $\forall r$, to derive the equation above.
 
We insert Eq.~\eqref{eq.sn} into the definition of the autocorrelation function in Eq.~\eqref{eq:auto_corr}. Using the properties~\cite{yeh2023decay} 
\begin{eqnarray}
    \text{Tr}\left(e^{-i\mathcal{L}_VT}\hat{A}\hat{B}\right) &=& \text{Tr}\left(\hat{A}e^{i\mathcal{L}_VT}\hat{B}\right),\nonumber\\
    \text{Tr}\left(e^{-2i\mathcal{L}_nTl}(i\mathcal{L}_{V_j})e^{-2i\mathcal{L}_nTm}S_{n,r}S_{n,r}\right) &=& 0,\qquad\qquad\qquad \forall\; j,l,m,
\end{eqnarray}
we can get the first three orders of the $T$-expansion for $C_{n,r} = C_{n,r}^{(0)} + C_{n,r}^{(1)} + C_{n,r}^{(2)} + \cdots$, as:
\begin{eqnarray}
    C_{n,r}^{(0)} &=& \frac{1}{2^L}\text{Tr}\left(e^{-2\ell T i\mathcal{L}_n}S_{n,r}S_{n,r}\right) = 1,\nonumber\\
    C_{n,r}^{(1)} &=& \frac{T^4}{2^L}\Bigg(\sum_{l\ge 1, m\ge 0}^{l+m=\ell-1}\text{Tr}\left(e^{-2i\mathcal{L}_nT(\ell-m-l)}(i\mathcal{L}_{V_1})e^{-2i\mathcal{L}_nTl}(i\mathcal{L}_{V_1})e^{-2i\mathcal{L}_nTm}S_{n,r}S_{n,r}\right)+ \sum_{l=0}^{\ell-1}\text{Tr}\left(e^{-2i\mathcal{L}_nT(\ell-l)}\mathcal{G}_{V_1}e^{-2i\mathcal{L}_nTl}S_{n,r}S_{n,r}\right)\nonumber\\
    &&+ \sum_{l=0}^{\ell-1}\text{Tr}\left(e^{-2i\mathcal{L}_nT(\ell-l)}(-\frac{1}{2}\mathcal{F}_{V_1})e^{-2i\mathcal{L}_nTl}S_{n,r}S_{n,r}\right)\Bigg),\nonumber\\
    C_{n,r}^{(2)} &=& \frac{T^5}{2^L}\Bigg(\sum_{l\ge 1,m\ge 0}^{l+m=\ell-1}\text{Tr}\left(e^{-2i\mathcal{L}_nT(\ell-l-m)}(i\mathcal{L}_{V_1}e^{-2i\mathcal{L}_nTl}i\mathcal{L}_{V_2}+i\mathcal{L}_{V_2}e^{-2i\mathcal{L}_nTl}i\mathcal{L}_{V_1})e^{-2i\mathcal{L}_nTm}S_{n,r}S_{n,r}\right) \nonumber\\
    &&+ \sum_{l=0}^{\ell-1}\text{Tr}\left(e^{-2i\mathcal{L}_nT(\ell-l)}\mathcal{G}_{V_{1,2}}e^{-2i\mathcal{L}_nTl}S_{n,r}S_{n,r}\right)+\text{Tr}\left(e^{-2i\mathcal{L}_nT(\ell-l)}(-\frac{1}{2}\mathcal{F}_{V_{1,2}})e^{-2i\mathcal{L}_nTl}S_{n,r}S_{n,r}\right)\Bigg).
\end{eqnarray}
In fact $C_{n,r}^{(1)}$ can be simplified as
\begin{eqnarray}
\label{eq:sm2_2}
    C_{n,r}^{(1)} &=& \frac{\ell T^4}{2^L}\left(\sum_{l=1}^{\ell-1}\left(1-\frac{l}{\ell}\right)\text{Tr}\left(e^{-2i\mathcal{L}_nTl}(i\mathcal{L}_{V_1}S_{n,r})(-i\mathcal{L}_{V_1}S_{n,r})\right) + \text{Tr}\left(V_1S_{n,r}V_1S_{n,r}\right)-\frac{1}{2}\text{Tr}\left(\{V_1^2, S_{n,r}\}S_{n,r}\right)\right)\nonumber\\
    &\overset{\ell\to\infty}{\longrightarrow}& -\frac{\ell T^4}{2^L}\left(\frac{1}{2}\text{Tr}\left(\dot{S}_{n,r}\dot{S}_{n,r}\right)+\sum_{l=1}^{\infty}\text{Tr}\left(e^{-2i\mathcal{L}_nTl}\dot{S}_{n,r}\dot{S}_{n,r}\right)\right)\nonumber\\
    & = & -\frac{\ell T^4}{2^L}\sum_{i,j}(\frac{1}{2}+\sum_{\ell=1}^{\infty}e^{-2i\ell T(\epsilon_i-\epsilon_j)})|\langle i|\dot{S}_{n,r}|j\rangle|^2\nonumber\\
    & = & -\frac{\ell T^3}{2^L}\sum_{i,j}| \langle i  | \dot{S}_{n,r} | j  \rangle|^2\frac{\pi}{2}\delta_F(\epsilon_i-\epsilon_j),
\end{eqnarray}
where we use $H_n|i\rangle = \epsilon_{i}|i\rangle$, $\dot{S}_{n,r} \equiv i\mathcal{L}_{V_1}S_{n,r}$, $\delta_F(\epsilon_i-\epsilon_j) = \sum_{p=-\infty}^{\infty}\delta\left(\epsilon_i-\epsilon_j+\frac{p\pi}{T}\right)$ and 
\begin{equation}
    \frac{1}{2}+\sum_{\ell=1}^{\infty}e^{-2i\ell T(\epsilon_i-\epsilon_j)} = \frac{1}{2}\sum_{\ell=-\infty}^{\infty}e^{-2i\ell T(\epsilon_i-\epsilon_j)} = \sum_{p=-\infty}^{\infty}\pi\delta(2T(\epsilon_i-\epsilon_j)+2p\pi) = \frac{\pi}{2T}\sum_{p=-\infty}^{\infty}\delta(\epsilon_i-\epsilon_j+\frac{p\pi}{T}).
\end{equation}
Since usually the correlation function $\text{Tr}\left(e^{-2i\mathcal{L}_nTl}(i\mathcal{L}_{V_1}S_{n,r})(-i\mathcal{L}_{V_1}S_{n,r})\right)$ decays rapidly as $l$ increases, here $(1-l/\ell)$ in the summation $\sum_{l=1}^{\ell-1}$ in ~\eqref{eq:sm2_2} can be approximated as 1 as $\ell\to \infty$ . We can also do a similar simplification to the $C_{n,r}^{(2)}$. 

Assuming 
$C_{n,r}(\ell T) = e^{-\Gamma_{n,r} \ell T}\approx1-\Gamma_{n,r} \ell T$, the heating rate $\Gamma_{n,r}$ is given by
\begin{equation}
\begin{aligned}
        \Gamma_{n,r} = \frac{T^2}{2^L}\sum_{i,j}|\left \langle i  | \mathcal{L}_{V_1}S_{n,r} | j  \right \rangle|^2\frac{\pi}{2}\delta_F(\epsilon_i-\epsilon_j) + \frac{T^3}{2^L}\sum_{i,j}\frac{1}{2}(\left \langle i  | \mathcal{L}_{V_1}S_{n,r} | j  \right \rangle\left \langle j  | \mathcal{L}_{V_2}S_{n,r} | i  \right \rangle \\ +
    \left \langle i  | \mathcal{L}_{V_2}S_{n,r} | j  \right \rangle\left \langle j  | \mathcal{L}_{V_1}S_{n,r} | i  \right \rangle)\frac{\pi}{2}\delta_F(\epsilon_i-\epsilon_j)+\mathcal{O}(T^4).
\end{aligned}
\label{eq:Gamma}
\end{equation}
The first and second terms correspond to the leading order and a cross term in the FGR results of the previous section, respectively.

When the auto-correlation function of $S_{n-1,r'}$ (i.e., the second plateau) is considered, we see that the first term and second term (the cross term corresponding to $C_{n,r}^{(2)}$) in Eq.~\eqref{eq:Gamma} vanishes
since $V_1$ preserves the symmetry $G_{n-1}$. Similarly, for a general symmetry generator $S_{q,r}$ with $1\le q\le n$ and $\forall r$, $[S_{q,r}, V_i] = 0$ for $i\le n-q)$, it follows that any combination of terms containing $\mathcal{L}_{V_i}$ in the expansion of $\Gamma_{q,r}$ vanish. The leading order in the decay rate $\Gamma_{q,r}$ is then 
\begin{equation}
    \Gamma_{q,r} = \frac{T^{2(n-q+1)}}{2^L}\sum_{i,j}|\left \langle i | \mathcal{L}_{V_{n-q+1}}S_{q,r} | j \right \rangle |^2 + \mathcal{O}(T^{2n-2q+3}).
\end{equation}
This result gives the same prethermal time scale $\tau_q\sim\mathcal{O}(T^{-2(n-q+1)})$, as in the former section, cf.~Eq.~\eqref{eq.FGR_rate}.

\section{ADDITIONAL RESULTS ON \HS IN SPIN SYSTEMS}
\label{sec:sm6} 

In this appendix, we first show the concrete form of the effective Hamiltonians for $\mathrm{SU}(2) \to \mathrm{U}(1) \to \mathbb{Z}_2 \to E$ in spin systems. We also present the results of the lower order \HS symmetry ladder (i.e., $\mathrm{U}(1)\to\mathbb{Z}_2\to E$) under the Floquet and random drives, where we introduce the participation entropy to show the distribution of the state population in different magnetization blocks. For randomly driven cases, we show numerically the decay of parity and illustrate that the $\mathbb{Z}_2$-plateau is also parametrically long lived by increasing the driving frequency.

\subsection{Effective Hamiltonian for $\mathrm{SU}(2) \to \mathrm{U}(1) \to \mathbb{Z}_2 \to E$}

In the main text, we construct four kick generators (Eq.~\eqref{eq.spinmodel}) preserving each symmetry in the symmetry ladder $\mathrm{SU}(2) \to \mathrm{U}(1) \to \mathbb{Z}_2 \to E$.
With the evolution operator defined in Eq.~\eqref{eq.specialconstruction}, the first three orders of the effective Hamiltonian are
\begin{eqnarray}
    Q^{(0)} &=& \frac{1}{7} H_3 ,\ \ Q^{(1)} = -\frac{iT}{98}[H_2, H_3], \nonumber\\
    Q^{(2)} &=& \frac{T^2}{2184}([H_3, [H_1, H_2]] - \frac{1}{2}[H_3,[H_3,H_2]] - [H_2, [H_2, H_3]]), \nonumber\\
    Q^{(3)}_{\pm} &=& \mathrm{terms\; containing\;} H_0^{\pm}, 
\end{eqnarray}
where $Q^{(0)}$ preserves the highest symmetry SU(2), $Q^{(1)}$ reduces SU(2) to U(1), $Q^{(2)}$ reduces U(1) to $\mathbb{Z}_2$ and higher-order terms break all symmetries.

\subsection{Effective Hamiltonian for $\mathrm{U}(1) \to \mathbb{Z}_2 \to E$ with Floquet drives}

In the main text, we also briefly mention the sequences $\mathrm{U}(1) \to \mathbb{Z}_2 \to E$. This level-2 \HS latter can be realized in spin systems using a Floquet drive. The kick generators are chosen as
\begin{equation}
    \begin{aligned}
       &H_2 = J\sum_{\langle i,j\rangle}\sigma_i^x\sigma_j^x+\sigma_i^y\sigma_j^y-\sigma_i^z\sigma_j^z,\ \  H_1 = -J\sum_{\langle i,j\rangle}\sigma_i^y\sigma_j^y-\sigma_i^z\sigma_j^z,\ \  H_0 = -\delta_x\sum_{i}\sigma_i^x,
    \end{aligned}
\end{equation}
where the $zz$-interaction is added to make the model non-integrable. Each generator preserves one symmetry in the symmetry ladder. To achieve a \HS structure, the time evolution operator is defined as
\begin{equation}
    U_F = U\left(-H_0, H_1, H_2 | \frac{T}{6}\right)U\left(H_0, -H_1, H_2 | \frac{T}{6}\right),
\end{equation}
where again we use $
    U(D_{1},\dots, D_{l}| T) {=} 
    e^{-iD_{1}T}\cdots e^{-iD_{l}T}.
$
Using the property $[H_0,H_1+H_2] = 0$, the leading three orders of effective Hamiltonian read
\begin{equation}
\begin{aligned}
   &Q^{(0)} =  \frac{1}{3}H_2,\ \ Q^{(1)} = -\frac{iT}{36}[H_1,H_2],\ \ Q^{(2)} = -\frac{T^2}{432}[H_0+H_1-H_2,[H_1,H_2]].
\end{aligned}
\end{equation}
The zeroth order of effective Hamiltonian $Q^{(0)}$ preserves the highest symmetry U(1), the first order term $Q^{(1)}$ reduces U(1) to $\mathbb{Z}_2$, while second-order term $Q^{(2)}$ breaks explicitly the $\mathbb{Z}_2$ symmetry. Therefore we expect the state of the system to go through two prethernal plateaus -- for the U(1) and $\mathbb{Z}_2$ symmetry, characterized by the $z$-magnetization and the parity [see main text], respectively. 

\begin{figure}[t!]
	\centering
\includegraphics[width=0.9\columnwidth]{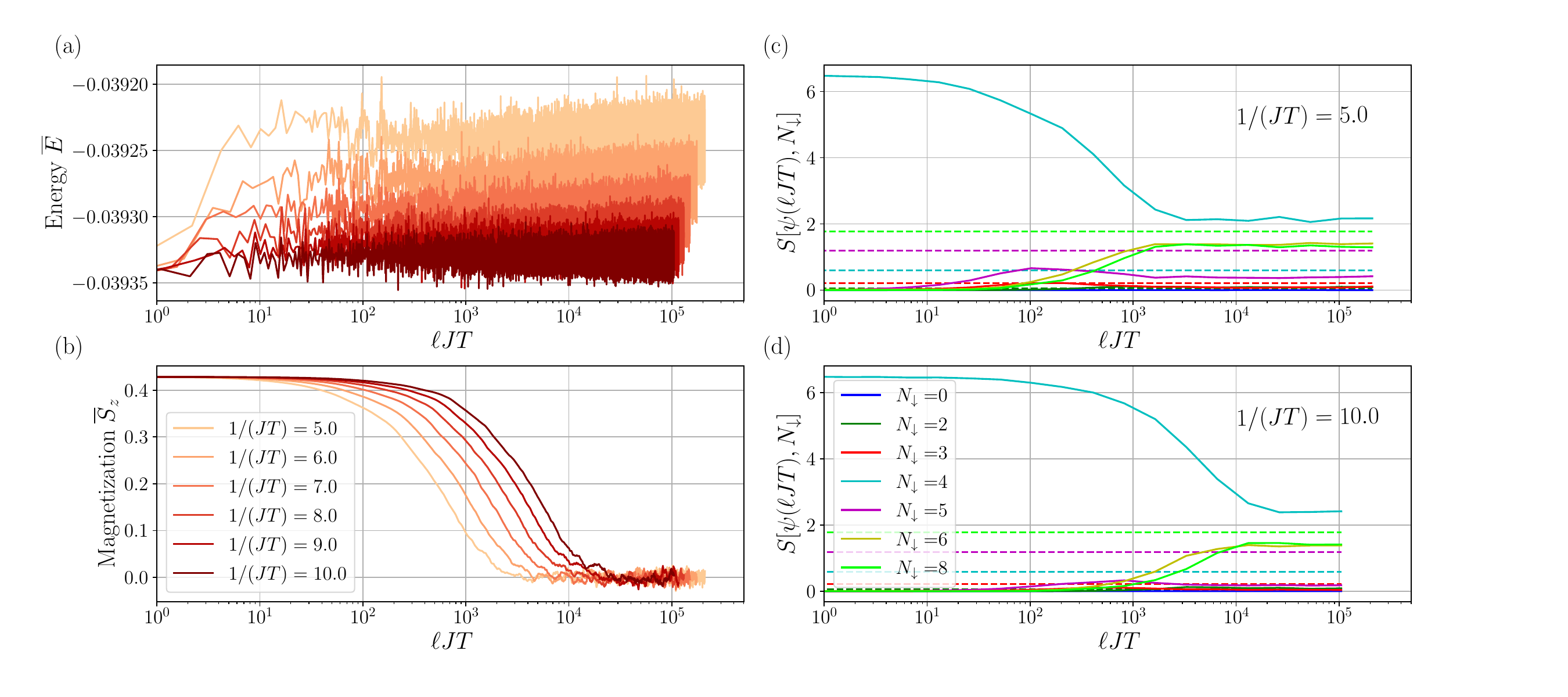}
	\caption{
 Dynamical detection of $U(1) \to \mathbb{Z}_2$ HS with Floquet drives.
 (a) Time evolution of the density of energy $\overline{E} = \overline{\left\langle Q^{(0)}\right\rangle}/L$. We show the energy prethermal plateau where energy is quasi-conserved. However, for a small system, it is difficult to fully thermalize under the Floquet drive even at long times due to finite-size effects.
 (b) Dynamics of the density of magnetization $\overline{S_{z}} = \overline{\left\langle \sum_{i}\sigma_i^z\right\rangle}/L$. The lifetime of $\mathrm{U}(1)$ plateau is prolonged with increasing driving frequency. (c),(d) The time evolution of the participation entropy for  $1/(JT) = 5.0, 10.0$. 
 $N_{\downarrow}$ represents number of downwards spins. The state has weight mainly in the even magnetization sector, showing the $\mathbb{Z}_2$ plateau. Dashed lines represent participation entropy at infinite temperature. The parameters are $L=14$ with symmetry-breaking perturbation strength $\delta_x/J = 10.0$. The numerical simulations are performed using exact diagonalization and 20 random realizations are used to compute the ensemble average over different initial states.} 
\label{fig:2}
\end{figure}

We perform numerical simulations for a system size of $L = 14$ spins. The initial state is a Haar-random state in the fixed magnetization sector $(N_{\downarrow},N_\uparrow) = (4,L-4)$, where $N_{\downarrow}/N_{\uparrow}$ denotes the number of up/down spins in the each basis configuration. We compute the energy density $\overline{E} = \overline{\left\langle Q^{(0)}\right\rangle}/L$ and the magnetization density along $z$-axis $\overline{S_{z}} = \overline{\left\langle \sum_{i}\sigma_i^z\right\rangle}/L$, where the overline means the average over different realizations of the initial state. To investigate the distribution of the evolved state in different magnetization blocks, we also compute the participation entropy ~\cite{luitz2020prethermalization}
\begin{equation}
    S_\text{part}[\psi, N_{\downarrow}] = -\sum_{|i\rangle\in\mathcal{H}|_{N_{\downarrow}}}|\left\langle i|\psi\right\rangle|^2\ln(|\left\langle i|\psi\right\rangle|^2).
    \label{equ:1}
\end{equation}
If the state of the system is only in a certain magnetization sector, its participation entropy on the other sector is 0. And for the thermal state, $S_{\text{part}}[\psi, N_{\downarrow}] = \dim(\mathcal{H}|_{N_{\downarrow}})L\ln2/2^L$, where $L$ is the system size.
As shown in Fig.~\ref{fig:2}(a), energy of the system is sufficiently close to zero, suggesting that the initial state corresponds to a very high temperature. 
Different colors correspond to different driving frequencies, and for all cases the energy density remains approximately the same as its initial value. We do not see notable changes in energy throughout the entire time evolution. It occurs possibly because the system size is not large enough and the driving frequency may be already comparable or even larger than the many-body band width, hence heating is significantly suppressed.

In contrast, dynamics of magnetization $\overline{S_{z}}$ is more sensitive to the variation of the driving frequency.
In Fig.~\ref{fig:2}(b), the density of magnetization decays after the first U(1) prethermal plateau since the first-order effective Hamiltonian reduces U(1) to $\mathbb{Z}_2$. To see the effect of the approximated $\mathbb{Z}_2$ symmetry, we further analyse the participation entropy shown in Fig.~\ref{fig:2}(c) and (d). After the initial transient period, we see that the participation entropy for the even magnetization sectors grows notably, verifying the existence of the remaining $\mathbb{Z}_2$ symmetry. However, participation entropy does not evolve to its infinite temperature value (dashed lines), suggesting that the $\mathbb{Z}_2$ symmetry does not appear to be strongly broken by the Floquet drive. We attribute this to finite-size effects. 

In Fig.~\ref{fig:2}(c,d), we also note that the $N_\downarrow=5$ (but also other odd-valued) magnetization sector acquires a finite population at intermediate times. This is because the sub-leading order in the heating rate, which contains the $\mathbb{Z}_2$ symmetry breaking term, affects the dynamics of the system for a sufficiently long time at small driving frequencies.
However, the peak will be suppressed as the driving frequency increases. 

\subsection{Effective Hamiltonian for $\mathrm{U}(1)\to\mathbb{Z}_2\to E$ with random drives}
\begin{figure}[t!]
	\centering
	\includegraphics[width= 1.0\linewidth]{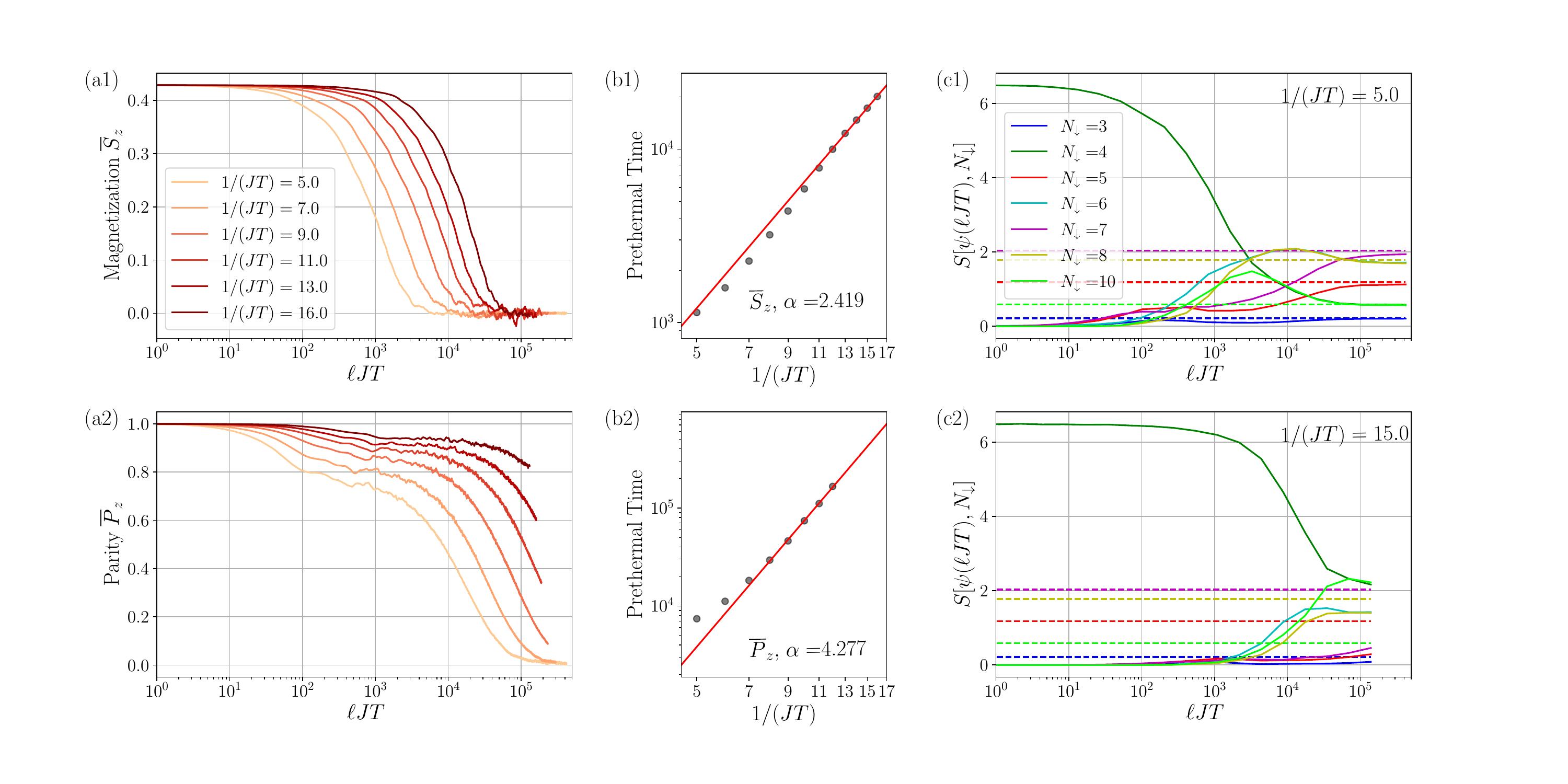}
	\caption{
 Dynamical detection of $U(1)\to\mathbb{Z}_2\to E$ HS with RMD.
 (a1) The time evolution of magnetization density $\overline{S_{z}} = \overline{\left\langle \sum_{i}\sigma_i^z\right\rangle}/L$. The quasi-conserved magnetization shows the $U(1)$ plateau.
 (a2) The time evolution of the expectation value of the parity $\overline{P_z} = \left\langle\prod_{i}\sigma_{i}^z\right\rangle$. The quasi-conserved parity shows the $\mathbb{Z}_2$ plateau. The breaking of early conservation is suppressed with increasing drive frequency.
 (b1),(b2) The scaling of the prethermal lifetime of the $U(1)$ plateau and the $\mathbb{Z}_2$ plateau $\tau_{m}\sim T^{-2.419},\; \tau_{p}\sim T^{-4.277}$ respectively, shows that the lifetime of HS can be dynamically prolonged by increasing the drive frequency. (c1),(c2) The time evolution of the participation entropy at drive frequencies $1/JT = 5.0$ and $1/JT = 15.0$, showing the distribution of state in each block. The Dashed lines represent the corresponding participation entropies at infinite temperature. The curves at low drive frequencies eventually coincide with the dashed lines, indicating that the system effectively heats up to an infinite-temperature state. The system size $L=14$, the symmetry-breaking perturbations are $\delta_x/J = 10.0, \epsilon/J = 6.0$. The numerical simulations are performed using exact diagonalization and 20 random realizations of the driving protocol and initial state are used to compute the ensemble average. The difference in simulation results between different realizations is very small, so we don't need too many realizations.}
\label{fig:s5}
\end{figure}

Finite size effect becomes less notable if we switch to random drives. We still use the same kick generators ($H_{2,1}$) as defined above in the Floquet case. However, the $\mathbb{Z}_2$ breaking term can occur randomly in time. More concretely, we consider
\begin{equation}
    H_0^{\pm} = (\delta_x\pm \epsilon)\sum_{i}\sigma_i^x.
\end{equation}
and the evolution operator is redefined as 
\begin{equation}
    U_{\pm} = U\left(-H_0^{\pm}, H_1, H_2| \frac{T}{6}\right)U\left(H_0^{\pm}, -H_1, H_2| \frac{T}{6}\right).
\end{equation}
The $U_{\pm}$ are then randomly aligned to form a stochastic drive sequence. Similar to the discussion in the Floquet case, the effective Hamiltonian $Q_{\pm}$ of the evolution operator $U_{\pm}$ has the same \HS structure; in particular, we have $Q_{-}^{(0),(1)} = Q_{+}^{(0),(1)}$, but the randomness affects the $\mathbb{Z}_2$ plateau since $Q_{-}^{(2)}\ne Q_{+}^{(2)}$.

Once again, we perform a numerical simulation for a system of $L = 14$ spins, starting from a Haar-random initial state within the $z$-magnetization sector $N_{\downarrow}=4$; this is a high-temperature (i.e., energy density) state. 
In Fig.~\ref{fig:s5}(a1) and (a2), the system shows two clear prethermal plateaus corresponding to the U(1) and $\mathbb{Z}_2$ symmetries.
We can clearly see the HS is well achieved dynamically through our protocol.
In Fig.~\ref{fig:s5}(c1) and (c2), we also show the $Z_2$ plateau exhibited by the participation entropy, during which the state is mainly distributed over the even magnetization sectors. At infinite times, the system will eventually fully thermalize, as indicated by the data in Fig.~\ref{fig:s5}(c1), and the curve of participation entropy will reach the dashed line corresponding to a fully thermalized state. 
Note that at small driving frequencies, the $Z_2$ symmetry breaking already kicks in at early times, but this effect can be pushed to parametrically longer times with increasing drive frequency. 
We define the lifetimes $\tau_{m}$ and $\tau_{p}$ as the time for $\Bar{S}_z$ and $\Bar{P}_z$ to decay to $e^{-1}$ of their initial values in Fig.~\ref{fig:s5}(a1) and (a2).
In Fig.~\ref{fig:s5}(b1) and (b2), we numerically show that the time scaling of the lifetimes $\tau_{m}$ and $\tau_{p}$ are approximately 
$
    \tau_{m} \sim T^{-2.419}, \; \tau_{p} \sim T^{-4.255}
$. 
The lifetime of the U(1) plateau approximately agrees well with the FGR scaling $T^{-2}$. 
The observed deviation is mainly due to the drive frequency not being large enough, where the $\mathbb{Z}_2$-breaking term is of comparable absolute magnitude to the leading order term $\mathbb{Z}_2$-preserving term (cf.~Sec.~\ref{sec:FGR}); thus, the next leading-order term will affect the dynamics during the U(1) plateau, which can be observed in the early-time dynamics of both the parity operator and the participation entropy. 
We emphasize that since the parity operator is a non-local operator, its dynamics may not be described by FGR. However, the numerical data clearly shows that the $\mathbb{Z}_2$-plateau lifetime can be parametrically controlled by modulating the drive frequency.

\subsection{Finite size effects}
\label{sec:fin_size}

\begin{figure}[t]
    \centering
    \includegraphics[width=0.4\linewidth]{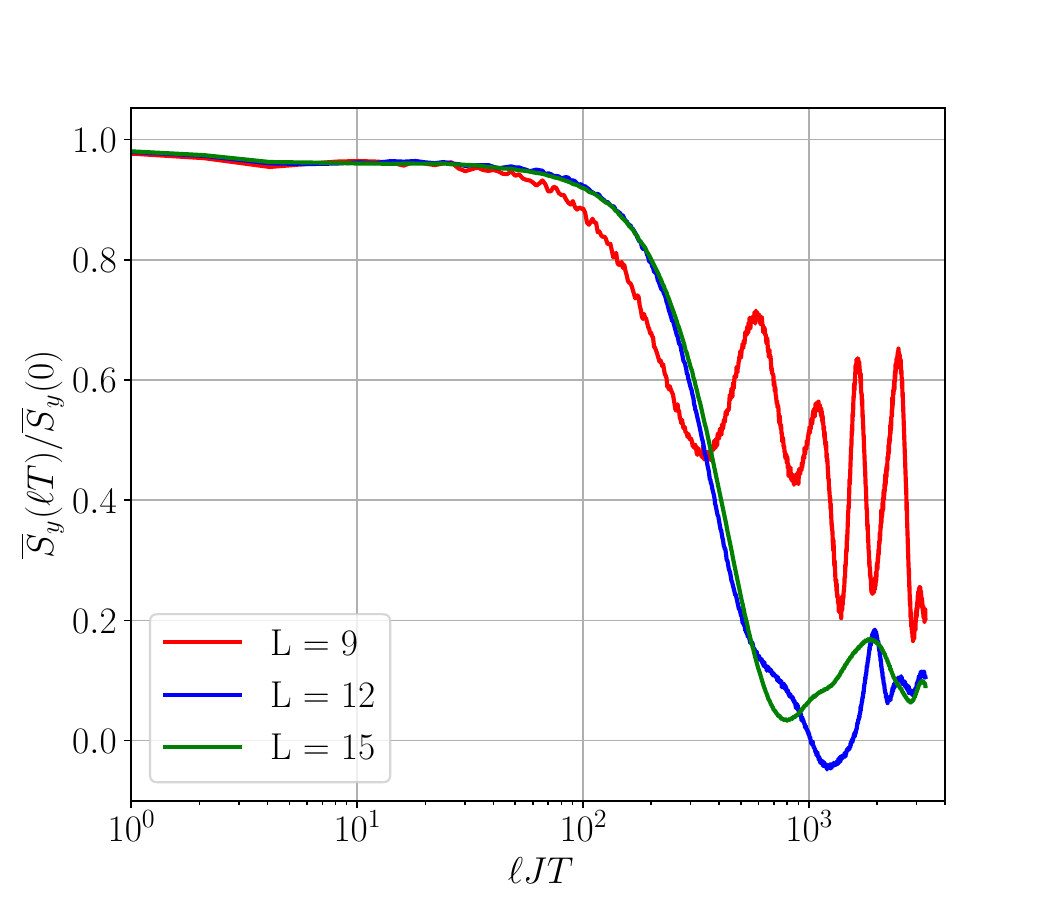}
    \caption{
    Numerical simulation of the 1D quantum spin chain model from Sec.~\ref{sec:SU2HSB} in the main text for different system sizes. We show the dynamics of the $\mathrm{SU(2)}$ order parameter $\overline{S}_y(\ell T)/\overline{S}_y(0)$. Finite-size effects are negligible for $L\ge12$ during the prethermal plateau; deviations are observable in the long-time thermal plateau value (which bears little relevance for the \HS phenomenon). The parameters are the same as in the main text.
    The initial state $|\psi(0)\rangle$ is a Haar random state in the $z$-magnetization sector containing $N_{\downarrow}{=}3,4,5$ down spins out of all $L{=}9,12,15$ sites respectively; it is then rotated around the $x$-axis by {$\prod_{i}e^{-i\pi/16\sigma_i^x}$} so that the initial magnetization density are same for each system size. The drive frequency is $1/(JT) = 10.0$.}
    \label{fig:finitesize}
\end{figure}

We expect that the prethermal lifetime, for a fixed driving period, should converge to a finite value in the thermodynamic limit. To verify this, we perform numerical simulations of quantum spin chains of different system sizes, although the simulations are inevitably limited to relatively small sizes by the exponential growth of the many-body Hilbert space. 
In Fig.~\ref{fig:finitesize} we plot the time evolution of the $y$-magnetization for $1/(JT)=10$ and different system sizes. The relaxation process depends on the system size, especially for small sizes, e.g., $L=9$. However, the time evolution of local observables starts converging already for $L\geq12$ for sufficiently long times, $lJT\approx 10^3$.
Finite size effects only become visible after this time scale~\cite{fleckenstein2021thermalization}, where the (normalized) order parameter already drops below $0.1$, which is smaller than the threshold values used to extract the prethermal lifetime in Fig.~\ref{fig:SU2}.

\section{\HS IN $Z_4$ QUANTUM CLOCK MODEL}
\label{sec:sm3}

In this appendix, we give the details of the effective Hamiltonian corresponding to the \HS ladder that implements the $\mathbb{Z}_4$ quantum clock model. 
Consider the two Hamiltonians $H_1$ and $H_0$ that preserve the $\mathbb{Z}_4$ and $\mathbb{Z}_2$ symmetry, respectively:
\begin{eqnarray}
        H_1 &=& \sum_{\langle i,j\rangle}J_{ij}\left(Z_i^2Z_j^2-\eta (e^{i\phi}Z_i^{\dagger}Z_j+\text{h.c.})\right)+ \sum_{i}h_{i}\left(Z_i^2-\frac{1}{2}\left(X_i+X_i^{\dagger}\right)\right) + \sum_{i}g_{i}X_i^2,\nonumber \\
        H_0 &=& \sum_{i=1}^L b_{i}(Z_i+Z_i^{\dagger}).
\end{eqnarray}
Following the driving protocol defined in Sec.~\ref{sec:4}, the effective Hamiltonian over four drive periods, $U_F^4 = e^{-iQ4T}$, is:
\begin{equation}
\begin{aligned}
& Q = Q^{(0)}+Q^{(1)}+Q^{(2)}+\mathcal{O}(T^3), \\
& Q^{(0)} = \frac{1}{2}\sum_{\langle i,j\rangle}J_{ij}\left(Z_i^2Z_j^2-\eta (e^{i\phi}Z_i^{\dagger}Z_j+e^{-i\phi}Z_iZ_j^{\dagger})\right)
+\frac{1}{2}\sum_{i}g_{i}X_i^2-\frac{1}{2}\sum_{i}h_{i}(X_i+X_i^{\dagger}),\\
& Q^{(1)} = \frac{iT}{16}\sum_{i} h_{i}^2[Z_i^2, X_i+X_i^{\dagger}],\\
& Q^{(2)} = \frac{T^2}{128}\left([Q^{(0)},[Q^{(0)}, \sum_{l}b_{l}((2-i)Z_l+(2+i)Z_l^{\dagger})]] + \frac{1}{2}[\sum_{l}h_{l}Z_l^2, [Q^{(0)}, \sum_{m}b_{m}((1-i)Z_m+(1+i)Z_m^{\dagger})]]\right).
\end{aligned}
\end{equation}
A Hamiltonian $\mathcal{O}$ is $Z_4$ and $Z_2$-preserving when $[\mathcal{O},\prod_{i}X_i] = 0$ and $[\mathcal{O},\prod_iX_i^2] = 0$. Clearly $[Q^{(0)}, \prod_{i}X_i]=0;\;[Q^{(1)}, \prod_{i}X_i]\ne0,\;[Q^{(1)}, \prod_{i}X_i^2]=0;\;[Q^{(2)}, \prod_{i}X_i]\ne0, \;[Q^{(2)}, \prod_{i}X_i^2]\ne0$, showing the $\mathbb{Z}_4\to\mathbb{Z}_2\to E$ HS.

\begin{figure}[t!]
	\centering
	\includegraphics[width= 1.0\linewidth]{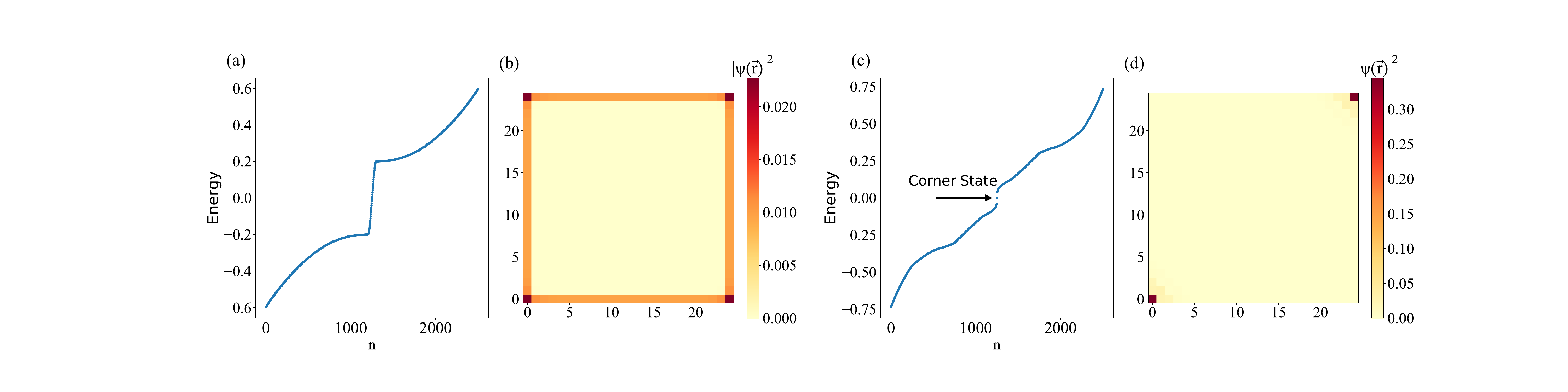}
	\caption{
 Energy spectrum and zero energy state of TI and HOTI. (a),(c) Energy spectrum of $Q^{(0)}$ and $Q^{(0)}+Q^{(1)}$ respectively. (b),(d) Density distribution corresponding to the zero energy mode of $Q^{(0)}$ and $Q^{(0)}+Q^{(1)}$ respectively (e.g., edge state and corner state). The parameters are $M = 1.0, t=1.0, \Delta_0 = 1.0, \Delta_1 = 7.0$, system size is $L=19$.}
\label{fig:5}
\end{figure}

\section{HIGH-ORDER TOPOLOGICAL INSULATOR FROM \HS}
\label{sec:HOTI}
\subsection{Hamiltonian in real space}
Here, we elaborate on the \HS setup implementing a change of topology from a TI to a HOTI. We give explicit expressions for the representation of the generating Hamiltonians in real space proposed in the main text for a possible experimental realization.
\begin{eqnarray}
     H_2 &=& \frac{M}{2}\sum_{\vec{r}, \alpha = 0,1}(-1)^{\alpha}c^{\dagger}_{\vec{r}, \alpha}c_{\vec{r},\alpha}
    +\sum_{\vec{r},\alpha = 0,1}\sum_{j = x,y}\left(\frac{J}{2}(-1)^{\alpha}c^{\dagger}_{\vec{r}+\vec{e}_j, \alpha}c_{\vec{r}, \alpha}
    +\frac{\Delta_0}{2i}c^{\dagger}_{\vec{r}+\vec{e}_j,\alpha+1}\sigma_j c_{\vec{r},\alpha}\right) + \text{h.c.}, 
    \nonumber\\
    H_1 &=& \Delta_1\sum_{\vec{r},\alpha=0,1}(-1)^{\alpha}c^{\dagger}_{\vec{r},\alpha}(\sigma_x+\sigma_y)c_{\vec{r},\alpha}, \nonumber\\
    H_1' &=& \Delta_1\sum_{\vec{r},\alpha=0,1}(-1)^{\alpha}c^{\dagger}_{\vec{r},\alpha}\sigma_zc_{\vec{r},\alpha},\nonumber\\
    H_0 &=& \Delta_2\sum_{\vec{r},\alpha = 0,1}c_{\vec{r},\alpha+1}^{\dagger}\sigma_yc_{\vec{r},\alpha},
\end{eqnarray}
where $c_{\vec{r},\alpha} = (c_{\vec{r},\alpha,\uparrow}, c_{\vec{r}, \alpha, \downarrow})$. We can also convert the fermionic orbital as well as the spin degrees of freedom equivalently into four sites in each unit cell $(c_{\vec{r},0,\uparrow},c_{\vec{r},0,\downarrow},c_{\vec{r},1,\uparrow}, c_{\vec{r},1,\downarrow})\to(c_{\vec{r},0},c_{\vec{r},1},c_{\vec{r},2}, c_{\vec{r},3})$. The protocol is thus also possibly realizable in ultra-cold atomic systems.

\begin{figure}[t!]
	\centering
	\includegraphics[width= 0.7\linewidth]{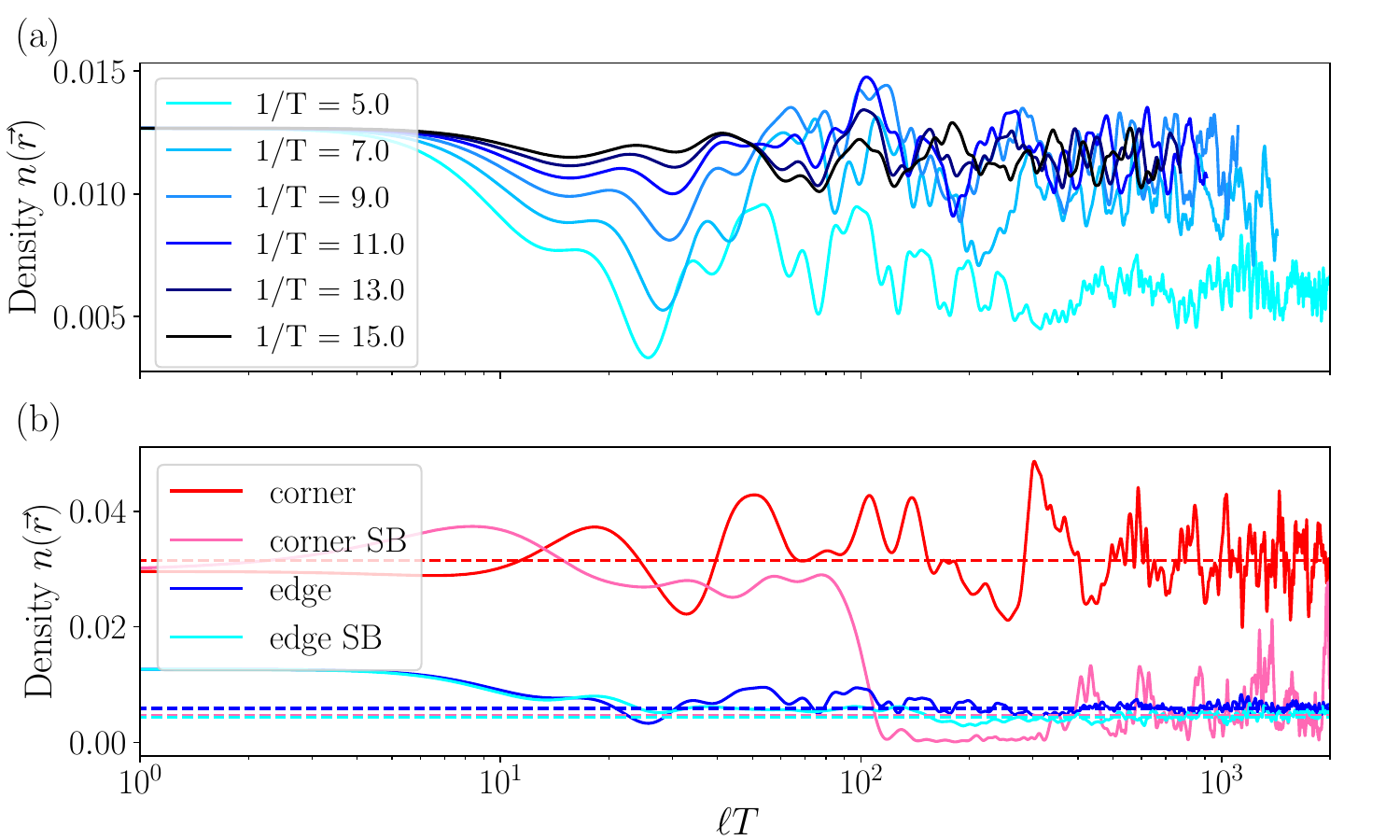}
	\caption{HOTI results when the initial state is the zero-energy eigenstate of $Q^{(0)}$ (i.e., edge state). (a) Time evolution of electron density at the edge without $\mathcal{I}$-breaking perturbation for different drive frequency $1/T$. The lifetime of the boundary state (i.e., the first plateau) is significantly prolonged with increasing driving frequency.(b) Time evolution of the electron density at corner and edge ($n_\text{corner} =(n{(0,0)}+n{(L,L)})/2, \; n_\text{edge} = 2\sum_{i = L/4}^{3L/4}n{(i,0)}/L$) without $\mathcal{I}$-breaking perturbation in $Q^{(1)}$ (marked by 'corner' and 'edge' labels in legend) and with $\mathcal{I}$-breaking perturbation ('corner SB' and 'edge SB' labels). When $\mathcal{I}$ is preserved in the first order effective Hamiltonian, density distribution at corners has a longer lifetime showing the existence of corner state.The system size is $L=19$ and the parameters are $M/J = 1.0$, $\Delta_0/J = 1.0$, $\Delta_1/J = 7.0$, and $\Delta_2/J = 12.0$. 
 }
\label{fig:HOTI_SM}
\end{figure}

With the evolution operator defined in Eq.~\eqref{eq:UF_HOTI}, the effective Hamiltonian $U_F\equiv e^{-iQT}$ is 
\begin{eqnarray}
    Q &=& Q^{(0)} + Q^{(1)} + \mathcal{O}(T^2),\ \ Q^{(0)} = \frac{1}{5}H_2,\ \ Q^{(1)} = -\frac{iT}{200}([H_1,H_1']+2[H_1+H_1', H_2]).
\end{eqnarray}
$Q^{(0)}\propto H_2$ is a standard Hamiltonian for a TI. As discussed in Appendix.~\ref{sec.manipulate}, the commutator of $H_1$ and $H_1'$ in $Q^{(1)}$ introduces an on-site $\mathcal{T}$-breaking but $\mathcal{I}$-preserving perturbation which opens the energy gap and leaves two degenerate zero-energy corner states. 
In Fig.~\ref{fig:5} we present the energy spectrum and the density distribution of the zero-energy eigenstates of $Q^{(0)}$ and $Q^{(0)}+Q^{(1)}$, respectively. We can clearly see the boundary states that characterize these two topological states of matter.

\subsection{Prethermal plateaus of the edge modes}
In the main text, we initialize the system as a product state with a large spatial support on the edge. Such a state can be experimentally prepared, for instance, by using cold atoms in optical lattices. We expect to observe a plateau of the particle density around the edge, which should be more pronounced by increasing the driving frequency, such that $Q^{(0)}$ dominates the early time evolution. However, technically, it is difficult to show a clear plateau because this product state can delocalize quickly into the bulk, and the remaining population around the edge is comparably weak, even in the absence of the first-order perturbation $Q^{(1)}$. To confirm the existence of this plateau, here we supply numerical simulation of the dynamics by starting from the zero-energy eigenstate of $Q^{(0)}$ (i.e., the edge states), such that only higher-order perturbations delocalizes the system.
As shown in Fig.~\ref{fig:HOTI_SM}(a), as we increase the drive frequency, the decay of the electron density at the boundary slows down accordingly, revealing the first plateau of $Q^{(0)}$. In Fig.~\ref{fig:HOTI_SM}.(b), we show the time evolution of the electron density at the corners and boundaries with and without an $\mathcal{I}$-breaking perturbation. In the presence of the perturbation, labeled by `corner SB' and `edge SB', the density distribution on the corners decays quickly and agrees with the density distribution on the edges at long times. 
However, without the $\mathcal{I}$-breaking perturbation, the lifetime of the density distribution at the corners is visibly prolonged by an order of magnitude.

\subsection{Topological invariants for high-order topological insulators}

We note that in the prototypical high-order topological insulator -- the Benalcazar–Bernevig–Hughes (BBH) model  
~\cite{benalcazar2017quantized} -- the HOTI property is characterized by a nested Wilson loop and the corresponding quantized polarization. By contrast, our protocol gives rise to a HOTI which is a fragile topological insulator with corner filling anomaly~\cite{ahn2019failure, benalcazar2019quantization} protected by inversion symmetry $\mathcal{I}$ but with very robust and localized corner charges; the topology of our system is captured by the expectation values of the symmetry indicators of the occupied bands~\cite{khalaf2018symmetry, schindler2022topological}
\begin{equation}
    C_B\!\!\!\! \mod 2 = \frac{1}{2}[1-\xi(\Gamma)\xi(X)\xi(Y)\xi(M)],
\end{equation}
where $\Gamma=(0,0), X=(0,\pi), Y=(\pi,0), M=(\pi, \pi)$ are layer-group points of time reversal invariant momentum (TRIM), and $\xi(\cdot)$ is the expectation value of parity $\mathcal{I} = \tau_z\sigma_0$. 
Since the expectation values of the parity of the two occupied bands share the same sign, we define $\xi(\mathbf{k}) = \text{sgn}\left(\langle \mathcal{I}\rangle_{1,\mathbf{k}}\right)
\langle \mathcal{I}\rangle_{1,\mathbf{k}}\langle \mathcal{I}\rangle_{2,\mathbf{k}}$, where $1,2$ label the occupied bands.
When $\mathcal{I}$ is not broken, $\xi(\mathbf{k})$ represents the sign of inversion eigenvalues of occupied bands at each momentum $\mathbf{k}$; if there are an odd number of minus signs at the TRIM points, an $\mathcal{I}$-symmetric and $\mathcal{T}$-broken 2D insulator will have anomalous corner charges, according to Refs.~\cite{benalcazar2019quantization, schindler2022topological}.

\begin{figure}[t!]
	\centering
	\includegraphics[width= 0.5\linewidth]{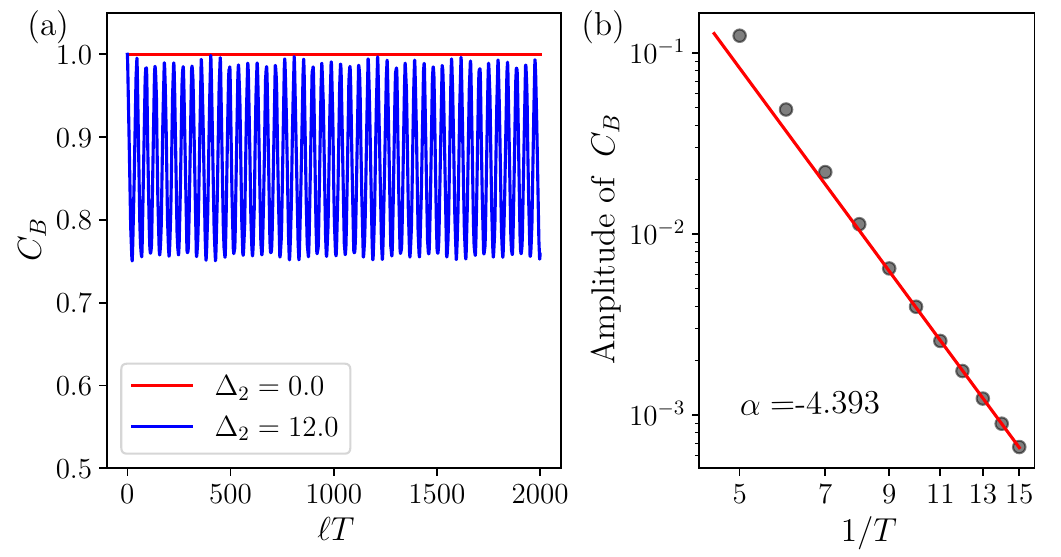}
	\caption{
    Dynamical topological invariant of HOTI from HS. 
    (a) Time evolution of the topological invariant $C_B$ without $\mathcal{I}$-breaking perturbation (red line) and with $\mathcal{I}$-breaking perturbation (blue line). If present, interactions will eventually cause the system to heat up to an infinite temperature state with a vanishing expectation value of $\mathcal{I}$, and $C_B$ will settle to the value $1/2$ [not shown]. The driving period is $JT = 0.2$.
    (b) The amplitude of the $C_B$ oscillations exhibits a power-law scaling with the drive frequency, $A\propto T^{-\alpha}$, with $\alpha \approx -4$, in accord with our perturbation theory analysis. 
    The parameters are $M/J = 1.0$, $\Delta_0/J = 1.0$ and $\Delta_1/J = 7.0$. We performed the simulation in momentum space using the four TRIM points [see text].
}
\label{fig:HOTI_SM_2}
\end{figure}

In practice, we prepare the initial state as the half-filling ground state of $H_2$ and we use the evolution operator $U_F$ defined in Eq.~\eqref{eq:UF_HOTI} to evolve it; we then measure the inversion operator $\mathcal{I}$ in the evolved state. As shown in Fig.~\ref{fig:HOTI_SM_2}(a), when there is no $\mathcal{I}$-breaking term in the evolution operator ($\Delta_2 = 0.0$, red line), $C_B$ remains equal to unity, revealing the topological property of the HOTI. When the $\mathcal{I}$-breaking term is introduced ($\Delta_2 = 12.0$, blue line), $C_B$ instead starts oscillating close to unity; in that case, we can still see a part of the fermion density localized to the corner for long times, as shown in Fig.~\ref{fig:6} and Fig.~\ref{fig:HOTI_SM}. In Fig.~\ref{fig:HOTI_SM_2}(b), we show the dependence of the oscillation amplitude of $C_B$ on the driving frequency, $A\propto T^{-\alpha}$, featuring the power-law exponent $\alpha\approx-4$, consistent with our perturbative analysis.

\section{ABSENCE OF SYMMETRY IN INSTANTANOUS
HAMILTONIANS }
\label{sec:Ap_F}

In the main text, Sec.~\ref{sec:2}, we propose a recursive construction for realizing arbitrary symmetry ladders. Although the overall Floquet operator does not preserve any symmetry, some generators $H_n$ in the protocol may preserve a higher-level symmetry. This can be challenging to realize in experimental implementations. In fact, preserving higher-level symmetries is not necessary for engineering a symmetry ladder, and in Sec.~\ref{sec:Exp} we illustrated one possible example where the instantaneous Hamiltonian does not preserves the highest SU(2) symmetry. Here, we provide more examples to demonstrate the existence of symmetry ladder when the instantaneous Hamiltonian preserves no symmetry. These protocols may be easier to realize experimentally compared to the general recursive protocol.

\subsection{Stepwise drive }

On some quantum simulators platforms, one may only have direct access to Hamiltonians that can be divided into several symmetric parts
\begin{equation}
    H = H_n + H_{n-1} + \cdots + H_{0},
\end{equation}
where $H_q$ preserves the symmetry group $G_q$ for $q=0,\dots, n$. Hence, the full Hamiltonian $H$ only preserves the lowest order symmetry group $G_0$. For this kind of system, if we are allowed to flip some of the signs in front of each Hamiltonian, for instance, by applying single-site $\pi/2$ gates, one can
realize a sequence of hierarchical
symmetries.

For the symmetry ladder $G_2\supset G_1 \supset G_0$, one possible driving protocol is
\begin{equation}
    U_F\equiv e^{-iTQ_2} = U\left(H_2+H_1+H_0, H_2-H_1-H_0, H_2+H_1-H_0, H_2-H_1+H_0\Bigg|\frac{T}{4}\right).
\end{equation}
The first two orders of the corresponding effective Hamiltonian, $Q_2=Q_2^{(0)}+Q_2^{(1)}+Q_2^{(2)}+\cdots$, are
\begin{eqnarray}
    Q_2^{(0)} &\propto& H_2,\nonumber\\
    Q_2^{(1)} &\propto& iT[H_2,H_1],\nonumber\\
    Q_2^{(2)} &\propto& \text{terms containing $H_0$}.
\end{eqnarray}

In addition, we can also realize the symmetry ladder $G_3\supset G_2 \supset G_1 \supset G_0$ with more restrictions on Hamiltonian.
The driving protocol is 
\begin{eqnarray}
  U_F\equiv e^{-iTQ_3} = U(H_3+H_2+H_1+H_0, H_3-H_2-H_1+H_0, H_3+H_2-H_1-H_0,H_3-H_2+H_1-H_0,\nonumber\\
  H_3+H_2+H_1-H_0, H_3-H_2-H_1-H_0, H_3+H_2-H_1+H_0, H_3-H_2+H_1+H_0|\frac{T}{8}).
  \label{eq:F4}
\end{eqnarray}
The first three orders of the corresponding effective Hamiltonian $Q_3=Q_3^{(0)}+Q_3^{(1)}+Q_3^{(2)}+Q_3^{(3)}\cdots$, in this ladder are
\begin{eqnarray}
    Q_3^{(0)} &\propto& H_3, \nonumber\\
    Q_3^{(1)} &\propto& iT[H_3,H_2], \nonumber\\
    Q_3^{(2)} &\propto& \text{terms not containing $H_0$} + T^2(\lambda [H_0,[H_1, H_0]] + \lambda'[H_3, [H_1, H_0]] + \lambda''[H_3, [H_3, H_0]]), \nonumber\\
    Q_3^{(3)} &\propto& \text{terms containing $H_0$}.
\end{eqnarray}
In order for $Q_3^{(2)}$ to preserve the symmetry $G_1$, two conditions need to be satisfied by the generator Hamiltonians:
\begin{equation}
    [H_1, H_0] = 0, \qquad [H_3, H_0] = 0.
\end{equation}
This can be satisfied, for instance, by the spin Hamiltonians $H_3 = \sum_{i,j}J(\sigma_i^x\sigma_j^x+\sigma_i^y\sigma_j^y+\sigma_i^z\sigma_j^z)$, $H_2 = \sum_{i,j}J'(\sigma_i^x\sigma_j^x+\sigma_i^y\sigma_j^y)$, $H_1 = \sum_{i,j}J''\sigma_i^x\sigma_j^x$, preserving $\mathrm{SU(2)}, \mathrm{U(1)}$ and $\mathbb{Z}_2$, respectively, and finally $H_0 = \sum_ih\sigma_i^x$. They can be used to realize the symmetry ladder $\mathrm{SU(2)}\to \mathrm{U(1)}\to \mathbb{Z}_2\to \mathrm{E}$ in the dynamics.
To appreciate the usefulness of the above expressions, consider a quantum simulator where we only have access to 
(i) evolution generated by a single $H=H_3+H_2+H_1+H_0$ which, say, corresponds to an XYZ model with a transverse field, and
(ii) the ability to change the sign of the different coupling strengths of $H_i$ 
Then we can still imprint \HS structure in the dynamics using Eq.~\ref{eq:F4}.

\subsection{Continuous drives with a few harmonics}

Here we also give examples with continuous driving protocols where a few harmonic drives are involved. This can be particularly useful for analog quantum simulators, where stepwise drive may be difficult to realize.

For the symmetry ladder $G_2\supset G_1\supset G_0$, we consider Hamiltonians $H_2,\; H_1,\;H_0$ which preserve the corresponding symmetries. Our continuous driving protocol is 
\begin{equation}
    U_F \equiv e^{-iTQ_2}= \mathcal{T}e^{-i\int_0^TH(t)dt},\; H(t) = H_2+H_1\sin(\Omega t)+H_0\cos(p\Omega t), 
\end{equation}
where $p\in \mathbb{N},\; p>1$ and the driving period is $T = 2\pi/\Omega$. The instantaneous Hamiltonian $H(t)$ only preserves the lowest-order symmetry $G_0$ whenever the prefactor of $H_0$ does not vanish.
The first two orders of the corresponding effective Hamiltonian $Q_2=Q_2^{(0)}+Q_2^{(1)}+Q_2^{(2)}+\cdots$ are
\begin{eqnarray}
    Q_2^{(0)} &\propto& H_2,\nonumber\\
    Q_2^{(1)} &\propto& iT[H_2,H_1],\nonumber\\
    Q_2^{(2)} &\propto& \text{terms containing $H_0$},
\end{eqnarray}
and hence $H_0$ only appears in $Q_2^{(2)}$ and higher orders.

For the more complex symmetry ladder $G_3\supset G_2 \supset G_1 \supset G_0$, we also need some restrictions on the Hamiltonian. For this case, we have found the driving protocol
\begin{equation}
    U_F \equiv e^{-iTQ_3}= \mathcal{T}e^{-i\int_0^TH(t)dt},\; H(t) = H_3+H_2\sin(\Omega t)+H_1\cos(2\Omega t)+H_0\cos(p\Omega t), 
\end{equation}
where this time $p\in \mathbb{N}, p> 4$. One can check that the first three orders of the corresponding effective Hamiltonian read as
\begin{eqnarray}
    Q_3^{(0)} &\propto& H_3, \nonumber\\
    Q_3^{(1)} &\propto& iT[H_3,H_2], \nonumber\\
    Q_3^{(2)} &\propto& \text{terms not containing $H_0$}+T^2(-\frac{9}{2(p^2-4)}[H_3,[H_0, H_1]]+\frac{4}{p^2}[H_0,[H_0, H_3]]+\frac{6}{p^2}[H_3,[H_3,H_0]]), \nonumber\\
    Q_3^{(3)} &\propto& \text{terms containing $H_0$}.
\end{eqnarray}
Similar to the former protocol, in order for $Q_3^{(2)}$ to preserve the symmetry $G_1$, the following two conditions need to be satisfied
\begin{equation}
    [H_1, H_0] = 0, \quad [H_3, H_0] = 0.
\end{equation} 

The above results demonstrate that the construction of \HS is not limited to step drives with symmetric instantaneous Hamiltonians. It is also accessible for experiments that can drive the system with continuous few-harmonic drives. Thus, the \HS toolbox we propose is versatile enough to suit both experimental platforms where step-like drives are native, as well as those where harmonic controls are preferred.

\section{POSSIBLE EXPERIMENTAL IMPLEMENTATION WITH GAUSSIAN PULSES}
\label{sec:Ap_exp}

\begin{figure}[t!]
	\centering
	\includegraphics[width= 0.6\linewidth]{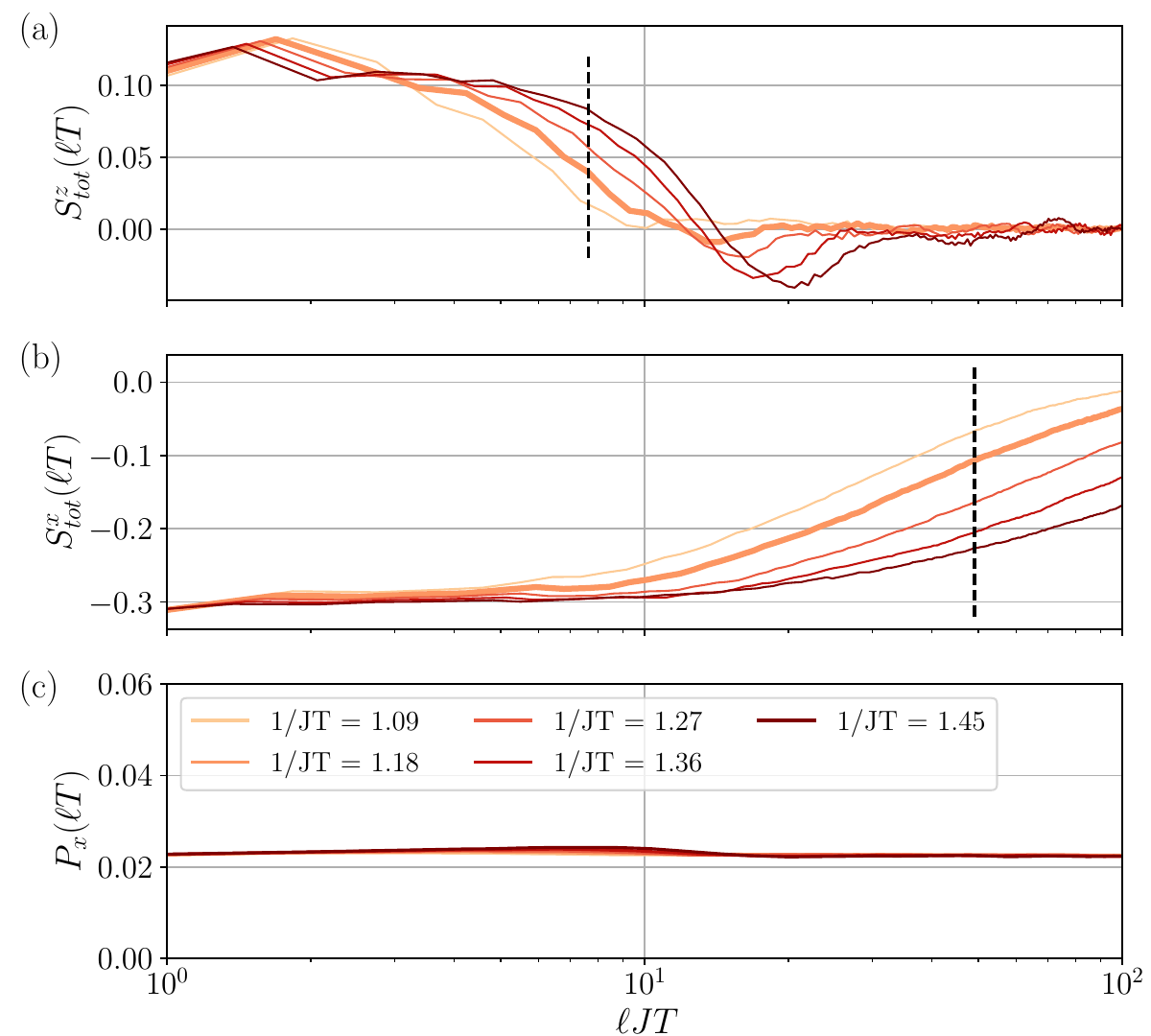}
	\caption{
Dynamical detection of $\text{SU(2)}\to\text{U(1)} \to \mathbb{Z}_2$ \HS with Gaussian drives.
(a),(b),(c) Dynamics of order parameters for the hierarchical quasi-conservation laws $\mathrm{SU(2)}: S^z_\text{tot}(\ell T) = \frac{1}{L}\langle\psi(\ell T)|\sum_i\sigma_i^z|\psi(\ell T)\rangle$, $\mathrm{U(1)}: S^x_\text{tot}(\ell T) = \frac{1}{L}\langle\psi(\ell T)|\sum_i\sigma_i^x|\psi(\ell T)\rangle$, $\mathbb{Z}_2: P_x(\ell T) = \frac{1}{L}\langle\psi(\ell T)|\prod_i\sigma_i^x|\psi(\ell T)\rangle$. Different lifetimes suggest that \HS emerges at different time scales. A hierarchy in lifetimes is clearly visible for moderate driving frequencies, e.g., $1/T{=}1.18$ (marked with a thick line). In (a) and (b), the vertical dash line corresponds to the evolve time to reach $e^{-1}S^z_{\mathrm{tot}}(0)$ and $e^{-1}S^x_{\mathrm{tot}}(0)$ respectively with driving frequency $1/JT=1.18$, which are the lifetime of SU(2) and U(1) plateau. We use $h {=} 90$ as the strength of the staggered magnetic field and the ratio of the pulse width to the drive period is $u {=} 0.1$. The system size is $L {=} 18$.}
 
\label{fig:prot3g}
\end{figure}

In Sec.~\ref{sec:Exp} we show that \HS can appear when the $\pi/4$-gate has a finite pulse width. There, to generate these gates we considered a strong uniform field over all sites of constant field strength during a short time window. 
In practice, the shape of the pulse is indeed tunable and 
 Gaussian pulses are one of the common choices. Here we will show that with  Gaussian pulses, \HS can also be clearly observed, as long as the pulse width is sufficiently small.
 
The Gaussian pulses have the following time-dependent form
\begin{equation}
    g(t) = \frac{\pi}{\sqrt{2\pi} \epsilon \tau}\exp\left({-8\frac{t^2}{\epsilon^2\tau^2}}\right),\qquad 
    f(h,t) = he^{-8\frac{t^2}{\epsilon^2\tau^2}}.
\end{equation}
To efficiently simulate the dynamics of the system numerically, we truncate the pulse beyond $t = \pm u\tau/2 ,\; u=2\epsilon$. Therefore, the evolution operator is
\begin{eqnarray}
    U_F &=& U_0P_x^+U_0U_x^+P_y^-U_0U_0P_y^+U_x^-U_0P_x^-U_0,\; U_0 \equiv  e^{-iH_0\tau}, \nonumber\\
    P_x^{\pm} &\equiv& \mathcal{T}\exp(-i\int_{-u\tau/2}^{u\tau/2}dt \left( H_0\pm g(t)\sum_i\sigma_i^x\right),\nonumber\\
    P_y^{\pm} &\equiv& \mathcal{T}\exp(-i\int_{-u\tau/2}^{u\tau/2}dt \left(H_0\pm g(t)\sum_i\sigma_i^y\right),\nonumber\\
    U_x^{\pm} &\equiv& \mathcal{T}\exp(-i\int_{-u\tau/2}^{u\tau/2}dt \left(H_0\pm f(h,t)H_x\right),
\end{eqnarray}
where $H_0{=}J\sum_{i}\sigma_i^x\sigma_{i+1}^x+\sigma_i^y\sigma_{i+1}^y$, $H_x = J\sum_i(-1)^i\sigma_i^x$
The driving period is $T = 6(1+u)\tau$ and $hu\sim \mathcal{O}(1)$ such that effectively the strength of $H_x$ 
is independent of $\epsilon$. Using this driving protocol we simulate the dynamics of a 1D spin chain with $L = 18$ sites. We use the same initial state $|\psi(0)\rangle$ as in Sec.~\ref{sec:Exp}, and measure the same order parameters to show that hierarchical symmetries persist even when using Gaussian pulses. As shown in Fig.~\ref{fig:prot3g}, we start with a relatively high-temperature state and within $\mathcal{O}(10^2)$ driving periods $\mathrm{SU(2)}$ and $\mathrm{U(1)}$ symmetries are hierarchically broken for moderate driving frequencies. Only a few tens of cycles are needed to observe this phenomenon, which should be already reachable on the state-of-the-art quantum simulators.

\end{document}